\documentclass[onecolumn,journal,romanappendices,11pt]{IEEEtran}

\usepackage[T1]{fontenc}		% Encoding
\usepackage[utf8]{inputenc}	% Write input

\usepackage{amsthm}
\usepackage{amsmath}
\usepackage{amssymb}
\usepackage{bm}
\usepackage{nicefrac}
\usepackage{xcolor}
\usepackage{relsize}
\usepackage{tikz}
\usepackage{subcaption}
\usetikzlibrary{decorations.pathreplacing}
\usetikzlibrary{arrows}
\usetikzlibrary{shapes}
\usetikzlibrary{arrows.meta}

\newcommand{\Ld}{\dot{{\bm{L}}}}
\newcommand{\Ldb}{\dot{{\bar{\bm{L}}}}}
\newcommand{\xd}{\dot{{\bm{x}}}}

\DeclareMathOperator{\DoF}{DoF}
\DeclareMathOperator{\Conv}{Conv}
\DeclareMathOperator{\round}{round}
\DeclareMathOperator{\sgn}{sgn}

\newcommand{\bL}{\bar{L}}

\newtheorem{thm}{Theorem}%[section]
\newtheorem{lem}{Lemma}%[section]
\newtheorem{prop}{Proposition}
\newtheorem{cor}{Corollary}

\newtheorem{exmp}{Example}%[section]
\newtheorem{property}{Property}

\newtheorem{rem}{Remark}
\newtheorem{obs}{Observation}
\theoremstyle{definition}
\newtheorem{defn}{Definition}%[section]
\newtheorem{fact}{Fact}

\newcommand{\dfn}{\triangleq}
\newcommand{\gamm}[1]{\mathlarger{\gamma}_{\mbox{\tiny $#1$}}}

\newcommand{\Up}{{\bm{L}}}

\newcommand{\Cb}[2]{C^{#1}_{#2}}
\newcommand{\Dwc}{\mathcal{D}_{wc}}
\newcommand{\RomanNumeralCaps}[1]
    {\MakeUppercase{\romannumeral #1}}
\def\Z{\bm{\mathcal{Z}}}

\def\Oplus{\mathlarger{\mathlarger{\oplus}}}
\DeclareMathOperator*{\argmin}{argmin}
\DeclareMathOperator*{\argmax}{argmax}

\newcommand{\CD}[0]{{\mathcal{D}}}

\newcommand{\CF}[0]{{\mathcal{F}}}

\newcommand{\CL}[0]{{\mathcal{L}}}

\newcommand{\CN}[0]{{\mathcal{N}}}

\newcommand{\CQ}[0]{{\mathcal{Q}}}
\newcommand{\CR}[0]{{\mathcal{R}}}
\newcommand{\CS}[0]{{\mathcal{S}}}
\newcommand{\CT}[0]{{\mathcal{T}}}
\newcommand{\CU}[0]{{\mathcal{U}}}

\newcommand{\CW}[0]{{\mathcal{W}}}
\newcommand{\CX}[0]{{\mathcal{X}}}

\newcommand{\CZ}[0]{{\mathcal{Z}}}

\newcommand{\Bd}[0]{{\mathbf{d}}}

\newcommand{\BL}[0]{{\mathbf{L}}}

\usepackage{hyperref}
\hypersetup{
	colorlinks = true,
	linkcolor  = blue!60!black,
	citecolor  = green!50!black,
	urlcolor   = black
}
\begin{document}
%%%%%%%%%%%%%% %%%%%%%%%%%%%% %%%%%%%%%%%%%% %%%%%%%%%%%%%% %%%%%%%%%%%%%% %%%%%%%%%%%%%% 
%%%%%%%%%%%%%% %%%%%%%%%%%%%% %%%%%%%%%%%%%% %%%%%%%%%%%%%% %%%%%%%%%%%%%% %%%%%%%%%%%%%% 

\title{{Fundamental Limits of Topology-Aware Shared-Cache Networks}}

\author{Emanuele~Parrinello,~\IEEEmembership{Student Member,~IEEE,}
        Antonio~Bazco-Nogueras,~\IEEEmembership{Member,~IEEE,}
        and~Petros~Elia,~\IEEEmembership{Member,~IEEE}% <-this % stops a space
\thanks{E. Parrinello and P. Elia are with the Communication Systems Department, EURECOM, France, e-mail: \{emanuele.parrinello; petros.elia\}@eurecom.fr}
\thanks{A. Bazco-Nogueras is with the IMDEA Networks Institute, Madrid, Spain, e-mail: antonio.bazco@imdea.org}% <-this % stops a space
\thanks{This work was supported by the European Research Council under the EU Horizon 2020 research and innovation program/ERC grant agreement no. 725929 (ERC project DUALITY). 
The work of A. Bazco-Nogueras was supported by the Regional Government of Madrid through the grant 2020-T2/TIC-20710 for Talent Attraction. 
This work was presented in part at the 2019 IEEE Information Theory Workshop (ITW)~\cite{ParEliaITW19}.
}}% <-this % stops a space

\maketitle

\begin{abstract}
    This work studies a well-known shared-cache coded caching scenario where each cache can serve an arbitrary number of users, analyzing the case where there is some knowledge about such number of users (i.e., the topology) during the content placement phase. 
Under the assumption of regular placement and a cumulative cache size that can be optimized across the different caches, we derive the fundamental limits of performance by introducing a novel cache-size optimization and placement scheme and a novel information-theoretic converse. The converse employs new index coding techniques to bypass traditional uniformity requirements,  thus finely capturing the heterogeneity of the problem, and it provides a new approach to handle asymmetric settings. 
The new fundamental limits reveal that heterogeneous topologies can in fact outperform their homogeneous counterparts where each cache is associated to an equal number of users. 
These results are extended to capture the scenario of topological uncertainty where the perceived/estimated topology does not match the true network topology. This scenario is further elevated to the stochastic setting where the user-to-cache association is random and unknown, and it is shown that the proposed scheme is robust to such noisy or inexact knowledge on the topology. 

\end{abstract}

\IEEEpeerreviewmaketitle

\section{Introduction}
\IEEEPARstart{C}{oded} caching is a communications technique proposed by Maddah-Ali and Niesen (MAN) in~\cite{MN14} which exploits cached content in order to reduce congestion in communication networks. This work in~\cite{MN14} revealed that a careful placement of content at the caches that are locally available to the users can substantially boost content delivery rates, by allowing for multicasting opportunities that enable the transmitter to serve multiple users at the same time through a shared link.
Specifically, the work in~\cite{MN14} considered a noiseless shared-link broadcast channel (BC) where a server, with access to a library of $N$ files, aims to communicate with $K$ users that are each equipped with a cache that can store a fraction $\gamma\in[0,1]$ of the library. The system consists of two distinct phases; a \emph{cache placement phase} during which --- without knowledge of the future requests of the users  --- portions of the files of the library are pre-stored at the users' cache, and a \emph{delivery phase} in which the users' demands are served.
A subdivision of the library files in many sub-files and a meticulous placement of these sub-files at the users' caches allow the server to simultaneously transmit to $K\gamma+1$ users during the delivery phase. This factor of $K\gamma+1$ describes the speedup in delivery rates due to \emph{coded} caching, and it is commonly referred to as the \emph{coding gain}, or as the \emph{global caching gain}, and it matches the well known \emph{Degrees-of-Freedom} (DoF). This DoF --- which was shown to be information-theoretically optimal within a gap of 2 in~\cite{YMA19IT} and exactly optimal under the assumption of uncoded cache placement in~\cite{WanTP15,YuMA16} --- scales with the cumulative cache capacity of the network~$K\gamma$.  

Since the introduction of coded caching, various works have extended the original MAN approach to various interesting scenarios. For example, the works in~\cite{NiesenMtit17Popularity,ZhangLW:18} addressed the average performance under the assumption of a non-uniform popularity distribution of the library files, while the work in~\cite{MND13} explored a \emph{decentralized} scenario where, during cache placement, the server is not aware of the number and identity of the users that will be present during the delivery phase. Similar concepts were also later considered  in~\cite{jin2019new,ZhangTaoLettershort}. Another interesting work can be found in~\cite{JiCM16D2D}, which extended the shared-link BC in~\cite{MN14} to the device-to-device (D2D) setting where users exchange messages in a peer-to-peer fashion to satisfy their requests. Furthermore, advances in coded caching proved to be also applicable in distributed computing through the development of coded distributed computing \cite{LiAliAvestimehrComputIT18,Lee2018,PLE:18a,reisizadeh2019coded, Ozfatura2022}.

An interesting direction included the study of coded-caching involving a server having multiple ($C$) transmitting antennas~\cite{ShariatpanahiMK16it} or equivalently the decentralized scenario with multiple transmitting servers~\cite{naderializadeh2017fundamental}. In both cases, the derived DoF of $K\gamma+C$ was shown  to be order optimal (cf.~\cite{naderializadeh2017fundamental}) and later to be exactly optimal (cf.~\cite{lampiris2018resolving}) under the assumption of linear one-shot schemes. Subsequently, finite-SNR studies of the multi-antenna setting can be found in~\cite{SharCaireKhal18IT,tolli2017multi,Zhao2021_asilomar,Zhao2022vectorCC}, the interplay of caching and Channel State Information (CSI) has been also analyzed in~\cite{ZE:17tit,Bazco2021_ITW}, as well as the impact of cache-less users~\cite{cachelessTIT} or uncacheable traffic~\cite{HamdiEleftTIT2021}. All the above aforementioned works, and in general most research on coded caching, focuses on the setting where each user is aided by its own dedicated cache. 
Recently though, the \emph{shared caches} paradigm has emerged as a much more realistic as well as powerful alternative to the traditional dedicated caches scenario. While in the latter scenario each user can have its own dedicated cache that can be drawn/filled independently of the other caches, this new shared caches approach asks that each cache can serve --- as argued below --- multiple users at the same time. As we will see, this shared caches approach captures several scenarios of interest.

\subsection{The shared cache model for coded caching}
The information theoretic study of the shared-cache coded caching scenario is strongly motivated by practical considerations and recent trends (see~\cite{WangIEEEAccessMobileEdgeNet17} and references therein). Moreover, it directly links our ability to exploit caching at the radio access network to a new ability to employ low-priced storage units at %an abundance of 
newly deployed macro and micro base stations,  each serving various sets of users. In our context, the central server represents a macro base station (MBS), whereas each cache represents a cache-aided micro base station (SBS) that can serve its nearby users at data rates that are significantly higher than those from a distant MBS.

The first extensive information-theoretic study that focuses on this shared-cache model can be found in~\cite{PUETIT2020}, which considered the scenario where each user has access to one of several caches at zero prefetching cost, and where each cache's occupancy (i.e., the number of users associated to each cache) can be arbitrary but \emph{unknown} during the cache placement phase. For this general setting, the work in~\cite{PUETIT2020} characterized the corresponding fundamental limits by deriving the exact optimal (under uncoded cache placement) worst-case normalized delivery time. An interesting subsequent work can be found in~\cite{KaratRajanISIT19}, which extends the setting in~\cite{PUETIT2020} to account for error-prone links. This work in~\cite{KaratRajanISIT19} proposes a scheme that employs the "leaders" approach from \cite{MND13} to extend the scheme in~\cite{PUETIT2020} to account for the scenario where multiple users may request the same file. Furthermore, the work in~\cite{Ibrahim2021tit} considered the shared-caches scenario with coded placement, which showed that the benefit of coded placement increases as the asymmetry in the number of users per cache increases, while a decentralized version of the scheme in~\cite{PUETIT2020} was proposed in~\cite{DuttaThomasComLett2021}. There exist additional interesting works for the scenario where each user can be associated to more than one cache~\cite{AsadiOngSPAWC19,sasi2021multiaccess,cheng2021novel,reddy2020structured,SerbParITW19,brunero2021fundamental}, although  this latter scenario is not considered in this current work.

It is worth noting that this shared cache model can be applied in a variety of settings that include the idealized model of cache-aided heterogeneous cellular networks, the multiple file request problem \cite{Ji2015CachingaidedCM,SenguptaT:17TCOMM,XuWang19Tcom}, and last but not least, this same shared cache approach can be effectively used to account for the omnipresent subpacketization bottleneck of coded caching~\cite{jin2019new,Shanmugam2013_femto,LampirisEliaJsac18}, which in essence forces different users to cache the same content. Interesting works on this latter matter can be found in~\cite{jin2019new,LampirisEliaJsac18}. In~\cite{jin2019new}, Jin et al. proposed a decentralized shared-cache scheme for the subpacketization-constrained (finite file size) scenario, while~\cite{LampirisEliaJsac18} revealed for the first time that in the presence of multiple antennas the shared-cache approach can dramatically alleviate the subpacketization problem while simultaneously exploiting both multiplexing and caching gains in their entirety. 
Recently,~\cite{Elizabath2022mixed} analyzed the case in which there exist both dedicated and shared caches, and~\cite{Merikhi2022} studied the case where the library files are correlated. Other related works can be found in \cite{AsadCaireISIT19,salehi2020lowcomplexity}. 

Moreover, an additional implication of the shared-cache setting was recently revealed in~\cite{ZhaoTWC2022}, this time linking this shared-cache approach to the well known worst-user bottleneck, where this bottleneck was previously thought to be fundamental to coded caching. In particular, the work in~\cite{ZhaoTWC2022} has provided a new method that exploits the often unavoidable need for shared caches (where users in predefined groups are forced to place identical content in their caches) in a manner than can entirely alleviate this worst-user bottleneck, thus proving that this bottleneck is not fundamental to coded caching. This method was also proved to overcome the worst-user effect in the presence of fast fading and different path losses in~\cite{Zhao2021_ISIT}. 

The above jointly suggest that the shared cache scenario can form a pivotal ingredient in any attempt to meaningfully employ coded caching. This motivates our information-theoretic study of this scenario.

%%%%%%%%%%%%%%%%%%%%%%%%%%%%%%%%%%%
\subsection{Memory allocation in cache-aided networks}
%%%%%%%%%%%%%%%%%%%%%%%%%%%%%%%%%%%
Optimizing memory allocation in cache-aided systems has been a critical topic of study in many works (see for example~\cite{VuChatziOtters17WCNCshort}).  This is particularly true in heterogeneous scenarios like the one we consider here, where one expects to find larger caches that serve the many users of large office buildings coexisting with smaller caches that serve smaller pockets of population. It is worth remarking that deriving fundamental limits for asymmetric settings is a challenging topic, as it has been shown in the recent work~\cite{Chang2022}, where the authors fully characterized the exact capacity of the 2-file/2-user setting simultaneously allowing for heterogeneous files sizes,  heterogeneous cache sizes, and user-dependent file popularities.

Various works have explored this problem of memory allocation in cache-aided communication systems. For example, in the context of traditional (non-coded) caching, the works~\cite{PengLetaiefICC16,LiuNiuGlobecom18} investigated this problem for the scenario of backhaul-limited cache-aided small-cell networks, while the work in \cite{LiaoYangTWC17} explored this problem for the scenario of a cache-enabled heterogeneous small-cell network, for which it proceeded to minimize the average backhaul load by optimizing --- subject to a cumulative cache capacity --- the distribution of cache sizes across the network. 
In the context of coded caching, \cite{brunero2022selfish} showed that unselfish memory allocation provides significant gains even in selfish settings where each user is interested only in a different subset of the library content. 

In this work, we consider {the memory allocation problem in a  cache-aided broadcast channel in the context of a shared-cache framework}, where the size of each cache can be optimized under the assumption of a sum cache size constraint. 
To the best of our knowledge, this memory allocation problem (i.e., where to allocate the memory resources) was first addressed in the context of coded caching in~\cite{TangLinkRatesTCOM18} and~\cite{WiggerBenefitsofcache19}, both for the setting with dedicated caches (or, equivalently, where each cache serves a single user), while here we consider {shared-cache framework}. In~\cite{TangLinkRatesTCOM18}, the authors proposed a delivery scheme with memory allocation optimization for a wireless backhaul link, which was modeled as a BC where the link rates are fixed and known and the cache sizes can be changed. The work in~\cite{WiggerBenefitsofcache19} considers the rate-memory trade-off in general cache-aided degraded broadcast channel scenario. In other words,~\cite{WiggerBenefitsofcache19} analyzes a scenario where each user has a distinct cache and under the assumption of having a \emph{degraded} channel, such that users can be sorted from the weakest to the stronger (whereas in this work we consider a (not degraded) broadcast channel where there exist $\Lambda \leq K$ different cache states and each user can store one of such states). That work proposes novel lower and upper bounds for the degraded BC, applicable to e.g. the erasure or the Gaussian BC. 

It is interesting to note that, even if the scenario considered in~\cite{TangLinkRatesTCOM18,WiggerBenefitsofcache19} and the one considered here are markedly dissimilar\footnote{In such works, the main parameter that defines the scenario is the relative channel strength between different users. In our case of study, the main parameter is the amount of users associated to each cache.}, and the placement, delivery schemes, as well as the novel proposed converse result are utterly different, the resulting memory optimization solutions have an analogous shape. {Nevertheless, both our work here and the results in~\cite{WiggerBenefitsofcache19} highlight the importance of a carefully designed memory allocation for systems with heterogeneous elements.} 

%%%%%%%%%%%%%%%%%%%%%
\subsection{Contributions}
%%%%%%%%%%%%%%%%%%%%%
In this work we consider a shared-cache setting where each cache serves an arbitrary number of users and where the size of each such cache can be optimized --- under a sum cache-size constraint --- as a function of the topology.\footnote{As it will become evident from the formal description of the system model, our derived schemes and results directly apply to the isomorphic coded caching problem where users having access to their own dedicated cache can request any number of files.} 
We focus on a {topology-aware} scenario where the cache occupancy is known during the memory-allocation phase. In particular: 
    
\begin{itemize}
    \item  We construct a novel converse bound based on index coding that allows us to overcome the challenges posed by the considered heterogeneous cache-aided setting. This converse bound is able to capture the complex influence of the asymmetry in the setting through a non-trivial and entangled combination of index coding bounds.   
    This comes at a time when most of the converse bounds for cache-aided networks build on mathematical machinery that depends heavily on symmetry properties. Such bounding techniques, while working well for idealized symmetric settings, can suffer in the presence of heterogeneity, which is known to cause additional challenges in achieving converse tightness. We believe that our proposed converse can shed light on this type of bounds and on their applicability on heterogeneous settings. 
    
    \item  For this setting, we will show that our proposed memory allocation, cache placement, and delivery scheme achieve the information theoretic optimal performance under some regularity assumptions. 

    \item As a by-product, we prove that asymmetry can be beneficial and asymmetric settings can outperform the well-studied symmetric setting. As a matter of example, for a given number of caches, total number of users and total cache size, we show that the uniform cache occupancy (where each cache serves the same number of users) is the one that results in the highest delivery time, whereas the lowest delivery time is achieved for the case where all caches but one serve a single user each, and the remaining cache serves the rest of the users. 

    \item
    Subsequently, motivated by the possibility of having a cache occupancy that varies with time, we consider a scenario of partial topology awareness where there is a mismatch between the perceived/estimated cache occupancy during the memory allocation phase and the true cache occupancy experienced during the actual delivery phase. This prior estimate of the topology, which will define the memory allocation, may for example reflect long-term statistical knowledge of this occupancy.  
    For this scenario of imperfect knowledge of topology, we will show that, under a specific memory allocation and cache placement that are both designed according to the \emph{expected} number of users connected to each cache, the proposed delivery scheme is exactly optimal. 
    Finally, in addition to providing exact information-theoretic optimality expressions, we also proceed to elevate our problem to the stochastic setting by studying the case where the number of users connected to each cache follows a Poisson distribution, for which  we compare our derived results with state-of-the-art schemes, namely the topology-agnostic scheme in~\cite{PUETIT2020} and~the scheme from~\cite{wanCaireFogRANTIT21}, which allows us to stress the fundamental importance of optimized memory allocation in such heterogeneous cache-aided networks. 
\end{itemize}

\subsection{Outline and Notation}
The rest of the paper is organized as follows. Section~\ref{sec:system_model} describes the system model, the formal problem definition, and various mathematical preliminaries, while Section~\ref{sec:example_shared_link} provides an illustrative example of the proposed scheme for the topology-aware scenario.  %is presented in Section~\ref{sec:example_shared_link}. 
After that, we present our main results in Section~\ref{sec:main_results}. For the topology-aware scenario, the achievable scheme is described in Section~\ref{sec:achiev}, while the converse bound can be found in Section~\ref{sec:converse}. Subsequently, we present in Section~\ref{sec:topology_partially_aware}  the achievable scheme and converse bound for the scenario with imperfect knowledge of topology, as well as  various numerical evaluations. Finally, our conclusions are discussed in Section~\ref{sec:conclusions}, while key mathematical derivations are presented in the appendices.

\subsubsection*{Notation}
For $\Lambda\in \mathbb{N}$, we use $[\Lambda] \dfn \{1,2,\dots,\Lambda\}$ and $[\Lambda]_0\dfn \{0,1,2,\dots,\Lambda\}$. For any set $\CT$, we define the set of all $k$-combinations of $\CT$ as $\Cb{\CT}{k}\dfn\{\tau : \tau\subseteq \CT, |\tau|=k \}$, while we denote the powerset of $\CT$ as $2^\CT$. For any $n\in \mathbb{N}$, we use $\mathcal{S}_n$ to denote the symmetric group of all permutations of $[n]$. For an ordered set $\tau$, we will refer to the $j$-th element of $\tau$ as $\tau(j)$. The XOR operation is here denoted by the symbol $\bigoplus$. We denote by $\round()$ the function rounding a real number to the nearest integer.
Furthermore, for a set $\CX$ and $k\in\CX$, and for $\{\mathrm{a}_k\}$ being a real-valued discrete sequence, we will use $\Conv_{k\in\CX}(\mathrm{a}_k)$ to denote the (real and continuous) lower convex envelope of the points $\{(k,\mathrm{a}_k)|k\in\CX\}$.

\section{System Model}\label{sec:system_model}
In this section, we present the system model for the two operating scenarios, as well as the corresponding performance metric. 
The system model will be fully described for the topology-aware scenario, while, for the subsequent scenario corresponding to imperfect knowledge of topology the description will focus on highlighting the differences with respect to the previous scenario. 

%%%%%%%%%%%%%%%%%%%%%%%%%%%%%%%%%%%%
\subsection{Topology-aware scenario}
%%%%%%%%%%%%%%%%%%%%%%%%%%%%%%%%%%%%
We consider a cache-aided network where a transmitter (TX) with access to a library of $N$ unit-sized files $W^{(1)},W^{(2)},\dots,W^{(N)}$ is connected via a shared-link broadcast channel to $K$ users~($N\geq K$), 
each of which connected to \emph{one} of $\Lambda$ different caches. The size of each cache $\lambda\in \{1,2,\dots,\Lambda\}$ is a design parameter denoted by $M_{\lambda}\in (0,N]$ (in units of file), adhering to a cumulative (sum) cache-size constraint defined by $M_{\Sigma}\triangleq\sum_{\lambda=1}^\Lambda M_{\lambda} $. 
We define the normalized size of cache $\lambda$ as $\gamm{\lambda}\triangleq \frac{M_{\lambda}}{N}$, such that the corresponding cache redundancy takes the form of the normalized sum cache-size constraint 
    \begin{equation}\label{eq:memconstr}
        t \triangleq \sum_{\lambda=1}^{\Lambda}\gamm{\lambda} =\frac{M_{\Sigma}}{N} .
    \end{equation} 
We consider the general scenario where the channel capacity is normalized to one file per unit of time, and we assume that users can access the content of their associated cache at zero cost.

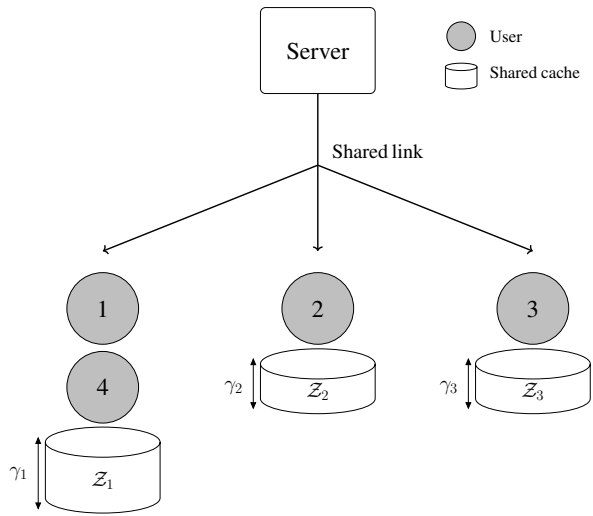
\begin{figure}[ht]\centering
    \resizebox{0.50\columnwidth}{!}{
          \begin{tikzpicture}[scale=0.5]
 \coordinate (Origin)   at (0,0);
    \coordinate (XAxisMin) at (-50,0);
    \coordinate (XAxisMax) at (50,0);
    \coordinate (YAxisMin) at (0,-50);
    \coordinate (YAxisMax) at (0,60);

	\coordinate (S1) at (-5,55);
	\coordinate (L1) at (8,54);
	
	\coordinate (A0) at (-5,51);
	\coordinate (Ac) at (-5,47);
	\coordinate (A1) at (-5,43);
	\coordinate (A2) at (-9,44);
	\coordinate (A3) at (-1,44);
	\coordinate (A4) at (-12,45);
	\coordinate (A5) at (2,45);

    \coordinate (S) at (0,48);
    \coordinate (BN) at (0,40);
    \coordinate (C1) at (-15,18);
    \coordinate (C2) at (0,24.2);
    \coordinate (C3) at (+15,24.2);

    \coordinate (U1) at (-15,30);
    \coordinate (U2) at (0,30);
    \coordinate (U3) at (+15,30);
    \coordinate (U4) at (-15,24.5);
    %\coordinate (U6) at (20,20);

\pgfdeclarelayer{bg}    % declare background layer
\pgfsetlayers{bg,main}  % set the order of the layers (main is the standard layer)

\begin{pgfonlayer}{main}    % select the background layer    
\fill [white] (25,25) rectangle (25,25);
\node[text width=5cm] at (6,41) {\huge Shared link};
%legend
\node[draw,circle,minimum size=1cm,fill=lightgray] at (10,49) {};
\node[text width=5cm] at (17,49) { \LARGE User};
\node[draw,cylinder,shape border rotate=90, draw,minimum height=0.9cm,minimum width=1.1cm] at (10,46) {};
\node[text width=5cm] at (17,46.5) { \LARGE Shared cache};
\end{pgfonlayer}

\node[draw,rectangle,rounded corners=5pt, minimum width=4cm,minimum height=3cm, name=S] at (S) {\Huge Server};

\node[draw,cylinder,shape border rotate=90, draw,minimum height=3cm,minimum width=4cm,name=C1] at (C1) {\huge $\mathcal{Z}_1$};
\node[draw,cylinder,shape border rotate=90, draw,minimum height=2cm,minimum width=4cm,name=C2] at (C2) {\huge $\mathcal{Z}_2$};
\node[draw,cylinder,shape border rotate=90, draw,minimum height=2cm,minimum width=4cm,name=C3] at (C3) {\huge $\mathcal{Z}_3$};

\node[draw,circle,minimum size=2.5cm,name=U1,fill=lightgray] at (U1) {\Huge 1};
\node[draw,circle,minimum size=2.5cm,name=U2,fill=lightgray] at (U2) {\Huge 2};
\node[draw,circle,minimum size=2.5cm,name=U3,fill=lightgray] at (U3) {\Huge 3};
\node[draw,circle,minimum size=2.5cm,name=U4,fill=lightgray] at (U4) {\Huge 4};

\draw [line width=0.5mm,-] (S) -- (BN) node[midway,above,color=black]{};

\draw [line width=0.5mm,->] (BN) -- (-15,34) node[midway,above,color=black]{};
\draw [line width=0.5mm,->] (BN) -- (-0,34) node[midway,above,color=black]{};
\draw [line width=0.5mm,->] (BN) -- (15,34) node[midway,above,color=black]{};

\draw[>=triangle 45, <->] (-19.5,21) -- (-19.5,16);
\node[text width=5cm] at (-16.5,18.5) { \huge $\gamma_1$}; 
\draw[>=triangle 45, <->] (-4.5,26.5) -- (-4.5,23);
\node[text width=5cm] at (-1.5,24.5) { \huge $\gamma_2$};
\draw[>=triangle 45, <->] (+10.5,26.5) -- (+10.5,23);
\node[text width=5cm] at (+13.5,24.5) { \huge $\gamma_3$};
  \end{tikzpicture}
    }\caption{The shared-cache setting with $\Up=(2,1,1)$.}
    \label{fig:schematic}
\end{figure}

Each cache $\lambda$ is connected to a disjoint set of users $\mathcal{U}_{\lambda}\subset [K]\triangleq\{1,\dotsc,K\}$, where these sets form a partition $\CU=\left\{\mathcal{U}_1,\dots,\mathcal{U}_\Lambda\right\}$ of $[K]$. 
The \emph{occupancy} of each cache $\lambda$ describes the number of users connected to this cache, and it is denoted by $L_{\lambda} = |\CU_\lambda| \in \mathbb{N}$. The corresponding \emph{cache occupancy} vector is denoted as 
\begin{equation}\label{eq:cache_ocup_vectot}
    \Up\triangleq (L_1,\dots,L_{\Lambda}),
\end{equation}
where naturally $\sum_{\lambda=1}^{\Lambda}L_\lambda=K$, and where we assume without loss of generality that 
{$L_1\geq L_2\geq\dots\geq L_\Lambda$}. 
With a slight abuse of notation, whenever needed, we will use $\Up$ in its set form, to represent $\{L_\lambda\}_{\lambda=1}^\Lambda$.
Figure~\ref{fig:schematic} provides a schematic representation of our setting. Hereinafter, we will denote this setting as the \emph{$(t,\Up)$ shared-cache BC network}, where $t$ and $\Up$ has been defined in~\eqref{eq:memconstr} and~\eqref{eq:cache_ocup_vectot}, respectively.

The system works in three different phases.
\begin{enumerate}
\item A \textit{memory allocation phase} during which the knowledge of the cache occupancy vector $\Up$ is used to allocate the total memory $M_{\Sigma}$ to the caches, yielding the allocated size set $\{\gamm{\lambda}\}_{\lambda=1}^{\Lambda}$.
\item A \textit{cache placement phase} during which each cache $\lambda$ --- of allocated size $\gamm{\lambda}$ --- is filled with content $\mathcal{Z}_{\lambda}$ from the library, according to a certain strategy $\Z_{\Up}=(\mathcal{Z}_{1},\dots,\mathcal{Z}_{\Lambda})$, where the lower index $\Up$ highlights the dependency of the cache placement on the cache occupancy vector $\Up$. Hereinafter, we will often omit the lower index $\Up$ whenever there is no possible ambiguity.  
In this work, as is often common, we focus only on \emph{uncoded cache placement schemes} 
where each packet stored in a cache can be traced back to the library (i.e., comes directly from the library, without any coding). Finally, let us again recall that $\Up$ is known during this phase.

\item A \textit{delivery phase} that starts with each user $k\in[K]$ requesting a single library file $W^{(d_k)}$. Once the demand vector~$\Bd\triangleq(d_1,d_2,\dots,d_K)$ of requested file indices is known, the server begins to deliver each requested file to its corresponding user. 
\end{enumerate}

%%%%%%%%%%%%%%%%%%%%%%%%%%%%%%%%%%%%%%%%%%%%%%%%%%%
\subsection{Scenario with imperfect knowledge of topology}
In this scenario, we consider a similar setting to the $(t,\Up)$ shared-cache BC network described before, with the only difference being that the number of users that will be connected to each cache during delivery phase is not known during the memory allocation phase and the subsequent cache placement phase. 
Instead of the knowledge of the true vector $\Up$, we assume that the first two phases are designed according to another (perhaps estimated) cache occupancy vector $\bar{\Up}=(\bar{L}_1,\bar{L}_2,\dots,\bar{L}_\Lambda)$ that is generally not equal to $\Up$. 
We will denote this scenario as the $(t,\bar{\Up},\Up)$ shared-cache BC network. 
For any cache $\lambda\in[\Lambda]$, the value of $\bar{L}_{\lambda}$ can represent an imperfect prediction or an expectation of the number of users that will be connected to this cache during the delivery phase. We also allow the sum $\sum_{\lambda=1}^\Lambda\bar{L}_\lambda$ to be arbitrary\footnote{We note that the integers $\bar{L}_\lambda,\lambda\in[\Lambda]$, are strictly positive.} and not necessarily equal to $\sum_{\lambda=1}^\Lambda L_\lambda = K$.

%%%%%%%%%%%%%%%%%%%%%%%%%%%%%%%%%%%%%%%%%%%
\subsection{Problem definition}
%%%%%%%%%%%%%%%%%%%%%%%%%%%%%%%%%%%%%%%%%%%
We consider the standard \emph{rate} metric that has been commonly used in coded caching literature \cite{MN14,YuMA16,WanTuninettiTIT2020}, which we hereinafter refer to as the \emph{delivery time}, and which we denote by~$T$. 
For any given \emph{uncoded cache placement} scheme $\Z$, any cache occupancy vector $\Up$, and any given demand $\Bd$, we define $T^*(\Z,\Bd,\Up)$ as the minimum delivery time (minimized over all delivery schemes) that guarantees delivery of the desired files $W^{(d_k)}$ to all users $k\in[K]$. Under the assumption of uncoded cache placement, our goal is to characterize 
the minimum worst-case delivery time over all memory-allocation strategies and all placement-and-delivery schemes, i.e., we aim to characterize
\begin{equation}\label{eq:goal}
    T^*(t,\Up) \triangleq \underset{\Z}{\min}~\underset{\Bd}{\max}~T^*(\Z,\Bd,\Up)
\end{equation}
as a function of $t~\text{and}~\Up$.
We omit hereinafter the dependence of  $T^*$ on $(t,\Up)$ when there is no possible ambiguity. For any cache occupancy vector $\Up$, we also define the optimal cache placement $\Z^*_{\Up}$ as $\Z^*_{\Up}\triangleq {\argmin}_{\Z}{\max}_{\Bd}~T^*(\Z,\Bd,\Up)$.
Next, we present the definition of what is often referred to as regular cache placement.
\begin{defn}\label{def:symmetric_placement}
A cache placement scheme is said to be \emph{regular} if each bit of the library is repeated the same number of times throughout the different caches.
\end{defn}
It is easy to see that any regular cache placement naturally implies that $t\in [\Lambda]$.
In the context of non-uniform $\Up$ and heterogeneous $\{\gamm{\lambda}\}_{\lambda=1}^{\Lambda}$, the concept of the sum-DoF in cache-aided networks \cite{ZE:17tit} naturally generalizes to
\begin{equation}
    \DoF \triangleq \frac{K-\sum_{\lambda=1}^{\Lambda}\gamm{\lambda}L_{\lambda}}{T},
\end{equation}
reflecting the rate of delivery of the non-cached desired information.
%%%%%%%%%%%%%%%%%%%%%%%%%%%%%%%%%%%%%%%%%%%

For the scenario with imperfect topology knowledge, where the memory allocation and cache placement is done on the basis of the available information $\bar{\Up}$,  we will characterize the optimal delivery time
\begin{equation}
   T^*(t,\Up,\bar{\Up})\triangleq \underset{\Bd}{\max}~T^*(\Z^*_{\bar{\Up}},\Bd,\Up)  
\end{equation}
as a function of $t$, $\Up$ and $\bar{\Up}$.

%%%%%%%%%%%%%%%%%%%%%%%%%%%%%%%%%%%%%%%%%%%
\subsection{Mathematical preliminaries}
%%%%%%%%%%%%%%%%%%%%%%%%%%%%%%%%%%%%%%%%%%%
Before presenting our main results, we include some mathematical preliminaries that will be useful for the derivation of our results. 
We first recall the well-known \emph{elementary symmetric functions}~\cite{Niculescu2000}, also known as elementary symmetric polynomials. These functions have already appeared in some works about cache allocation for coded caching~\cite{TangLinkRatesTCOM18,WiggerBenefitsofcache19} and  will be used extensively in our proofs. We recall that we denote the set of all $k$-combinations of a set $\CX$ as $\Cb{\CX}{k}\dfn\{\tau : \tau\subseteq [\CX], |\tau|=k \}$. 

\begin{defn}[Elementary symmetric functions]\label{def_elemSymF}
    For any $n$-set $\CX=\{x_1,x_2,\dots,x_n\}$ and any $k\in\{1, \dotsc, n\}$, the $k$-th \emph{elementary symmetric function} $e_k(\CX)$ is defined as  
 			\begin{align}
 				e_k(\CX) \triangleq \sum_{q\in \Cb{\CX}{k}}^{} \prod_{j=1}^{k} x_{q(j)}
 			\end{align}				
where $e_0(\CX) \triangleq 1$.
 	\end{defn}
Furthermore, we present some known (or otherwise here derived) basic properties of such elementary symmetric functions. 
\begin{property}\label{prop:recursive_e_t} 
    For any elementary symmetric function $e_k(\CX)$, $k\in\{1,2,\dots,n\}$, on any set $\CX \triangleq \{x_1,\dots, x_n\}$, it holds that 
     	\begin{align}\label{eq:eq_prop1}
     		e_k(\CX)=e_k(\CX\setminus\{x_i\})+x_i e_{k-1}(\CX\setminus\{x_i\}),
     	\end{align}		
\end{property}
\begin{proof}
    The proof of Property~\ref{prop:recursive_e_t}  is straightforward and follows from the definition of the elementary symmetric functions (cf. Def.~\ref{def_elemSymF})~\cite{Niculescu2000}, since~\eqref{eq:eq_prop1} just represents that the sum of all products of $k$ distinct elements in set $\CX$ (i.e., $e_k(\CX)$) can be split in two terms: $(i)$ The sum of all such products of $k$ distinct elements that \emph{do not include} $x_i$ (i.e., $e_k(\CX\setminus\{x_i\})$), and $(ii)$ the sum of all products of $k$ distinct elements that \emph{do include} $x_i$, where the latter can be written as $x_i e_{k-1}(\CX\setminus\{x_i\})$.
\end{proof}

\begin{property}\label{prop:prop3on_elementary} For any elementary symmetric function $e_k(\CX)$, $k\in\{1,2,\dots,n\}$, on any set $\CX \triangleq \{x_1,\dots, x_n\}$, the following equality holds for any subset $\phi \subset \CX$.
 	\begin{align}\label{eq:prop3on_elementary}
 	\sum_{x_i\in \{\CX\setminus \phi\}}x_i\cdot e_{k-1}(\CX\setminus\{x_i\}) = \sum_{q\in \Cb{\CX}{k}}\prod_{j=1}^{k} x_{q(j)} \cdot |\{q\setminus \phi\}|,
 	\end{align}		
\end{property}	
    \begin{proof} The proof is relegated to Appendix~\ref{app:elemfunc_proof}.
    \end{proof}
\begin{cor}\label{prop:ripetitive_e_t} 
For any elementary symmetric function $e_k(\CX)$, $k\in\{1,2,\dots,n\}$, on any set $\CX \triangleq \{x_1,\dots, x_n\}$, it holds that 
 	\begin{align}\label{eq:property_ripetitive}
 	    \sum_{i\in [|\CX|]}x_i	e_{k-1}(\CX\setminus\{x_i\}) = k\cdot e_k(\CX),
 	\end{align}		
\end{cor}

\begin{proof}
    Corollary~\ref{prop:ripetitive_e_t} follows directly from Property~\ref{prop:prop3on_elementary} after setting $\phi=\emptyset$.
\end{proof}

\section{Memory-Allocation, Placement and Delivery: an Illustrative Example}
\label{sec:example_shared_link}
Before presenting the main results, we provide an example that illustrates the main ideas behind the proposed general scheme in the topology-aware scenario.

Consider an instance of the $(t,\Up)$ shared-cache BC network (see Fig.~\ref{fig:schematic}) with $N=6$ files, $\Lambda=3$ caches, and $K=6$ users associated to the different caches according to the cache occupancy vector $\Up=(3,2,1)$. We assume that a sum memory of $M_{\Sigma}=12$ units of file --- corresponding to $t=2$ times the size of the library --- is available to be allocated across the caches. In the first phase, we use the knowledge of $\Up$ to allocate fractions
\begin{align}
\gamm{1}&=\frac{L_1L_2+L_1L_3}{L_1L_2+L_1L_3+L_2L_3}=\frac{9}{11}\\
\gamm{2}&=\frac{L_1L_2+L_2L_3}{L_1L_2+L_1L_3+L_2L_3}=\frac{8}{11}\\
\gamm{3}&=\frac{L_1L_3+L_2L_3}{L_1L_2+L_1L_3+L_2L_3}=\frac{5}{11}
\end{align}
of the library to caches $1,2,3$, respectively. Seeing that $\frac{9}{11}+\frac{8}{11}+\frac{5}{11} = t=2$ allows us to verify that the cumulative cache size is not exceeded. In the subsequent caching phase, we split each file $W^{(n)}$ ($n\in[6]$) into $11$ equally-sized subfiles which we label as $W^{(n)}_{\tau,m_\tau}$, where each pair $(\tau,m_\tau)$ is taken from the following set:\footnote{For the sake of being concise, we have used a compact notation in \eqref{eq:indices} and throughout this example, such that we use for example $(12,1)$ instead of the less concise notation $\left(\{1,2\},1\right)$.}
\begin{equation}\label{eq:indices}
    \{(12,1),(12,2),(12,3),(12,4),(12,5),(12,6),(13,1),(13,2),(13,3),(23,1),(23,2)\}.
\end{equation}
In the above subfile labeling, the first index $\tau$ represents the set of caches that will store the associated subfile. On the other end, the second index $m_\tau$ is a mere counter that helps us differentiate subfiles with the same first index $\tau$. Hence, for example, $W^{(n)}_{12,4}$ is the fourth subfile out of the subfiles of $W^{(n)}$ that are exclusively stored in caches $1$ and $2$ ($6$ subfiles in this case).
  Then, for each $n\in[6]$, each cache $\lambda$ stores those subfiles whose first index $\tau$ includes $\lambda$. Consequently, in our example, the content of each cache is: 
\begin{align*}
&\CZ_1=\{W^{(n)}_{12,1},W^{(n)}_{12,2},W^{(n)}_{12,3},W^{(n)}_{12,4},W^{(n)}_{12,5},W^{(n)}_{12,6},W^{(n)}_{13,1},W^{(n)}_{13,2},W^{(n)}_{13,3}: n\in[6]\},\\
&\CZ_2=\{W^{(n)}_{12,1},W^{(n)}_{12,2},W^{(n)}_{12,3},W^{(n)}_{12,4},W^{(n)}_{12,5},W^{(n)}_{12,6},W^{(n)}_{23,1},W^{(n)}_{23,2}: n\in[6]\},\\
&\CZ_3=\{W^{(n)}_{13,1},W^{(n)}_{13,2},W^{(n)}_{13,3},W^{(n)}_{23,1},W^{(n)}_{23,2}: n\in[6]\},
\end{align*}
which adheres to the aforementioned memory allocation $\gamm{1}=\frac{9}{11},\gamm{2}=\frac{8}{11},\gamm{3}=\frac{5}{11}$. 

In the delivery phase, we consider the demand vector $\Bd=(1,2,3,4,5,6)$ where $W^{(1)},W^{(4)}$ and $W^{(6)}$ are each requested by one of the three users associated to cache~$1$, $W^{(2)}$ and $W^{(5)}$ by the two users associated to cache~$2$, and $W^{(3)}$ by the user associated to cache~$3$. For the sake of a more understandable exposition, we re-denote the ordered set of files $\{W^{(n)}\}_{n=1}^{6}$ as $\{A,B,C,D,E,F\}$.  We can see that each set $\CR_\lambda$ of uncached subfiles wanted by the users of cache $\lambda$ takes the form
\begin{center}
\begin{tabular}{ |c c c| }
\hline
\text{\quad$\CR_1$\quad} & \text{\quad$\CR_2$\quad} & \text{\quad$\CR_3$\quad}\\ 
\hline
 $A_{23,1}$ & $B_{13,1}$ & $C_{12,1}$\\
 \hline
 $D_{23,1}$ & $E_{13,1}$ & $C_{12,2}$\\
 \hline
 $F_{23,1}$ & $B_{13,2}$ & $C_{12,3}$\\
 \hline
 $A_{23,2}$ & $E_{13,2}$ & $C_{12,4}$\\
 \hline
 $D_{23,2}$ & $B_{13,3}$ & $C_{12,5}$\\
 \hline
 $F_{23,2}$ & $E_{13,3}$ & $C_{12,6}$\\
 \hline
\end{tabular}
\end{center}
We notice that as a consequence of the proposed heterogeneous memory allocation and cache placement --- which we will present in detail in Section~\ref{sec:achiev} --- the number of equi-sized subfiles jointly requested from all the users of each cache remains the same. This fact is key to allowing the transmission of all data during the delivery phase in the form of multicast messages {serving $t+1$ users at a time}. In our example, these messages will contain information for $t+1=3$ different users at a time. The subsequently transmitted $6$ XORs that deliver all requested subfiles take the form
\begin{gather}
X_{123}(1)=D_{23,1}\Oplus B_{13,1}\Oplus C_{12,1}\label{eq:example_X_1}\\ 
X_{123}(2)=A_{23,1}\Oplus E_{13,1}\Oplus C_{12,2}\\
X_{123}(3)=A_{23,2}\Oplus B_{13,2}\Oplus C_{12,3}\\
X_{123}(4)=F_{23,1}\Oplus E_{13,2}\Oplus C_{12,4}\\
X_{123}(5)=F_{23,2}\Oplus B_{13,3}\Oplus C_{12,5}\\
X_{123}(6)=D_{23,2}\Oplus E_{13,3}\Oplus C_{12,6}\label{eq:example_X_last}
\end{gather}
and each XOR can be easily decoded in the classical manner described in \cite{MN14}. Consequently, the total delay is $T=\frac{6}{11}$, which will be shown to be exactly optimal under the assumption of uncoded cache placement. On the other hand, if we had forced equal-sized caches, the best possible performance under the same assumptions would be $T=1$ (cf. \cite{PUETIT2020}), which almost doubles the delay of the new scheme. 
We recall that the work in \cite{PUETIT2020} has shown that, without knowledge of the topology during the caching phase, the heterogeneity in the cache occupancy numbers results in an unavoidable reduction in the multicasting gain, which reduces below $t+1$ as the skewness of the cache occupancy vector increases. On the other hand, the current knowledge of the topology allows our scheme to optimize the cache sizes, yielding a symmetry that in turn allows for a constantly full multicasting gain of $t+1$ while also allowing for a local caching gain that, as we will explore later on, interestingly increases with the skewness of the cache occupancy vector.

\section{Main Results}\label{sec:main_results}
We present in this section our main contributions. Let us start by presenting {an information-theoretic converse (lower bound)} on the delivery time under uncoded cache placement for the topology-aware scenario described before. 

\begin{thm}\label{thm:lower_bound_total}
Under the assumption of uncoded cache placement, the optimal normalized delivery time of the $(t,\Up)$ shared-cache BC network satisfies
    \begin{align}\label{eq:general_lower_bound0}
        T^*(t,\Up)\geq \mathcal{T}^{(\bar{t},\Up)}_{low}(t) 
    \end{align}
where $\bar{t}\triangleq\round(t)$, and  $\mathcal{T}^{(\bar{t},\Up)}_{low}(x)$ is defined as
    \begin{align}\label{eq:general_lower_bound}
     \mathcal{T}^{(\bar{t},\Up)}_{low}(x) &\triangleq 
		\Conv_{j\in[\Lambda]_0}\left(\frac{%
				    	\sum_{q\in \Cb{[\Lambda]}{\bar{t}+1}}^{}\frac{\bar{t}+1-|q\cap \tau^\star_j|}{j+1-|q\cap \tau^\star_j|}\prod_{i=1}^{\bar{t}+1} L_{q(i)} 
					}{%
						\sum_{\ell\in \Cb{[\Lambda]}{\bar{t}}} \prod_{i=1}^{\bar{t}}L_{\ell(i)} %
							}\right),
    \end{align}
where $\tau^\star_j$ is given by 
	\begin{align}
				 \tau^\star_j  =
				\begin{cases}
						 \{\emptyset\} & \text{ if } j = 0\\ 
						 \{\Lambda -j + 1, \ \Lambda -j + 2,\ \dots,\  \Lambda\} & \text{ if } 1\leq j < \bar{t}\\ 
						\ \{1,\ 2,\ \dots,\  j\} & \text{ if } j \geq \bar{t}
				\end{cases}\nonumber 
		\end{align}
\end{thm}

\begin{proof}
The proof is presented in Section~\ref{sec:converse}. 
\end{proof}
The heterogeneity of the cache occupancy vector $\Up$ and the fact that this vector is known during the placement phase jointly introduce a new important challenge in the derivation of the converse bound. 
As briefly demonstrated in~\cite{ParEliaITW19}, a direct application of the traditional index coding techniques (see.~\cite{WanTuninettiTIT2020}, or equivalently see the genie-aided approach of \cite{YuMA16}) would result in very loose bounds. 
The reason for which the known bounds do obtain a loose result is because they generally rely on derivations that require some \emph{symmetry} in the topology of the scenario.   
What the bound in~\eqref{eq:general_lower_bound} achieves is to render unnecessary the previously generally employed symmetries, thanks to involved combinatorial derivations, thus allowing us to capture the heterogeneity of the system. 

In conjunction with the developed achievable coded caching scheme of Section~\ref{sec:achiev}, this new converse becomes exactly tight under the basic assumptions  of regular and uncoded cache placement (cf.~Definition~\ref{def:symmetric_placement}), which are common properties of the content placement for most of the known coded caching schemes~\cite{MN14,YuMA16,LampirisEliaJsac18,SalehiTolliICC2020linsubpack,Mital2020}. 
Let us now present the converse result for the case where we restrict ourselves to the common \emph{regular} placement from Definition~\ref{def:symmetric_placement}. 

\begin{thm}\label{thm:lower_bound_regular}
Under the assumption of regular and uncoded cache placement, the optimal normalized delivery time of the $(t,\Up)$ shared-cache BC network satisfies
    \begin{align}\label{eq:regular_lower_bound0}
        T^*(t,\Up)\geq \mathcal{T}^{(t,\Up)}_{low,\; reg}(t) 
    \end{align}
where $\mathcal{T}^{({t},\Up)}_{low,\; reg}(x)$ is defined as
    \begin{align}\label{eq:regular_lower_bound}
     \mathcal{T}^{({t},\Up)}_{low, reg}(x) &\triangleq 
		\frac{\sum_{q\in \Cb{[\Lambda]}{t+1}}\prod\limits_{j=1}^{t+1} L_{q(j)}}{ \sum_{q\in \Cb{[\Lambda]}{t}}\prod\limits_{j=1}^{t}L_{q(j)}}=\frac{e_{t+1}(\Up)}{e_{t}(\Up)}.
    \end{align}
\end{thm}
\begin{proof}
    The proof is presented in Section~\ref{sec:converse}. 
\end{proof}

Note that for the standard scenario with dedicated caches, where $L_i=1$ for any $i\in[\Lambda]$ and $\Lambda = K$,~\eqref{eq:regular_lower_bound} reduces to the optimal delivery time from~\cite{YuMA16}. 
Next, we present the achievable delivery {time} of our proposed scheme, which will be described in detail in Section~\ref{sec:achiev}. 

\begin{lem}\label{lem:achiev}
    For the $(t,\Up)$ shared-cache BC network  %$K$-user BC 
    with $\Lambda$ shared caches, normalized sum-cache constraint $t$, and  cache occupancy vector $\Up$, the worst-case delivery time
        \begin{equation}\label{eq:Tachiev}
            T_{}(t,\Up)=\Conv_{t\in[\Lambda]_0}\left({\frac{\sum_{\lambda=1}^{\Lambda}L_{\lambda}(1-\gamm{\lambda})}{t+1}}\right)
        \end{equation}
    is achievable, where the memory allocation $\{\gamm{\lambda}\}_{\lambda=1}^{\Lambda}$ is given by
        \begin{equation}\label{eq:gammas}
            \vspace{-4pt}
            \gamm{\lambda}=\frac{L_\lambda\cdot e_{t-1}(\Up\setminus{\{L_\lambda\}})}{e_t(\Up)}.
        \end{equation}
\end{lem}
\vspace{+5pt}
\begin{proof}
The placement and delivery schemes are presented in Section~\ref{sec:achiev}. 
\end{proof}

 We immediately notice that the delay in~\eqref{eq:Tachiev} directly implies that $\DoF =t+1$, which is an improvement over the case where $\Up$ is unknown to the placement phase. 
 In fact, we know from~\cite{PUETIT2020} that, without knowledge of $\Up$ during the (uncoded) cache placement, $\DoF=t+1$ can be achieved only in the uniform case where we have $\frac{K}{\Lambda}$ users per cache, and that any non-uniformity in $\Up$ {strictly} forces a $\DoF$ penalty. The above lemma shows that knowledge of the profile $\Up$ allows for a redesigned and skewed memory allocation that, in turn, simultaneously allows a better local caching gain and a higher sum-DoF. To clarify this, a strategy that allocates more memory to more loaded caches automatically allows for higher local caching gains than a uniform memory allocation across the caches. At the same time, such heterogeneous allocation allows for multicasting messages that always serve $t+1$ users at a time. {These observations lead to the surprising fact that, for a fixed number of users $K$ and $\Lambda$ non-empty caches, the uniform cache occupancy vector $\Up=(\frac{K}{\Lambda},\frac{K}{\Lambda},\dots,\frac{K}{\Lambda})$ is the one that results in the highest delivery time, while the lowest delivery time is achieved for $\Up=(K-\Lambda+1,1,1,\dots,1)$.}

 \begin{obs}\label{obs:relation}
 It is interesting to observe that, for any given normalized total cache size $t$, the memory allocation $\{\gamm{\lambda}\}_{\lambda=1}^\Lambda$ given in Lemma~\ref{lem:achiev} coincides with the one of the scheme proposed in \cite{TangLinkRatesTCOM18} (see also \cite{WiggerBenefitsofcache19}) for a cache-aided setting with fixed unequal channel capacities. In particular, the work in \cite{TangLinkRatesTCOM18} by Tang et al. considers a wireless broadcast channel where the link between the transmitter and cache-aided receiver $\lambda$ has normalized capacity $R_\lambda$, and where these capacities are known during the cache allocation and placement phases. Assuming that each user requests only one file, the authors proposed a scheme which requires an unequal cache size allocation that coincides with the one in \eqref{eq:gammas}, where $L_\lambda$ would be replaced by the inverse of the rate of user $\lambda$, i.e. $R_\lambda=\frac{1}{L_\lambda},\forall \lambda\in[\Lambda]$. Despite the different nature of these two problems, we can observe that in the unequal link rate setting a user $i$ connected to the server through a link of capacity $R_i$ is reminiscent of the cache serving $L_i$ users in our shared-cache setting. {Similar analogies could be done between our contribution and the aforementioned work in~\cite{WiggerBenefitsofcache19}, where they studied the cache-aided degraded broadcast channel that includes the setting with fixed unequal channel capacities of~\cite{TangLinkRatesTCOM18}. 
 }
\end{obs}

We have presented the proposed lower bound and achievable scheme. In the next theorem, we show that they match and that the achievable delivery time is exactly optimal under the assumption of regular placement. 

 \begin{thm}\label{th:optimality}
    For the $(t,\Up)$ shared-cache BC network, 
    the achievable delivery time $T_{}(t,\Up)$ in~\eqref{eq:Tachiev} is exactly optimal under the assumption of uncoded and regular cache placement.
\end{thm}
\begin{proof}
    First, we note that we can write the lower-bound in Theorem~\ref{thm:lower_bound_regular} as 
            $\mathcal{T}^{(t,\Up)}_{low,\; reg}(t) 
	        =\frac{e_{t+1}(\Up)}{e_{t}(\Up)}$.
    Then, it follows  that
    \begin{equation}
        \begin{aligned}\label{eq:proof_relation_Tachiev}
        T^*(t,\Up)&\geq\frac{e_{t+1}(\Up)}{e_{t}(\Up)}\\ %\nonumber
           &\overset{(a)}{=}\frac{1}{t+1}\frac{\sum_{\lambda=1}^{\Lambda}L_\lambda\cdot e_{t}(\Up\setminus{\{L_\lambda\}})}{e_t(\Up)}\\%\nonumber
           &\overset{(b)}{=}\frac{1}{t+1}\frac{\sum_{\lambda=1}^{\Lambda}L_\lambda\cdot (e_t(\Up)-L_\lambda e_{t-1}\Up\setminus\{L_\lambda\})}{e_t(\Up)}\\%\nonumber
           &\overset{(c)}{=}\frac{\sum_{\lambda=1}^{\Lambda}L_{\lambda}(1-\gamm{\lambda})}{t+1}=T_{}(t,\Up),%\nonumber,
        \end{aligned}
    \end{equation}
    where $(a)$ follows from using Corollary~\ref{prop:ripetitive_e_t}, $(b)$ follows from employing Property~\ref{prop:recursive_e_t}, and $(c)$ is because of~\eqref{eq:gammas}.  $T_{}(t,\Up)$ represents the achievable delivery time of Lemma~\ref{lem:achiev}, which concludes the proof of Theorem~\ref{th:optimality}.
\end{proof}
These optimality results are extended beyond the assumption of regular placement in Section~\ref{sec:converse}. 

%%%%%%%%%%%%%%%%%%%%%%%%%%%%%%%%%%%%%%%%%%%%%%% Scenario 2
The previous results show how the knowledge of the cache occupancy vector $\Up$ considerably impacts the performance of coded caching, and that we can derive  optimal schemes that leverage this knowledge to improve considerable both local and global caching gains.   
Of course, there may be scenarios where having perfect knowledge of the topology during the placement phase is not feasible, and instead one has to rely on some noisy, imperfect, or average information about the said topology. 
Because of that, we now shift the focus to the scenario in which such topology knowledge is imperfect or noisy.

For this setting, the next theorem describes the optimal delivery time for the scenario where, during the memory allocation and cache placement phases, the assumed cache occupancy vector does not match the actual vector that materializes during the subsequent delivery phase. 

\begin{thm}\label{thm:thmL'}
For the $(t,\bar{\Up},\Up)$ scenario with imperfect topology knowledge, the delivery time  
\begin{equation}
    T^*(t,\bar{\Up},\Up)=\max_{\sigma\in S_{\Lambda,\Lambda-t}}\sum_{\lambda=1}^{\Lambda-t}L_{\sigma(\lambda)}\sum_{q\in\Cb{[\Lambda]\setminus\{\sigma(1),\dots,\sigma(\lambda)\}}{t}}\frac{\dot{\bar{L}}_q}{e_t(\bar{\Up})}
\end{equation}
is exactly optimal under the assumption that the memory allocation and content placement is applied under the premise that  the topology during the delivery phase matches the available knowledge~$\bar{\Up}$.
\end{thm}
\begin{proof}
    The achievable delivery scheme and the converse bound are presented in Section~\ref{sec:topology_partially_aware}.
\end{proof}
Theorem~\ref{thm:thmL'} refers to the case in which, during the first 2 phases, the server assumes that users will be connected to the caches in the delivery phase according to cache occupancy vector $\bar{\Up}$, but it turns out that the actual cache occupancy vector will finally be $\Up$. It is easy to conclude that such delivery time is higher than the optimal delivery time for the scenario where users actually show up according to $\bar{\Up}$ (as expected by the server) when the total number of users is the same. This results is stated in the following corollary.
\begin{cor}\label{cor:TLL'vsTL}
    For the case in which $\sum_{\lambda=1}^\Lambda \bar{L}_\lambda=\sum_{\lambda=1}^\Lambda L_\lambda$, the scenario where $\Up =\bar{\Up}$ achieves the lowest delivery time, i.e., $T^*(t,\bar{\Up},\Up)\geq T^*(t,\Up)$.
\end{cor}
The above corollary is  a direct derivation from the definitions of $T^*(t,\bar{\Up},\Up)$ and $T^*(t,\Up)$ in the system model.  
Note that, in scenarios where only long-term statistical information is available, {e.g. when expected values are known}, it is common to consider the strategy of acting as if this information was perfect and true for any realization. 
In Section~\ref{sec:numerical_results}, we will show through some numerical evaluations how $T^*(t,\bar{\Up},\Up)$ behaves on average when $\Up$ is a realization of $\Lambda$ independent Poisson random variables.
{Finally, as it will be clear from Section~\ref{sec:topology_partially_aware}, we note that the scheme achieving the performance in Theorem~\ref{thm:thmL'} creates XORs that do not always serve $t+1$ users. This is clearly a drawback of the fact that the memory allocation and cache placement phases are designed according to a cache occupancy vector that is different that the one that materializes in the delivery phase.}

%%%%%%%%%%%%%%%%%%%%%%%%
\subsection*{Comparison with related scenarios}
%%%%%%%%%%%%%%%%%%%%%%%%
The aforementioned analogies with other settings (cf.~Observation~\ref{obs:relation}), such as the cache-aided degraded BC or the BC where each user can demand several files, raise the questions of whether our results can be applied to such scenarios and, conversely, whether there is any overlap of results. In the following, we provide a brief discussion on this topic. 

First, the shared-cache scenario here considered is a one-to-one mapping to the standard dedicated-cache BC as in~\cite{MN14} where users are allowed to demand more than one file and no file is requested twice. Thus, all results here presented apply to that scenario. To prove this, we refer to Fig.~\ref{fig:schematic} and we note that, in our setting, two users sharing the same cache content will never be served simultaneously, as that would create interference and both users could not decode their packets. Because of that, it also holds that each user is able to obtain the files of all the other users with which it shares the cache content. Consequently, the transmission to the subset of users connected to cache~$i$, $\mathcal{U}_i$, who demand files $\{W^{(d_u)}\}_{u\in\mathcal{U}_i}$, is equivalent to the transmission to a single user with dedicated cache~$i$ that requested all files in $\{W^{(d_u)}\}_{u\in\mathcal{U}_i}$, and the lower and upper bounds here presented can be applied to that setting just by defining $L_i$ as the number of files that user~$i$ requests. 

With respect to the unequal link strength scenario from~\cite{TangLinkRatesTCOM18}, the authors present an achievable scheme whose memory allocation matches the one of our achievable scheme (as mentioned in~Observation~\ref{obs:relation}), although the process and design of the algorithms considerably differs. The converse result in~\cite{TangLinkRatesTCOM18} is a loose bound, whose gap is proportional to $12\frac{R_{\max}}{R_{\min}}$, where  $R_{\max}, R_{\min}$ denote the maximum and the minimum link capacities, respectively. Applying our converse approach to such scenario would close this gap; however, the derivation of the converse is not direct, as we require to prove that the models describing both scenarios are analogous. 

The most interesting work is~\cite{WiggerBenefitsofcache19}. They consider the cache-aided degraded BC with dedicated caches. In fact, the model in~\cite{WiggerBenefitsofcache19} contains the scenario analyzed in~\cite{TangLinkRatesTCOM18}. The authors in~\cite{WiggerBenefitsofcache19} derive lower and upper bounds for the rate-memory trade-off. While a similar memory allocation strategy arises in the achievable scheme, the converse results and derivation are utterly different from our contributions. Furthermore, their generic bound does not have a close-form solution, in the sense that it is required to find  the optimal choice of auxiliary variables to find the best bound, for each possible subset of coefficients. Both converse results are also difficult to compare, since~\cite{WiggerBenefitsofcache19} measures rate/capacity while we measure delivery time. The found analogies motivate further analysis on whether the tools derived in this work are applicable to such unequal-rate BC scenarios, although this fails out of the scope of this manuscript.

\section{Achievable Scheme}
\label{sec:achiev} 
In this section, we present our caching and delivery scheme, and we provide an analysis of its performance. This analysis allows us to prove Lemma~\ref{lem:achiev} from the characterization of the achievable delivery time. We recall that the said achievable delivery time is in turn proven optimal in Theorem~\ref{th:optimality}. 
    
In this section, we present the scheme for integer values of $t$,  $t\in\{1,2\dots,\Lambda\}$, while the case with non-integer $t$ is optimally handled by memory-sharing (cf. \cite{MN14}), and it is presented in Appendix~\ref{app:proof_lemma1_achiev_nonInt}.
%%%%%%%%%%%%%%%%%%%%%%%%%%%%%%%%%%%%%%%%%%
\subsection{Memory Allocation and Cache Placement}\label{subsec:cacheplacementBC}
%%%%%%%%%%%%%%%%%%%%%%%%%%%%%%%%%%%%%%%%%%
We first split each file $W^{(n)}, n\in[N]$, into
\begin{equation} \label{eq:subpacketization}
    S=e_t(\Up)=\mathlarger{\sum}\limits_{\tau\in \Cb{[\Lambda]}{t}}\prod\limits_{j=1}^{t}L_{\tau(j)},
\end{equation}
subfiles of equal size, such that $W^{(n)}$ is partitioned as 
\begin{equation*}
W^{(n)}=\Big\{W^{(n)}_{\tau,1},W^{(n)}_{\tau,2},\dots,W^{(n)}_{\tau,|A_{\tau}|}\mid\tau\in\Cb{[\Lambda]}{t} \Big\}
\end{equation*}
where $A_{\tau}\triangleq\{1,2,\dots,\prod\nolimits_{j=1}^{t}L_{\tau(j)}\}$. Afterwards, each cache $\lambda\in[\Lambda]$ stores in its memory all subfiles $W^{(n)}_{\tau,m_\tau}, m_\tau\in A_{\tau}$, whose first subscript $\tau$ includes $\lambda$, which results in the following  cache content.
$$\mathcal{Z}_\lambda=\Big\{W^{(n)}_{\tau,m_\tau}\mid W^{(n)}_{\tau,m_\tau}\in W^{(n)},\tau\ni\lambda,m_\tau\in A_\tau,n\in[N]\Big\}.$$
This automatically yields the memory allocation
\begin{equation}\label{eq:gamma_lambda}
    \gamm{\lambda}=\frac{L_\lambda\cdot e_{t-1}(\Up\setminus{\{L_\lambda\}})}{e_t(\Up)}, ~~\lambda\in[\Lambda].
\end{equation}
A detailed explanation on how to obtain~\eqref{eq:gamma_lambda} is presented in Appendix~\ref{app:proof_eq_gamma}.

 This same placement also assures that each subfile is cached in exactly $t$ caches (because each $\tau$ satisfies $|\tau| = t$), which guarantees the sum memory constraint in \eqref{eq:memconstr}. This memory constraint can also be verified by noting that
 \begin{align}
     \sum_{\lambda=1}^{\Lambda}\gamm{\lambda}=\sum_{\lambda=1}^{\Lambda}\frac{L_\lambda\cdot e_{t-1}(\Up\setminus{\{L_\lambda\}})}{e_t(\Up)}=t
 \end{align}
 where the last step follows directly from Property~\ref{prop:ripetitive_e_t} of the elementary symmetric functions.
Also, this placement yields an interesting property --- described in the following proposition %lemma 
--- that will be instrumental in the design and performance of the delivery phase.
\begin{prop}\label{prop:Pq}
    For any $(t+1)$-tuple $\CQ\subset [\Lambda]$, and for any specific cache $\lambda\in\CQ$, the total number of subfiles with first subscript $\tau=\CQ\setminus{\{\lambda\}}$ that are missing from all the users associated to cache $\lambda$ is the same for any $\lambda\in\CQ$ and it equals
    \begin{equation}\label{eq:P_Q}
        P_{\mathcal{Q}}\triangleq\prod\limits_{j=1}^{t+1}L_{\mathcal{Q}(j)}.
    \end{equation}
\end{prop}
\begin{proof}
For any $(t+1)$-tuple $\CQ\subset [\Lambda]$, consider cache $\lambda\in\CQ$ and let $\tau=\CQ\setminus{\{\lambda\}}$. There are $L_{\lambda}$ requested files from the users $\mathcal{U}_{\lambda}$ of cache $\lambda$, each having $\prod\nolimits_{j=1}^{t}L_{\tau(j)}$ subfiles with first index $\tau$. This means that the total number of subfiles that need to be sent to serve users in $\mathcal{U}_{\lambda}$ is $L_{\lambda}\prod\nolimits_{j=1}^{t}L_{\tau(j)}=\prod\nolimits_{j=1}^{t+1}L_{\CQ(j)}$, which does not depend on which~$\lambda\in\CQ$ is selected.
\end{proof}

%%%%%%%%%%%%%%%%%%%%%%%%%%%%%%%%
\subsection{Delivery phase}\label{subsec:delivery_BC}
%%%%%%%%%%%%%%%%%%%%%%%%%%%%%%%%
For ease of presentation, we will use $\Bd_{\lambda}$ to denote the vector of indices of the files requested by the users in $\mathcal{U}_{\lambda}$. 
For a fixed $(t+1)$-tuple $\CQ$ and any $\lambda\in\mathcal{Q}$, consider the set of subfiles 
\[\{W^{(\Bd_{\lambda}(j))}_{\tau,m}:j\in[L_{\lambda}], m\in A_{\tau}\}\] with first subscript $\tau=\CQ\setminus{\{\lambda\}}$, where these subfiles are desired by the users in $\mathcal{U}_{\lambda}$. Recalling from Proposition~\ref{prop:Pq} that the cardinality of this set is $P_\CQ$ (cf. \eqref{eq:P_Q}), we relabel the subfiles of the set as
\[\{F^{(\lambda)}_{\tau,j}:j\in[P_{\mathcal{Q}}]\}.\]
Because of the design of the cache placement phase in Section~\ref{subsec:cacheplacementBC}, we note that, for any $(t+1)$-tuple $\mathcal{Q}$ and any $j\in[P_{\mathcal{Q}}]$, the set of subfiles
\begin{equation}\label{eq:clique}
F^{(\lambda)}_{\mathcal{Q}\setminus{\{\lambda\}},j}, \ \forall \lambda\in \mathcal{Q},
\end{equation}
forms a clique of $t+1$ nodes. By Proposition~\ref{prop:Pq}, for any $(t+1)$-tuple $\mathcal{Q}\in[\Lambda]$, we have $P_{\mathcal{Q}}$ cliques as in \eqref{eq:clique}, all containing $t+1$ nodes. 
Consequently, we transmit, for each $(t+1)$-tuple $\mathcal{Q}\subseteq[\Lambda]$, the following $P_{\mathcal{Q}}$ XORs:
\begin{equation}
X_{\mathcal{Q}}(j)=\underset{\lambda\in \mathcal{Q}}{\Oplus}F^{(\lambda)}_{\mathcal{Q}\setminus{\{\lambda\}},j}, \ \ \forall j\in[P_{\mathcal{Q}}],
\end{equation}
whose structure allows for clique-based decoding as in~\cite{MN14}.

%%%%%%%%%%%%%%%%%%%%%%%%%%%%%%%%%%
\subsection{Performance of the scheme}\label{sec:perf_scheme}
%%%%%%%%%%%%%%%%%%%%%%%%%%%%%%%%%%
The fact that there are $P_{\mathcal{Q}}$ XORs for each $(t+1)$-tuple $Q$ implies a total of
\[
\mathlarger{\sum}\limits_{\CQ\in \Cb{[\Lambda]}{t+1}}P_{\CQ}=
\mathlarger{\sum}\limits_{\CQ\in \Cb{[\Lambda]}{t+1}}~\prod\limits_{j=1}^{t+1}L_{\mathcal{Q}(j)}=e_{t+1}(\Up)
\]
transmissions, and a corresponding delivery time of
\begin{equation} \label{eq:Delay3}
T(t,\Up)=\frac{e_{t+1}(\Up)}{e_{t}(\Up)},
\end{equation}
where the denominator $e_t(\Up)$ is due to \eqref{eq:subpacketization}. In the proof of Theorem~\ref{th:optimality} (cf.~\eqref{eq:proof_relation_Tachiev}), we have seen that the above achievable delivery time in~\eqref{eq:Delay3} can be written in the more standard form
\begin{equation}\label{eq:T_ach}
    T(t,\Up)=\frac{\sum_{\lambda=1}^{\Lambda}L_{\lambda}(1-\gamm{\lambda})}{t+1},
\end{equation}
where $\{\gamm{\lambda}\}_{\lambda=1}^\Lambda$ is the memory allocation obtained in~\eqref{eq:gamma_lambda} and that leads to~\eqref{eq:gammas} in Lemma~\ref{lem:achiev}.

\section{Information Theoretic Converse}
\label{sec:converse}
In this section, we present a converse bound on the optimal delivery time $T^*(t,\Up)$, which will serve as a proof for Theorem \ref{thm:lower_bound_total}. 
We will also prove Theorem~\ref{thm:lower_bound_regular} by restricting the cache placement scheme to be regular as in Definition~\ref{def:symmetric_placement}, which implies that $t$ is integer.

This converse result builds on a different approach with respect to previous bounds. We remark that, in the scenario here considered, we are deriving the optimal cache placement \emph{and} the optimal memory allocation, i.e., the size of each cache memory. This diverges from previous results, where the optimal placement was derived for a homogeneous memory allocation~\cite{PUETIT2020}. Indeed, as we will prove in the following, the results are against the intuition from \cite{PUETIT2020} that more homogeneous profiles $\Up$ would lead to better performances, as it turns out to be the opposite. 

In what follows, we denote the set of {demand vectors having} distinct file requests by $\Dwc$, such that $\Dwc\triangleq\{\Bd\in[N]^K:d_j\neq d_i,~\forall~ i\neq j\}$. Finally, we will use the notation $W^{(i)}_{\tau}$ to refer to the part of file $W^{(i)}$ exclusively stored in the caches in set $\tau$.

%%%%%%%%%%%%%%%%%%%%%%%%%%%%%%%%%
\subsection{Lower bounding  $T^*(\Z,\Bd,\Up)$}
%%%%%%%%%%%%%%%%%%%%%%%%%%%%%%%%%
We first present a \emph{generic} lower bound on the delivery time $T^*(\Z,\Bd,\Up)$ as a function of the cache permutation $\sigma\in\CS_\Lambda$, where $\mathcal{S}_n$ denotes the symmetric group of all permutations of $[n]$. 
This result was first stated in~\cite{PUETIT2020} (cf.~equation (51)). However, in~\cite{PUETIT2020}, this partial result was not stated as a lemma and/or proposition, and we reproduce it here for the sake of completeness.

\begin{lem}\label{lem:bound_mais} 
Consider the delivery phase of a shared-cache network with a cache placement described by $\Z$, demand vector $\Bd$ and cache occupancy vector $\Up$. Then, under the assumption of uncoded cache placement, the optimal delivery time $T^*(\Z,\Bd,\Up)$ can be lower bounded by the quantity
\begin{equation} \label{eq:TLB}
    T_{lb,\sigma}(\Z,\Bd,\Up)\triangleq  \sum_{\lambda=1}^{\Lambda}\sum_{\ell=1}^{L_{\sigma(\lambda)}}\sum_{\tau_{\lambda}\subseteq [\Lambda]\setminus \{\sigma(1),\dots,\sigma(\lambda)\}}\!\!\!\!|W^{(\Bd_{\sigma(\lambda)}(\ell))}_{\tau_{\lambda}}|,
\end{equation}
where $\sigma$ denotes an arbitrary permutation of the set of caches $[\Lambda]$.
\end{lem}
\begin{proof}
    The proof\footnote{The proof of this lemma is fully presented in~\cite{PUETIT2020}. Therefore, we omit in this work the detailed derivation and restrict ourselves to provide the main aspects of the proof, whereas we refer to Section~\RomanNumeralCaps{5} of~\cite{PUETIT2020} for a detailed proof.} builds on index coding arguments and is an adaptation of Corollary~1 in~\cite{ArbaWang13} to the considered caching problem, in a similar manner as how it has been done in~\cite{WanTuninettiTIT2020}. As described in \cite{PUETIT2020}, for any cache placement $\Z$, any demand vector $\Bd$, and any cache occupancy vector $\Up$, the caching problem considered here can be converted into an index coding problem {and its associated \emph{side information graph}}. Then, Lemma 1 in \cite{PUETIT2020} can be used to obtain a lower bound on $T^*(\Z,\Bd,\Up)$ by identifying any acyclic subgraph of the side-information graph. Next, in the Appendix Section~\RomanNumeralCaps{7}-B of~\cite{PUETIT2020} it is proved that, for any cache permutation $\sigma\in\CS_{\Lambda}$, an acyclic subgraph can be identified and used in conjunction with Lemma~$1$ in~\cite{PUETIT2020} to prove our above lemma.
\end{proof}

Next, an \emph{adaptable} lower bound on $T^*(\Z,\Bd,\Up)$ can be constructed as a weighted average of the $\Lambda!$ possible lower bounds 
that stem from~\eqref{eq:TLB}. Thus, it holds that
\begin{equation}\label{eq:weighted_average}
    T^*(\Z,\Bd,\Up)\geq \sum_{\sigma\in\CS_\Lambda}w_\sigma T_{lb,\sigma}(\Z,\Bd,\Up),
\end{equation}
where the weights $\{w_\sigma\}$ satisfy $\sum_{\sigma\in\CS_\Lambda}w_\sigma=1$. 
Any of such possible sets of weights $\{w_\sigma\}$ provides a valid lower-bound for our problem.

\begin{rem}
Generally, the approach to construct lower bounds on the delivery time for coded caching problems in works that follow the index coding approach originally proposed in \cite{WanTuninettiTIT2020} (and the similar genie-aided approach used in \cite{YuMA16}) is based on generating symmetry, e.g., by averaging over all possible permutations, or based on a certain cache permutation $\sigma$  (see for example \cite{YuMA16,WanISIT16,HamdiEleftTIT2021, PUETIT2020}). 
This method has been shown to work well in settings that are uniform in terms of number of users per cache and sizes of the caches. However, whenever the system model is affected by some heterogeneity, this approach can easily fail to meet the goal. For the scenario here considered, we have shown in~\cite{ParEliaITW19} that the uniform average (i.e. $w_\sigma=\frac{1}{\Lambda!}$) leads to a loose bound, which proved our achievable performance to be optimal within a gap that scales linearly with the normalized total cache size~$t$.

A key contribution of this work is to show that the limitations of index coding bounds in heterogeneous settings are not fundamental and can be overcome by an asymmetric combination of lower bounds, where the combination depends on the topology of the setting. For that, the use of the weighted average and a careful choice of the weights in~\eqref{eq:weighted_average} is crucial to the construction of a tight bound. 
This approach is utterly different from previous solutions inasmuch as  before the goal was generally to \emph{avoid} asymmetry, as it was thought that otherwise the exponential complexity induced by the combinatorial nature of the problem would make unfeasible to find a solution. Conversely, we take the opposite direction and we \emph{enforce} asymmetry, but  a structured asymmetry that allows us to map the heterogeneity of the setting to the bound. 
We believe that this approach can be helpful to derive lower bounds for other generic heterogeneous coded caching problems.   
\end{rem}

In our derivation, the weights $w_\sigma$ depend on a parameter $p$. In particular, for any $p\in [\Lambda]_0$, the choice of the weights $w_\sigma$ is taken as
\begin{equation}\label{eq:weights}
    w_\sigma^{(p)}\triangleq\frac{\prod\limits_{j=1}^{p}L_{\sigma(\Lambda-p+j)}}{\sum_{\sigma'\in\CS_\Lambda}\prod\limits_{j=1}^{p}L_{\sigma'(\Lambda-p+j)}},
\end{equation}
where we have used the upper index $^{(p)}$ to highlight the dependency of the value of the weights on the choice of the parameter~$p$, and we recall that $\sigma(n)$ denotes the $n$-th element of any ordered set $\sigma$. For $p=0$ we define $w^{(0)}_\sigma\triangleq \frac{1}{|\CS_\Lambda|}$.

\subsection{Lower bound on $T^*$}

We now proceed to derive the lower bound on the optimal delivery time $T^*$. In this respect, we start by bounding from below the worst-case delay for a fixed cache placement $\Z$, where the bound is obtained as the average rate over all demands with distinct requests as
    \begin{equation}\label{eq:bound_by_average}
        T^*(\Z,\Up)\triangleq\max_{\Bd}~T^*(\Z,\Bd,\Up)\geq \frac{1}{|\Dwc|}\sum_{\mathbf{d}\in \Dwc}T^*(\Z,\Bd,\Up).
    \end{equation}
Combining \eqref{eq:bound_by_average} and \eqref{eq:weighted_average}, for any $p\in[\Lambda]_0$,  yields 
    \begin{align}\label{eq:TLBcompact}
        T^*(\Z,\Up)
        &\geq\frac{1}{|\Dwc|}\sum_{\Bd\in \Dwc}\sum_{\sigma\in\CS_\Lambda}w_\sigma^{(p)}T_{lb,\sigma}(\Z,\Bd,\Up)\\
        &\overset{(a)}{\geq}\underbrace{\frac{1}{|\Dwc|}\sum_{\Bd\in \Dwc}\sum_{\sigma\in\CS_\Lambda}w_\sigma^{(p)}\sum_{\lambda=1}^{\Lambda}\sum_{\ell=1}^{L_{\sigma(\lambda)}}\sum_{\tau_{\lambda}\subseteq [\Lambda]\setminus \{\sigma(1),\dots,\sigma(\lambda)\}}\!\!\!\!|W^{(\Bd_{\sigma(\lambda)}(\ell))}_{\tau_{\lambda}}|}_{\triangleq T_{lb,p}(\Z,\Up)},\label{eq:lb_long}
    \end{align}
where in $(a)$ we have used~\eqref{eq:TLB}. 

Next, we rewrite the right-hand side of \eqref{eq:lb_long}, which we denote by $T_{lb,p}(\Z,\Up)$, in the more compact form
	\begin{equation}\label{eq:TLBcmin}
		T_{lb,p}(\Z,\Up)
		        = \sum_{n=1}^N\sum_{\tau\in 2^{[\Lambda]}}c_{\tau,n}^{(p)}\frac{|W^{(n)}_\tau|}{N},
	\end{equation}
where the value of $c_{\tau,n}^{(p)}$ is expressed in the following lemma.
Before presenting the lemma, let us introduce the  notation $\Ld_{q}\triangleq \prod_{j=1}^{|q|} L_{q(j)}$ for any subset $q\subseteq [\Lambda]$  for the sake of readability. 

    \begin{lem}\label{lem:long_coefficient}
        The value of $c_{\tau,n}^{(p)}$ does not depend on the file index $n$ and it takes the form
        \begin{equation}\label{eq:value_coefficient}
            c_{\tau}^{(p)}= \frac{%
								1
							}{%
								\sum_{\ell\in \Cb{[\Lambda]}{p}} \Ld_{\ell} %
							}
						\left(
							\sum_{q\in \Cb{[\Lambda]}{p+1}}^{}\Ld_{q} \frac{p+1-|q\cap \tau|}{|\tau|+1-|q\cap \tau|}%
							+
							\frac{1}{|\tau|+1}
								\Ld_{\tau}
								\sum_{s\in
								\Cb{[\Lambda]\backslash\{\tau\}}{p-j}
								}
								\left(
								\Ld_s \cdot \sum_{i=1}^{p-j}L_{s(i)}\right)
						\right)	.
        \end{equation}
    \end{lem}
    \begin{proof}
        The proof of this lemma is presented in Appendix~\ref{subsec:proof_of_lemma_long_coeff}.
    \end{proof}
We can tighten the bound on $T^*(\Z,\Up)$ by selecting the most restricting $p$, such that
\begin{equation}\label{eq:max_p}
    T^*(\Z,\Up)\geq \max_{p\in [\Lambda]_0}T_{lb,p}(\Z,\Up).
\end{equation}

From the definition of the optimal delay $T^*(t,\Up)$ in \eqref{eq:goal}, and from~\eqref{eq:TLBcmin} and~\eqref{eq:max_p}, 
{we get}
\begin{align}
    T^*(t,\Up)&=\min_{\Z}T^*(\Z,\Up)\\
       &\geq\min_{\Z}\max_{p\in [\Lambda]_0} \sum_{n=1}^N\sum_{\tau\in 2^{[\Lambda]}}c_{\tau}^{(p)}\frac{|W^n_\tau|}{N}\nonumber\\
	   &=\min_{\Z}\max_{p\in [\Lambda]_0} \sum_{\tau\in 2^{[\Lambda]}} c_{\tau}^{(p)} a_{\tau}\label{eq:full_bound_interm_step},
\end{align}
where we have introduced the notation %used
$a_{\tau}\triangleq\frac{1}{N}\left(|W^{(1)}_{\tau}|+|W^{(2)}_{\tau}|+...+|W^{(N)}_{\tau}|\right)$. Now, by considering the library size and the sum cache size constraints, a lower bound on the optimal delay $T^*(t,\Up)$ can be obtained from the solution of the following linear program
\begin{equation}\label{eq:general_LP_bound}
    \begin{aligned}
    &\min_{a_\tau}\max_{p\in [\Lambda]_0}\quad  \sum_{\tau\in 2^{[\Lambda]}} c_{\tau}^{(p)} a_{\tau}\\
    & \text{subject to} \sum_{\tau\in 2^{[\Lambda]}}a_{\tau}=1,\\
    & \quad \quad \quad \quad  \sum_{\tau\in 2^{[\Lambda]}}|\tau|a_{\tau}=t,\\
    & \quad \quad \quad \quad  a_\tau\geq 0, ~~~\forall \tau \in 2^{[\Lambda]}.
    \end{aligned}
\end{equation}
Let us now focus on the proof of the general lower bound in Theorem~\ref{thm:lower_bound_total}, and later we will consider the proof for the case with regular placement of Theorem~\ref{thm:lower_bound_regular}, to conclude with an optimality result for the case where we are not restricted to regular placement.

%%%%%%%%%%%%%%%%%%%%%%%%%%%%%%%%%%%%
\subsection{Proof of Theorem \ref{thm:lower_bound_total}}
%%%%%%%%%%%%%%%%%%%%%%%%%%%%%%%%%%%%
In what follows, we further lower-bound the constructed lower bound in \eqref{eq:general_LP_bound}. 
% First of all, it holds that
First of all, let us introduce some useful notation. 
We {first} define $\Tilde{c}^{(p)}_\tau$ as 
   $$\Tilde{c}^{(p)}_\tau\triangleq 
    \frac{%
			\sum_{q\in \Cb{[\Lambda]}{p+1}}^{}\Ld_{q} \frac{p+1-|q\cap \tau|}{|\tau|+1-|q\cap \tau|}
	}{%
		\sum_{\ell\in \Cb{[\Lambda]}{p}} \Ld_{\ell} %
	}.$$ 
Then, we define the subset of cardinality $j$ that minimizes $\Tilde{c}^{(p)}_\tau$ as $\tau^\star_j$, i.e., $\tau^\star_j \triangleq \argmin_{\tau\in 2^{[\Lambda]}: |\tau|=j}\Tilde{c}_{\tau}^{(p)}$, and we define $\Bar{a}_j$ as $\Bar{a}_j\triangleq \sum_{\tau\in 2^{[\Lambda]}:|\tau|=j}a_\tau$.
Then, it holds that
\begin{align}
		\sum_{\tau\in 2^{[\Lambda]}} c_{\tau}^{(p)} a_{\tau}
				&\overset{(a)}{\geq} \sum_{\tau\in 2^{[\Lambda]}} \Tilde{c}_{\tau}^{(p)} a_{\tau}
				 \overset{(b)}{\geq} \sum_{j=0}^{\Lambda}\Tilde{c}_{\tau^\star_j}^{(p)}\Bar{a}_j,\label{eq:last_Tlbp}
	\end{align}	
where in $(a)$ we have applied the fact that $\Tilde{c}_{\tau}^{(p)}\leq c_{\tau}^{(p)}$ (cf.~\eqref{eq:value_coefficient}), and in $(b)$ we have used the definitions of  $\tau^\star_j$ and~$\Bar{a}_j$.

We now provide a lemma that provides the value of the optimal $\tau^\star_j$, for any  $p\in[\Lambda]_0$ and $j\in[\Lambda]_0$.
    \begin{lem}\label{lem:optimal_tau}
		Let us consider that the caches are sorted such that $L_1\geq L_2\geq\dots\geq L_\Lambda$. Then, for any cardinality $|\tau|=j$, $j\in[\Lambda]$, it holds that
		\begin{align}
				\argmin_{\substack{\tau\subseteq [\Lambda] \\ |\tau|=j}} \Tilde{c}_{\tau}^{(p)} 
				\triangleq 
				\begin{cases}
						\tau^\star_j  = \{\emptyset\} & \text{ if } j = 0\\ 
						\tau^\star_j  = \{\Lambda -j + 1, \ \Lambda -j + 2,\ \dots,\  \Lambda\} & \text{ if } 1\leq j < p\\ 
						\tau^\star_j  =  \{1,\ 2,\ \dots,\  j\} = [j] & \text{ if } j \geq p.\\				
				\end{cases}
		\end{align}
		Note that for $j=p$, $\Tilde{c}_\tau^{(p)} $ is the same for every $\tau$ such that $|\tau| = j$.  
		This means that $\Tilde{c}_{\tau^\star_j}^{(p)} = \Tilde{c}_{\tau}^{(p)}$ $\forall \tau : |\tau| = j$.  
    \end{lem}
    \begin{proof}
        The proof is relegated to Appendix~\ref{app:optimal_tau}.
    \end{proof}
Now, jointly employing \eqref{eq:last_Tlbp} in \eqref{eq:full_bound_interm_step} and using the max-min inequality yields
\begin{equation}
    \min_{\Z}\max_{p\in [\Lambda]_0} \sum_{\tau\in 2^{[\Lambda]}} c_{\tau}^{(p)} a_{\tau}\geq \max_{p\in [\Lambda]_0}  \min_{\Z}  \sum_{j=0}^{\Lambda}\Tilde{c}_{\tau^\star_j}^{(p)}\Bar{a}_j
\end{equation}
which implies that
    \begin{equation}\label{eq:final_LP}
        \begin{aligned}
			T^*(t,\Up)\geq & \max_{p\in \{0,1,\dots,\Lambda\}} \min_{\bar{a}_j} \sum_{j=0}^{\Lambda}\Tilde{c}_{\tau^\star_j}^{(p)}\Bar{a}_j\\
				&\quad\text{subject to} \quad ~
				\sum_{j=0}^\Lambda \bar{a}_j = 1\\
				&~\quad\qquad \quad \quad \quad\sum_{j=0}^\Lambda j \bar{a}_j = t \\
				&~\quad\quad\qquad \quad \quad \bar{a}_j \geq 0,\quad\forall j\in[\Lambda]_0. 
		\end{aligned}	
    \end{equation}
    We present now a result that will be instrumental in establishing the following step in the derivation. 
    \begin{prop}\label{lem:monotone_c}
		The sequence $\big\{\Tilde{c}_{\tau^\star_j}^{(p)}\big\}$ is a decreasing sequence in $j\in [\Lambda]_0$.
    \end{prop}
    \begin{proof}
    The proof is relegated to Appendix~\ref{app:monotone_c_proof}.
    \end{proof}
We now focus on the inner optimization problem in \eqref{eq:final_LP} for any fixed %any
$p\in [\Lambda]_0$, and we follow the same steps as in \cite{YuMA16} to solve this problem analytically. 
In this respect, we know from Proposition~\ref{lem:monotone_c} that $\big\{\Tilde{c}_{\tau^\star_j}^{(p)}\big\}$ is a decreasing sequence in $j\in [\Lambda]_0$, and thus 
its convex envelope is a decreasing and convex sequence. Thus, applying Jensen inequality, we obtain that
\begin{equation}
  T^*(t,\Up)\geq \max_{p\in [\Lambda]_0} \mathcal{T}^{(p,\Up)}_{low}(t)      
\end{equation}
where 
\begin{equation}
   \mathcal{T}^{(p,\Up)}_{low}(t)\triangleq \Conv_{j\in[\Lambda]_0}\left(\Tilde{c}_{\tau^\star_{j}}^{(p)}\right).
\end{equation}
Then, Theorem \ref{thm:lower_bound_total} simply follows from the fact that
\begin{equation}
    \max_{p\in \{0,1,\dots,\Lambda\}} \mathcal{T}^{(p,\Up)}_{low}(t) \geq \mathcal{T}^{(\bar{t},\Up)}_{low}(t),
\end{equation}
where $\bar{t}=\round(t)$. Consequently, we have proved Theorem~\ref{thm:lower_bound_total}.

%%%%%%%%%%%%%%%%%%%%%%%%%%%%%%%%%%%%%%%%%%%%%%%%
\subsection{Proof of Theorem~\ref{thm:lower_bound_regular}}
%%%%%%%%%%%%%%%%%%%%%%%%%%%%%%%%%%%%%%%%%%%%%%%%
Since under the regular assumption of Definition~\ref{def:symmetric_placement} it holds that $t$ is integer (i.e., $t\in [\Lambda]_0$), we first note that $\bar{t}=t$. Furthermore, this assumption implies that $\Bar{a}_t=1$ and $\Bar{a}_j=0$ for any $j\in\{[\Lambda]_0\backslash t\}$. 
We can lower bound~\eqref{eq:final_LP} by fixing $p$ to $p=t$, which reduces~\eqref{eq:final_LP} to
\begin{equation}\label{eq:silly_LP}
\begin{aligned}
				&\quad \min_{\bar{a}_j}\quad \quad\Tilde{c}_{\tau^\star_t}^{(t)}\Bar{a}_t\\
					&\text{subject to} \quad 
					\bar{a}_t = 1.
			\end{aligned}	
\end{equation}
It is easy to verify that, for all $\tau\in[\Lambda]:|\tau|=t$, with $p=t$, it holds that
\begin{equation}\label{eq:the2_proof_ele}
    \Tilde{c}_{\tau}^{(t)}=\Tilde{c}_{\tau^\star_t}^{(t)}=\frac{\sum_{q\in \Cb{[\Lambda]}{t+1}}\prod\limits_{j=1}^{t+1} L_{q(j)}}{ \sum_{q\in \Cb{[\Lambda]}{t}}\prod\limits_{j=1}^{t}L_{q(j)}} = \frac{e_{t+1}(\Up)}{e_{t}(\Up)} = \mathcal{T}^{(t,\Up)}_{low,\; reg}(t),
\end{equation}
where the last step follows from the definition of $\mathcal{T}^{(t,\Up)}_{low,\; reg}(t)$ in~\eqref{eq:regular_lower_bound}. 
This, together with \eqref{eq:silly_LP}, directly results in
\begin{equation}
    T^*(t,\Up)\geq \Tilde{c}_{\tau^\star_t}^{(t)} = \mathcal{T}^{(t,\Up)}_{low,\; reg}(t),
\end{equation}
which concludes the proof of Theorem~\ref{thm:lower_bound_regular}. \qed

%%%%%%%%%%%%%%%%%%%%%%%%%%%%%%%%%%%%%%%%%%%%%%%%
\subsection{Optimality beyond regular placement}
%%%%%%%%%%%%%%%%%%%%%%%%%%%%%%%%%%%%%%%%%%%%%%%%
Apart from the optimality results presented in Theorem~\ref{th:optimality} for the case with regular placement, we present in the following theorem a new tight bound for the case where no assumption on regular placement is taken, and for which the achievable delivery time in~\eqref{eq:Tachiev} is exactly optimal.

 \begin{thm}\label{th:optimality2}
    For integer values of $t\in [\Lambda]_0$, the achievable delivery time  $T_{}(t,\Up)$ in \eqref{eq:Tachiev} is exactly optimal under the assumption of uncoded cache placement when the sequence $\{\Tilde{c}_{\tau^\star_j}^{(t)}\}_{j\in[\Lambda]_0}$ is convex in $j$, where 
                \begin{align}\label{eq:def_ctj}
                    \Tilde{c}_{\tau^\star_j}^{(t)}\triangleq\frac{%
    				    \sum_{q\in \Cb{[\Lambda]}{t+1}}^{}\frac{t+1-|q\cap \tau^\star_j|}{j+1-|q\cap \tau^\star_j|}\prod_{i=1}^{t+1} L_{q(i)} 
    		         }{%
    			        \sum_{\ell\in \Cb{[\Lambda]}{t}} \prod_{i=1}^{t+1}L_{\ell(i)}. %
    		        }.
                \end{align}
    \end{thm}
\begin{proof}
    From~\eqref{eq:def_ctj}, we can write $\mathcal{T}^{({t},\Up)}_{low}(x)$ (defined in~\eqref{eq:general_lower_bound}) as
    \begin{align}\label{eq:general_lower_boundbis}
     \mathcal{T}^{(t,\Up)}_{low}(x) &= 
		\Conv_{j\in[\Lambda]_0}\left(\Tilde{c}_{\tau^\star_j}^{(t)}\right).
    \end{align}
    For any convex sequence, its lower convex envelope contains all its elements. 
    Thus, if $\{\Tilde{c}_{\tau^\star_j}^{(t)}\}_{j\in[\Lambda]_0}$ is convex, then   $\mathcal{T}^{(t,\Up)}_{low}(t)=\Tilde{c}_{\tau^\star_t}^{(t)}$ for any integer point $t\in [\Lambda]_0$. 
    
    Moreover, as explained in~\eqref{eq:the2_proof_ele} for the proof of Theorem~\ref{thm:lower_bound_regular}, it holds that, for $j=t$, 
        $\Tilde{c}_{\tau^\star_t}^{(t)}
	        =\frac{e_{t+1}(\Up)}{e_{t}(\Up)}$, 
    and also that (cf.~\eqref{eq:proof_relation_Tachiev})
    \begin{equation}
        \begin{aligned}\label{eq:proof_relation_Tachiev_app}
        T^*(t,\Up)&\geq\frac{e_{t+1}(\Up)}{e_{t}(\Up)}
           {=}\frac{\sum_{\lambda=1}^{\Lambda}L_{\lambda}(1-\gamm{\lambda})}{t+1}=T_{}(t,\Up),%\nonumber,
        \end{aligned}
    \end{equation}
    which concludes the proof of the optimality of Theorem~\ref{th:optimality2}.
\end{proof}

Next, we present an example of a setting for which our scheme is optimal under the constraint of uncoded cache placement without assuming regular placement.
\begin{exmp}
Consider the cache-aided network of the example in Section~\ref{sec:example_shared_link} with $\Up=(3,2,1)$ and $t=2$. The sequence $\big\{\Tilde{c}_{\tau^\star_j}^{(2)}\big\}_{j\in\{0,1,2,3\}}$ takes the values $\{\nicefrac{18}{11},\nicefrac{12}{11},\nicefrac{6}{11},0\}$, which is a convex sequence. From Theorem~\ref{th:optimality2}, this implies that for the considered example in Section~\ref{sec:example_shared_link} the achievable delivery time $T^*(2,(3,2,1))=\nicefrac{6}{11}$ is information-theoretically optimal under the assumption of uncoded cache placement. 
\end{exmp}

\section{The scenario with imperfect topology knowledge}\label{sec:topology_partially_aware}
In this section, we present the achievable scheme and the matching converse for the {scenario with imperfect topology knowledge} previously described. 
We remind the reader that our setting entails only partial knowledge about the cache occupancy vector $\Up$, and that this available knowledge is denoted by $\bar{\Up}$.  
This analysis allows us to characterize the impact of the degree of knowledge about the network topology on the performance of coded caching. 
In this scenario, we recall that the memory allocation and placement phases assume that the future cache occupancy vector during the delivery phase matches the available information $\bar{\Up}$, although eventually this cache occupancy vector turns out to be $\Up$ at the delivery phase. 

We first present the proposed scheme for this scenario and its performance, followed by the converse analysis to provide a bound. We then prove that the presented scheme attains the converse result to prove Theorem~\ref{thm:thmL'}, and finally we provide some numerical examples to better illustrate the impact of the topology knowledge. 

%%%%%%%%%%%%%%%%%%%%%%%%%%%%%
\subsection{Achievable scheme}\label{subsec:ach_topology-partially-aware}
%%%%%%%%%%%%%%%%%%%%%%%%%%%%%

With the knowledge of the cache occupancy vector $\bar{\Up}$ at hand  --- which  as we recall can represent the expected cache occupancy vector in the delivery phase --- the server designs the memory allocation $\{\gamma_\lambda\}_{\lambda=1}^\Lambda$ and cache placement $\Z^*_{\bar{\Up}}$ as described in Section~\ref{subsec:cacheplacementBC}. For the subsequent delivery phase with topology described by $\Up$, the following fact holds.
\begin{prop}\label{prop:PQ2}
For any $(t+1)$-tuple $\mathcal{Q}\subset[\Lambda]$, any scalar $\lambda\in\mathcal{Q}$ and tuple $\tau=\mathcal{Q}\setminus\{\lambda\}$ ($|\tau|=t$), the total number of subfiles of the form $W^{(n)}_{\tau,m_\tau}$ %with first subscript
that are missing from all users associated to any specific cache $\lambda\in\mathcal{Q}$ is equal to
\begin{equation}
P_{\lambda_\mathcal{Q}}=L_\lambda \prod_{j=1}^{t}\bar{L}_{\tau(j)}.    
\end{equation}

\end{prop}

\begin{proof}
Considering cache $\lambda$, there are $L_\lambda$ users requesting $L_\lambda$ files. For each of these files there are $|A_{\tau}| = \prod_{j=1}^{t}\bar{L}_{\tau(j)}$ subfiles with first index  $\tau=\mathcal{Q}\setminus\{\lambda\}$.  
\end{proof}

In what follows, we will use $\Bd_{\lambda}$ to denote the vector of indices of the files requested by the users in $\mathcal{U}_{\lambda}$.
Similarly to the delivery scheme in section~\ref{subsec:delivery_BC}, for a fixed $(t+1)$-tuple $\CQ$ and any $\lambda\in\mathcal{Q}$, let us consider the set of subfiles with first subscript $\tau=\CQ\setminus{\{\lambda\}}$ that are requested from users in $\mathcal{U}_{\lambda}$, i.e.,
\[\{W^{(\Bd_{\lambda}(j))}_{\tau,m}:j\in[L_{\lambda}], m\in A_{\tau}\},\]
where we recall that $A_{\tau}=\big\{1,2,\dots,\prod\nolimits_{j=1}^{t}\bar{L}_{\tau(j)}\big\}$. From Proposition \ref{prop:PQ2}, we know that the cardinality of this set is $P_{\lambda_\CQ}$, and thus we can relabel the set of these subfiles with successive integer indexes as
\[\CF_{\lambda_\CQ}=\{F^{(\lambda)}_{\tau,j}:j\in[P_{\lambda_\CQ}]\}.\]

Let us now define the quantity $P^{(max)}_{\CQ}\triangleq\underset{\lambda\in\CQ}{\max}~P_{\lambda_\CQ}$, and let us note that, for each $\lambda\in\CQ$, it holds that  $F^{(\lambda)}_{\tau,j}\triangleq\emptyset$ for any index~$j$ such that $P_{\lambda_\CQ}<j\leq P_{\CQ}^{(max)}$.
Because of the design of the cache placement phase, we notice that for any $(t+1)$-tuple $\mathcal{Q}$ and any $j\in[P^{(max)}_{\mathcal{Q}}]$, the set of subfiles
\begin{equation}\label{eq:clique_partially}
F^{(\lambda)}_{\mathcal{Q}\setminus{\{\lambda\}},j}, \quad \forall \lambda\in \mathcal{Q}
\end{equation}
forms a clique of $t+1$ nodes. For any $(t+1)$-tuple $\mathcal{Q}\in[\Lambda]$, we have $P^{(max)}_{\mathcal{Q}}$ cliques as in \eqref{eq:clique_partially}, all corresponding to  $t+1$ nodes. Consequently, we transmit the following $P^{(max)}_{\mathcal{Q}}$ XORs for each $(t+1)$-tuple $\mathcal{Q}\subseteq[\Lambda]$:
\begin{equation}
X_{\mathcal{Q}}(j)=\underset{\lambda\in \mathcal{Q}}{\Oplus}F^{(\lambda)}_{\mathcal{Q}\setminus{\{\lambda\}},j}, \ \ \forall j\in[P^{(max)}_{\mathcal{Q}}],
\end{equation}
whose structure allows for clique-based decoding as in~\cite{MN14}.

\subsubsection*{Delay Evaluation} 
For each $(t+1)$-tuple $\mathcal{Q}$, the server transmits 
$$P^{(max)}_{\mathcal{Q}}=\max_{\lambda\in \mathcal{Q}}P_{\lambda_\mathcal{Q}}$$
different XORs. The total number of XORs sent through the channel is
    \begin{align}\label{eq:total_xor}
        \sum_{\mathcal{Q}\in \Cb{[\Lambda]}{t+1}} \max_{\lambda\in \mathcal{Q}}P_{\lambda_\mathcal{Q}}=\sum_{\mathcal{Q}\in \Cb{[\Lambda]}{t+1}}\max_{\lambda\in \mathcal{Q}}L_\lambda \prod_{j=1}^{t}\bar{L}_{\tau_\lambda(j)}
    \end{align}
where $\tau_\lambda\triangleq\CQ\setminus\{\lambda\}$. 
The subpacketization applied at cache placement $\Z^*_{\bar{\Up}}$ (cf.~\eqref{eq:subpacketization})  and~\eqref{eq:total_xor} imply that the normalized delivery time of the achievable scheme for any $t,\Up$ and $\bar{\Up}$ is 
\begin{equation}\label{eq:ach_delayL'}
    T(t,\Up,\bar{\Up})=\frac{\sum_{\CQ\in \Cb{[\Lambda]}{t+1}}\max_{\lambda\in \CQ}L_\lambda \dot{\bar{\BL}}_{\tau_\lambda}}{e_t(\bar{\Up})},
\end{equation}
where we have applied the notation  $\dot{\bar{\BL}}_{\tau_\lambda}\triangleq \prod_{j=1}^t\bar{L}_{\tau_\lambda(j)}$, and we recall that~$e_t(\bar{\Up})\triangleq\sum_{\tau_\lambda\in\Cb{[\Lambda]}{t}}\prod_{j=1}^t\bar{L}_{\tau_\lambda(j)}$.

%%%%%%%%%%%%%%%%%%%%%%%%%%%%%%%%%%%%
\subsection{Converse bound}
%%%%%%%%%%%%%%%%%%%%%%%%%%%%%%%%%%%%
To develop the lower bound for $T^*(t,\Up,\bar{\Up})$, we immediately observe that the applied placement is determined and given by $\Z^*_{\bar{\Up}}$, i.e., the optimal placement for $\bar{\Up}$. 
Before proceeding with the proof, we note that, under the cache placement $\Z^*_{\bar{\Up}}$, {we can simplify the notation by considering as a single subfile $W^{(n)}_\tau$ the set of subfiles stored exactly in the caches in set $\tau$, for any set $\tau\subset [\Lambda]$ of cardinality $|\tau|=t$,  such that $W^{(n)}_\tau=\{W^{(n)}_{\tau,m_\tau}:m_\tau\in A_\tau\}$ for any $n\in[N]$.}

Under the cache placement $\Z^*_{\bar{\Up}}$, Lemma \ref{lem:bound_mais} also holds for the considered scenario with imperfect topology knowledge, such that $T^*(\Z^*_{\bar{\Up}},\Bd,\Up)$ can be lower bounded as
\begin{align}
   T^*(\Z^*_{\bar{\Up}},\Bd,\Up)
   &\geq   \sum_{\lambda=1}^{\Lambda}\sum_{\ell=1}^{L_{\sigma(\lambda)}}\sum_{\tau_{\lambda}\subseteq [\Lambda]\setminus \{\sigma(1),\dots,\sigma(\lambda)\}}\!\!|W^{(\Bd_{\sigma(\lambda)}(\ell))}_{\tau_{\lambda}}|,\\
    &=\sum_{\lambda=1}^{\Lambda}L_{\sigma(\lambda)}\sum_{q\in\Cb{[\Lambda]\setminus\{\sigma(1),\dots,\sigma(\lambda)\}}{t}}\!\!\frac{\dot{\bar{\BL}}_q}{e_t(\bar{\Up})},\label{eq:eqmais2}
\end{align}
where \eqref{eq:eqmais2} follows directly from the fact that, under the cache placement $\Z^*_{\bar{\Up}}$, we have that
$$|W^{(n)}_{\tau}|=\begin{cases} 
      \frac{\dot{\bar{\BL}}_\tau}{e_t(\bar{\Up})} & \text{if }|\tau|=t \\
      0 & \text{otherwise}
         \end{cases}~~~\forall n \in[N].$$
Now, we first note that \eqref{eq:eqmais2} does not depend on the specific demand $\Bd$. From this fact, we proceed to maximize over all possible user caches permutations $\sigma$ to obtain
\begin{align}
    T^*(t,\Up,\bar{\Up})
    &=\underset{d}{\max}~T^*(\Z^*_{\bar{\Up}},\Bd,\Up)\\
    &\geq \underset{\sigma\in S_{\Lambda}}{\max} \sum_{\lambda=1}^{\Lambda}L_{\sigma(\lambda)}\sum_{q\in\Cb{[\Lambda]\setminus\{\sigma(1),\dots,\sigma(\lambda)\}}{t}}\!\!\frac{\dot{\bar{\BL}}_q}{e_t(\bar{\Up})}\\
    &= \underset{\sigma\in S_{\Lambda,\Lambda-t}}{\max} \sum_{\lambda=1}^{\Lambda-t}L_{\sigma(\lambda)}\sum_{q\in\Cb{[\Lambda]\setminus\{\sigma(1),\dots,\sigma(\lambda)\}}{t}}\!\!\frac{\dot{\bar{\BL}}_q}{e_t(\bar{\Up})}\label{eq:con_L'}
\end{align}
where $S_{\Lambda,\Lambda-t}$ is the set of $(\Lambda-t)$-permutations of $[\Lambda]$.

%%%%%%%%%%%%%%%%%%%%%%%%%%%%%%%%%%%%%%%%%%
%%%%%%%%%%%%%%%%%%%%%%%%%%%%%%%%%%%%%%%%%%
\subsection{Proof of Theorem \ref{thm:thmL'}}
%%%%%%%%%%%%%%%%%%%%%%%%%%%%%%%%%%%%%%%%%%
%%%%%%%%%%%%%%%%%%%%%%%%%%%%%%%%%%%%%%%%%%

\label{subsec:proofoftheoremscenario2}
In order to prove Theorem~\ref{thm:thmL'}, we have to prove that the achievable delivery time in \eqref{eq:ach_delayL'} matches  the lower bound in \eqref{eq:con_L'}. To do so, we first notice that \eqref{eq:ach_delayL'} and \eqref{eq:con_L'} have the same denominator, thus leaving us to prove that the numerator of the achievable delivery time
is exactly equal to the numerator of the bound, i.e., to prove that
\begin{equation}\label{eq:equality_Ach_bound}
    \sum_{\tau\in \Cb{[\Lambda]}{t+1}}\max_{\lambda\in \tau}L_\lambda \Ldb_{\tau\setminus\lambda}=\max_{\sigma\in S_{\Lambda,\Lambda-t}}\sum_{\lambda=1}^{\Lambda-t}L_{\sigma(\lambda)}\sum_{\tau\in\Cb{[\Lambda]\setminus\{\sigma(1),\dots,\sigma(\lambda)\}}{t}}\Ldb_\tau.
\end{equation}
Proving~\eqref{eq:equality_Ach_bound} is challenging, and it exemplifies one of the main challenges arising when dealing with asymmetric settings, that we have to operate with asymmetric combinatorial expressions. 
To prove~\eqref{eq:equality_Ach_bound}, we start by constructing the set
\begin{equation}
    \Phi\triangleq\Big\{\max_{i\in\tau}L_i\Ldb_{\tau\setminus\{i\}}:\tau\subset[\Lambda],\ |\tau|=t+1\Big\},
\end{equation}
which is comprised of the addends of the left-hand-side of \eqref{eq:equality_Ach_bound}.
We naturally have that $|\Phi|= {\Lambda\choose t+1}$.

Let us introduce the term \emph{cache leader} in a specific way.
In the following, we say that cache $j$ is a \emph{leader} of a  set of caches  $\tau\subset[\Lambda]$ if $j=\argmax_{i\in\tau}L_i\Ldb_{\tau\setminus\{i\}}$.
In other words, cache~$j$ is a \emph{leader} of a set $\tau\subset[\Lambda]$ if it is the cache in $\tau$ that maximizes the geometric mean of  the tuple $\{\bL_i\}_{i\in\tau}$ when one of these $\bL_i$  is substituted by the corresponding $L_i$ from the same cache. 
The next lemma shows that the caches that act as leaders follow a particular structure.

\begin{lem}
For any $j\in[\Lambda]$, let us denote the  number of times that $j$ is a leader in $\Phi$ as $\CN_j$. Then, $\CN_j$ satisfies that
\begin{equation}
    \CN_j\in\left\{0,{t\choose t},{t+1\choose t},\dots,{\Lambda-1\choose t}\right\}.
\end{equation}
\end{lem}

\begin{proof}
Without loss of generality, let us assume that $j$ is such that $L_j\bL_i\geq L_i\bL_j~\forall i\in \phi$ for some $\phi\subset [\Lambda]\setminus\{j\}, |\phi|=m$, and for some $m\in \{t,t+1,\dots,\Lambda-1\}$. Notice that we must have $m\geq t$, since, for $m<t$, $j$ cannot be a leader in $\Phi$. The fact that there exist $m\choose t$ $t-$combinations of $\phi$ implies that $j$ is a leader in $\Phi$ at least $m\choose t$ times, since $L_j\bL_{\tau}\geq L_i \bL_{\{j\}\cup\tau\setminus\{i\}},\forall i\in\tau,\forall \tau\in \Cb{\phi}{t}$ by assumption. 
This consideration and the fact that $j$ might not be a leader in $\Phi$ complete the proof.
\end{proof}

Let us now consider the set of all the leaders $\ell_1,\ell_2,\ell_3,\dots,\ell_\Lambda$ in $\Phi$, and let us sort them such that, without loss of generality, we assume that $L_{\ell_j}\bL_{\ell_{j+1}}\geq L_{\ell_{j+1}}\bL_{\ell_j}$ for any $j\in[\Lambda-1]$. It can be easily verified that this order of the leaders implies that
	\begin{align}
			L_{\ell_j}\bL_{\ell_{i}} \geq L_{\ell_{i}}\bL_{\ell_j} ~\qquad \forall i >j. \label{eq:proof_lprim0}
	\end{align}
	
Let us consider $j=1$. From the above, we have that $L_{\ell_1}\bL_{\ell_{i}} \geq L_{\ell_{i}}\bL_{\ell_1}$ for any $i >1$. 
We can multiply both sides of the inequality by $\Ldb_{\eta}$ for any $\eta\subseteq [\Lambda]\setminus\{\ell_1,\ell_i\}$ of cardinality $|\eta|=t-1$, such that we can write that %.  This writes as 
	\begin{align}
			L_{\ell_1}\bL_{\ell_{i}}\Ldb_{\eta} \geq L_{\ell_{i}}\bL_{\ell_1}\Ldb_{\eta} ~\qquad \forall \eta\subseteq [\Lambda]\setminus\{{\ell_1},{\ell_i}\},\quad \forall i >1. \label{eq:proof_lprim1}
	\end{align}
There are $\binom{\Lambda-2}{t-1}$ such subsets for each~$i$, thus a total of $(\Lambda-1)\binom{\Lambda-2}{t-1}$ different inequalities. For $\ell_1$ to be the leader of a set $\tau$, $|\tau| = t+1$, we need $L_{\ell_1} \bL_{\ell_i}\Ldb_{\tau\setminus\{\ell_1,\ell_i\}}\geq L_{\ell_i} \bL_{\ell_1}\Ldb_{\tau\setminus\{\ell_1,\ell_i\}}$ for any $i:\ell_i\in\tau\setminus \ell_1$. That is, we need $t$ different inequalities among those in~\eqref{eq:proof_lprim1}, each one from a different $i$. Then, the set of $(\Lambda-1)\binom{\Lambda-2}{t-1}$ different inequalities in~\eqref{eq:proof_lprim1} imply that $\ell_1$ is a leader of $\frac{\Lambda-1}{t}\binom{\Lambda-2}{t-1} = \binom{\Lambda-1}{t}$ different sets $\tau\subseteq[\Lambda]$, $|\tau| = t+1$. Indeed, that amounts to all the possible sets in which $\ell_1$ appears.

After having considered $\ell_1$ for the sake of comprehension, let us now consider a general $\ell_j$. Let us now multiply~\eqref{eq:proof_lprim0} by $\Ldb_{\eta}$ for any $\eta\subseteq [\Lambda]\setminus\{\{\ell_{k}\}_{k\leq j},\ell_i\}$, $|\eta|=t-1$. Note that now we have only considered the subsets that do not include neither $\ell_i$ nor any $\ell_k$ for $k\leq j$.
Hence, we have $\binom{\Lambda-j-1}{t-1}$ such subsets for each $i$, thus a total of $(\Lambda-j)\binom{\Lambda-j-1}{t-1}$ different inequalities. 

Now, let us denote by $\Omega_j$ the set of subsets of cardinality $t+1$ which contain $\ell_j$ but do not contain any $\ell_k$ such that $k<j$, i.e., $\Omega_j\triangleq \{\tau : |\tau| = t+1,\ \tau \ni j, \{\tau\setminus \ell_j\} \subseteq [\Lambda]\setminus\{\ell_{k}\}_{k\leq j}\}$. 
Note that within the previous set of inequalities, i.e., within
	\begin{align}
			L_{\ell_j}\bL_{\ell_{i}}\Ldb_{\eta} \geq L_{\ell_{i}}\bL_{\ell_j}\Ldb_{\eta} ~\qquad \forall \eta\subseteq [\Lambda]\setminus\{\{\ell_{k}\}_{k\leq j},i\},\quad \forall i >j, \label{eq:proof_lprim2}
	\end{align}
we can find the $t$ inequalities required for $\ell_j$ to be the leader of any subset $\tau$ in $\Omega_j$ (since, for an arbitrary $\tau\in\Omega_j$, we need $L_{\ell_j} \bL_{\ell_i}\Ldb_{\tau\setminus\{\ell_j,\ell_i\}}\geq L_{\ell_i} \bL_{\ell_j}\Ldb_{\tau\setminus\{\ell_j,\ell_i\}}$ for any $i:\ell_i\in\tau\setminus \ell_j$).
Consequently, the set of $(\Lambda-j)\binom{\Lambda-j-1}{t-1}$ different inequalities in~\eqref{eq:proof_lprim2} imply that $\ell_j$ is a leader of $\frac{\Lambda-j}{t}\binom{\Lambda-j-1}{t-1} = \binom{\Lambda-j}{t}$ different sets $\tau\subseteq[\Lambda]$, $|\tau| = t+1$. 
Indeed, that amounts to all the possible sets in which $\ell_j$ appears and no $\ell_k$, $k<j$, appears.

Interestingly, this implies that each $\ell_j$ is the leader of all the sets $\tau$ in which it appears \emph{and} none of the previous $\{\ell_{k}\}_{k< j}$ appears. Summing up all the sets for which $\ell_1,\dotsc,\ell_{\Lambda-t}$ are leaders yields 
	\begin{align}
			\sum_{n=1}^{\Lambda-t}{\Lambda-n\choose t}={\Lambda\choose t+1}, \label{eq:proof_lprim3}
	\end{align}
which matches  the cardinality of $\Phi$. Hence, there are only $\Lambda-t$ leaders.\footnote{Note that, for $\ell_{\Lambda-t+1}$, the number of possible subsets of cardinality $t+1$  in $[\Lambda]$ which do not contain any element in $\{\ell_k\}_{k\in[\Lambda-t+1]}$ is zero because the set $[\Lambda]\setminus\{\ell_k\}_{k\in[\Lambda-t]}$ has cardinality $t$. } 

Finally, by considering all possible $(\Lambda-t)$-combinations of $[\Lambda]$ as all the possible set of leaders, we can conclude that 
\begin{equation}\label{eq:proof_th3_lasteq}
    \sum_{q\in \Cb{[\Lambda]}{t+1}}\max_{\lambda\in q}L_\lambda \Ldb_{q\setminus\lambda}=\max_{\sigma\in S_{\Lambda,\Lambda-t}}\sum_{\lambda=1}^{\Lambda-t}L_{\sigma(\lambda)}\sum_{q\in\Cb{[\Lambda]\setminus\{\sigma(1),\dots,\sigma(\lambda)\}}{t}}\Ldb_q,
\end{equation}
which concludes the proof of Theorem \ref{thm:thmL'}.\qed

%%%%%%%%%%%%%%%%%%%%%%%%%%%%%%%%%%%%%%%%%%
%%%%%%%%%%%%%%%%%%%%%%%%%%%%%%%%%%%%%%%%%%
\subsection{Performance comparison for different degrees of topology knowledge}\label{sec:numerical_results}
%%%%%%%%%%%%%%%%%%%%%%%%%%%%%%%%%%%%%%%%%%
%%%%%%%%%%%%%%%%%%%%%%%%%%%%%%%%%%%%%%%%%%
{In order to provide some insights about the previously derived expressions, we illustrate the derived results for a particular instance of the problem, and we present a comparison with some state-of-the-art schemes to show the extent of the derived results.} 

We assume in the delivery phase that the actual cache occupancy vector $\Up$ is a realization of a collection of independent random variables $\boldsymbol{\CL}=(\CL_1,\CL_2,\dots,\CL_\Lambda)$ with expected value $\bar{\Up}=(\bar{L}_1,\bar{L}_2,\dots,\bar{L}_\Lambda)$, i.e., where the cache occupancy vector \emph{assumed} in the placement phase is such that $\bar{L}_\lambda=\mathbb{E}[\CL_\lambda]$, $\lambda\in[\Lambda]$, and hence $K = \sum_{\lambda=1}^\Lambda \mathbb{E}[\CL_\lambda]$. Let us first present the following proposition, which shows that the expected minimum delivery time in the scenario with imperfect knowledge of the topology $\bar{\Up}$ is lower-bounded by the minimum delivery time of the setting where the topology is perfectly known and it matches $\bar{\Up}$.

\begin{prop}\label{prop_lowerbound_imperfect}
    For the ($\boldsymbol{\CL}$,t) {scenario with imperfect topology knowledge}, the expected delivery time over  $\boldsymbol{\CL}$ satisfies
    \begin{equation}
        \mathbb{E}_{\boldsymbol{\CL}}[T^*(t,\Up,\bar{\Up})]\geq T(t,\bar{\Up}).
    \end{equation}
\end{prop}

\begin{proof}
    We have that
    \begin{align}
        \mathbb{E}_{\boldsymbol{\CL}}[T^*(t,\Up,\bar{\Up})]
        &=\mathbb{E}_{\boldsymbol{\CL}}\left[\max_{\sigma\in S_{\Lambda,\Lambda-t}}\sum_{\lambda=1}^{\Lambda-t}L_{\sigma(\lambda)}\sum_{q\in\Cb{[\Lambda]\setminus\{\sigma(1),\dots,\sigma(\lambda)\}}{t}}\frac{\dot{\bar{L}}_q}{e_t(\bar{\Up})}\right]\nonumber\\
        &\geq \max_{\sigma\in S_{\Lambda,\Lambda-t}}\mathbb{E}_{\boldsymbol{\CL}}\left[\sum_{\lambda=1}^{\Lambda-t}L_{\sigma(\lambda)}\sum_{q\in\Cb{[\Lambda]\setminus\{\sigma(1),\dots,\sigma(\lambda)\}}{t}}\frac{\dot{\bar{L}}_q}{e_t(\bar{\Up})}\right]\label{eq:Exp_inequality1}\\
        &= \max_{\sigma\in S_{\Lambda,\Lambda-t}}\sum_{\lambda=1}^{\Lambda-t}\mathbb{E}_{\boldsymbol{\CL}}\left[L_{\sigma(\lambda)}\right]\sum_{q\in\Cb{[\Lambda]\setminus\{\sigma(1),\dots,\sigma(\lambda)\}}{t}}\frac{\dot{\bar{L}}_q}{e_t(\bar{\Up})}\label{eq:Exp_inequality2}\\
        &=\frac{\sum_{\CQ\in \Cb{[\Lambda]}{t+1}}\max_{\lambda\in \CQ}\mathbb{E}_{\boldsymbol{\CL}}\left[L_{\lambda}\right] \dot{\bar{\BL}}_{\CQ\setminus\{\lambda\}}}{e_t(\bar{\Up})}\label{eq:Exp_inequality3}\\
        &=\frac{\sum_{\CQ\in \Cb{[\Lambda]}{t+1}}\max_{\lambda\in \CQ}\bar{L}_{\lambda} \dot{\bar{\BL}}_{\CQ\setminus\{\lambda\}}}{e_t(\bar{\Up})}\nonumber\\
        &=\frac{e_{t+1}(\bar{\Up})}{e_t(\bar{\Up})}%\nonumber\\ &
         =T(t,\bar{\Up}),
    \end{align}
    where \eqref{eq:Exp_inequality2} follows from the independence of the random variables $\{\CL_\lambda\}$ and {where} \eqref{eq:Exp_inequality3} has been proven in Section~\ref{subsec:proofoftheoremscenario2} (cf.~\eqref{eq:proof_th3_lasteq}).
\end{proof}

For the above setting, we will {compare} three different schemes. The first one is the topology-agnostic scheme in \cite{PUETIT2020}, which does not exploit the knowledge of $\bar{\Up}$ for the cache placement, {and which instead uses the} MAN cache placement corresponding to a uniform cache-size allocation. The second scheme, which we will refer to as \emph{ITK ECS scheme} (imperfect-topology-knowledge equal-cache-size scheme), is the one achieving the delivery time in equation (24a) of \cite{wanCaireFogRANTIT21} for the case when there are no cache-less users (i.e., when $K_{\rm mbs} = 0$ following the notation from~\cite{wanCaireFogRANTIT21}). We notice that \cite{wanCaireFogRANTIT21} assumes that all the caches have the same size, which cannot be optimized. The cache placement of the aforementioned scheme partitions the set of caches in $G$ groups such that all the caches in the same group store the same content, and it applies MAN placement for $G$ caches/users. If the cache occupancy vector $\Up$ is known in  advance during placement, the best partition is chosen by leveraging $\Up$ in order to minimize the delivery time. 
In the {scenario with imperfect topology knowledge}, however, $\Up$ is not precisely known in advance, and thus the partition in the \emph{ITK ECS scheme} is selected with respect to the available cache occupancy vector knowledge $\bar{\Up}$. Delivery is performed by means of the multi-round scheme in \cite{PUETIT2020,jin2019new}. 
Finally, the third scheme is the proposed scheme achieving the delivery time in Theorem \ref{thm:thmL'} for the {scenario with imperfect topology knowledge}. 

\begin{figure}[t!]
\centering
\includegraphics[width=0.6\columnwidth]{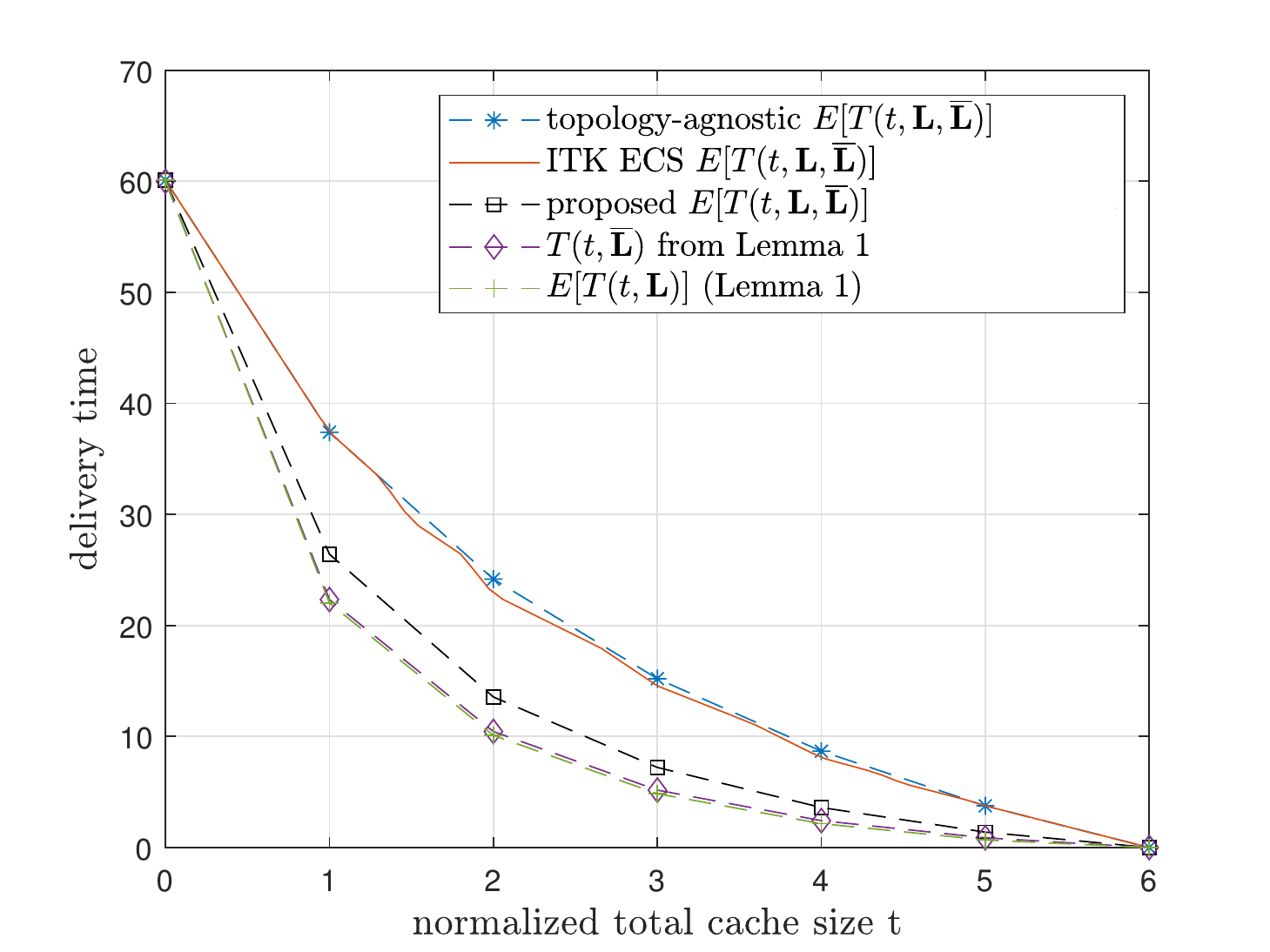}
\caption{Delivery time comparison for $\bar{\Up}=(20,20,8,6,4,2)$}
\label{Fig:con}
\end{figure}

We assume that each random variable $\CL_\lambda$ follows a Poisson distribution with mean $\bar{L}_\lambda$, i.e. $\CL_\lambda \sim \text{Poiss}(\bar{L}_\lambda)$, and we consider the 
scenario where the expected number of users per cache is $\bar{\Up}=(20,20,8,6,4,2)$. Figure~\ref{Fig:con} shows the average delivery time $\mathbb{E}_{\boldsymbol{\CL}}[T(t,\Up,\bar{\Up})]$ of the  three said schemes, as well as:
    \begin{itemize}
        \item the memory-rate curve $(t,T(t,\bar{\Up}))$ from Lemma \ref{lem:achiev}, which would be achieved if ${\boldsymbol{\CL}}$ was deterministic, equal to $\bar{\Up}$, and perfectly known during placement (i.e., in the topology-aware setting), represented by the diamond purple line, and 
        \item the average memory-rate curve $(t,\mathbb{E}_{\boldsymbol{\CL}}[T(t,\Up)])$, which would be achieved on average for any ${\boldsymbol{\CL}}$, $\CL_\lambda \sim \text{Poiss}(\bar{L}_\lambda)$, if, for each realization, $\Up$ is known at placement (represented by the cross green line).
    \end{itemize} It is evident that the proposed scheme with optimized shared caches largely outperforms the other two schemes, thus highlighting the importance of proper memory allocation. The plot also confirms Proposition~\ref{prop_lowerbound_imperfect}, {interestingly showing that the loss of performance due to the randomness in the number of users per cache is not so important thanks to the use of our proposed scheme.
    Furthermore, it shows that the optimal performance of the topology-aware scenario where the  cache occupancy vector $\bar{\Up}$ is such that $\bar{L}_\lambda=\mathbb{E}[\CL_\lambda], \lambda\in[\Lambda]$, (i.e., assuming that the cache occupancy vector available information is the mean value of $\boldsymbol{\CL}$), which is denoted by  $T(t,\bar{\Up})$, is a good approximation of the expected performance over $\boldsymbol{\CL}$,  
    i.e. $T(t,\bar{\Up})\approx \mathbb{E}_{\boldsymbol{\CL}}[T(t,\Up)]$.}

\section{Discussion and Conclusions}\label{sec:conclusions}
This work explores the shared-cache coded caching problem under the well known bottleneck of having an asymmetric user-to-cache association. Such asymmetry was previously shown to inflict substantial performance degradation in coded caching systems.  
This work reveals, that --- under any cumulative cache-size constraint --- a carefully optimized cache-size allocation, together with a novel cache-placement and delivery scheme, not only entirely alleviate this bottleneck but in fact turn this bottleneck into an advantage  compared with the symmetric case. 
The new optimized allocation and scheme jointly allow for the maximal coding gain~$t+1$ as well as for a boosted local caching gain, and thus for a reduced overall delivery time. 
Together with the novel cache size allocation and cache content placement, a main contribution of this work is the novel information theoretic converse that proves the information theoretic optimality of the achieved performance under simple and practical assumptions. 
Crucial to the tightness of the new converse is a novel careful combination of several MAIS bounds (cf. \cite{ArbaWang13}) which,  deviating from classical approaches that consist of averaging them uniformly, is one of the first converse results that captures the heterogeneity of the system and hits the exact optimal delivery time.

{In a setting where asymmetry generally resulted in very substantial DoF losses, the new approach manages to exploit knowledge of the number of users connected to each cache to substantially increase the performance compared to topology-agnostic scenarios.} 
Our work has offered tools that can help in the construction of other converses in the presence of asymmetry and has shed more light on how placement can be changed to work together with arbitrary cache sizes.

One of the crucial outcomes of our work is that the aforementioned asymmetry bottleneck can be substantially alleviated even if we are only partially aware of the user-to-cache association. A new scheme here proposed manages to substantially alleviate the bottleneck even in the presence of significant uncertainty of the user-to-cache association. 
This is one if the first studies exploring the connection between coded caching and the degree of knowledge of the network topology. 
We have also shown that, for any given the memory allocation and cache placement, this scheme is optimal. 

All these results show the decisive importance of memory allocation in coded caching, since not considering it leads to the collapse of the coded caching multiplicative gains. Furthermore, this allocation is in fact very impactful even in the presence of noisy knowledge of the topology.

This work has focused on deriving the optimal coded caching gains that can be achieved in the considered heterogeneous setting. An interesting direction for future works would be to study the subpacketization-constrained version of the problem, which might limit the actual gains in some practical applications.  Also, while in this paper we assume that the cost of fetching data from the caches is negligible, an interesting extension of this work could explore the performance of such networks where the access to a cache implies a certain performance cost.

%%%%%%%%%%%%%%%%%%%%%%%%%%%%%%%%%%%%%%%%%%%%%%%%%%%%%%%%%%%%%%%%%%%%
%%%%%%%%%%%%%%%%%%%%%%%%%%%%%%%%%%%%%%%%%%%%%%%%%%%%%%%%%%%%%%%%%%%%
\appendices
%%%%%%%%%%%%%%%%%%%%%%%%%%%%%%%%%%%%%%%%%%%%%%%%%%%%%%%%%%%%%%%%%%%%
%%%%%%%%%%%%%%%%%%%%%%%%%%%%%%%%%%%%%%%%%%%%%%%%%%%%%%%%%%%%%%%%%%%%

\renewcommand{\thesectiondis}[2]{\Roman{section}:}

%%%%%%%%%%%%%%%%%%%%%%%%%%%%%%%%%%%%%%%%%%%%%%%%%%%%%%%%
\section{Proof of Property~\ref{prop:prop3on_elementary}}\label{app:elemfunc_proof}
%%%%%%%%%%%%%%%%%%%%%%%%%%%%%%%%%%%%%%%%%%%%%%%%%%%%%%%%
    Let us  define $\xd_\theta = \prod_{j=1}^{k} x_{\theta(j)}$. By Property~\ref{prop:recursive_e_t}, we have that 
        \begin{align}\label{eq:proof_prop2_1}
            \sum_{x_i\in \{\CX\setminus \phi\}}x_i\cdot	e_{k-1}(\CX\setminus\{x_i\}) = |\CX\setminus \phi |\cdot e_k(\CX) \ - \sum_{x_i\in \{\CX\setminus \phi\}}e_{k}(\CX\setminus\{x_i\}).
        \end{align} 
     We can then write that $e_{k}(\CX\setminus\{x_i\}) = \sum_{\substack{q\subseteq \CX, \ |q|=k, \  x_i\notin q}} \xd_q$. 
     Hence, a particular set $q\subseteq \CX, \ |q|=k$, will appear in $e_{k}(\CX\setminus\{x_i\})$ if and only if $x_i\notin q$, 
     which implies that a particular $q$ will appear in $\sum_{x_i\in \{\CX\setminus \phi\}}e_{k}(\CX\setminus\{x_i\})$ only for the addends $x_i$ belonging to $\{\CX\setminus\{\phi,q\}\}$. 
     Consequently, it follows that
        \begin{align}\label{eq:proof_prop3_3}
            \sum_{x_i\in \{\CX\setminus \phi\}}e_{k}(\CX\setminus\{x_i\}) =  \sum_{\substack{q\subseteq \CX \\ |q|=k}}  \xd_q\cdot |\CX\setminus\{\phi,q\}|.
        \end{align}         
     Applying~\eqref{eq:proof_prop3_3} into~\eqref{eq:proof_prop2_1} yields
     \begin{align}\label{eq:proof_prop2_1b}
            \sum_{x_i\in \{\CX\setminus \phi\}}x_i\cdot	e_{k-1}(\CX\setminus\{x_i\}) &= \sum_{q\in \Cb{\CX}{k}}\prod_{j=1}^{k} x_{q(j)} \cdot (|\CX\setminus \phi | -  |\CX\setminus\{\phi,q\}|)\\
            &\overset{(a)}{=}\sum_{q\in \Cb{\CX}{k}}\prod_{j=1}^{k} x_{q(j)}\cdot|q\setminus \phi|,
        \end{align}
        where in $(a)$ we have applied that 
        $$|\CX\setminus \phi | -  |\CX\setminus\{\phi,q\}|=|\CX|-|\phi|-(|\CX|-|\phi\cup q|)=|q\setminus \phi|,$$
     which concludes the proof of Property~\ref{prop:prop3on_elementary}.

%%%%%%%%%%%%%%%%%%%%%%%%%%%%%%%%%%%%%%%%%%%%%%%%%%%%%%%%
\section{Proofs for the achievability results}
%%%%%%%%%%%%%%%%%%%%%%%%%%%%%%%%%%%%%%%%%%%%%%%%%%%%%%%%
\label{app:achiev_proofs} 

%%%%%%%%%%%%%%%%%%%%%%%%%%%%%%%%%%%%%%%%%%%%%
\subsection{Extension of Lemma~\ref{lem:achiev} to non-integer values of $t$}\label{app:proof_lemma1_achiev_nonInt}
%%%%%%%%%%%%%%%%%%%%%%%%%%%%%%%%%%%%%%%%%%%%%
When the normalized total memory $t$ has a non-integer value, we apply memory sharing along the same lines as in \cite{MN14}. We write $t$ as $t=\alpha\lfloor t \rfloor  + (1-\alpha) \lceil t \rceil, ~\text{for}~ \alpha\in[0,1]$, and we split each file $W^{(n)}$ of the library in two parts ${W^{(n),1},W^{(n),2}}$, where $|W^{(n),1}|=\alpha$ and $|W^{(n),2}|=(1-\alpha)$, such that the library remains partitioned in two sub-libraries
$$\CW_1=\{W^{(n),1}|n\in[N]\},~~~\CW_2=\{W^{(n),2}|n\in[N]\}.$$
Afterwards, we first employ the cache placement scheme in Section~\ref{subsec:cacheplacementBC} for sub-library $\CW_1$ with a total sum-cache constraint $\lfloor t \rfloor$, and then we do the same for sub-library $\CW_2$ with a total sum-cache constraint $\lceil t \rceil$. 
The delivery phase now consists of 2 rounds, each as in Section~\ref{subsec:delivery_BC}: the first round employs XORs of order $\lfloor t \rfloor +1$ to serve files $\{W^{(d_k),1}|k\in[K]\}$, whereas the second round employs XORs of order $\lceil t \rceil +1$ to serve  $\{W^{(d_k),2}|k\in[K]\}$.
This scheme clearly results in the following delivery time
\begin{equation}\label{eq:memory_sharing_delay}
    T(t,\Up)=\alpha T(\lfloor t \rfloor,\Up)+(1-\alpha)T(\lceil t \rceil,\Up).
\end{equation}

We now present a lemma that is instrumental for the proof of Lemma~\ref{lem:achiev} for non-integer values of~$t$.
	\begin{lem}\label{lemma_decreasing_ct_t}
		The sequence $\left\{\frac{e_{t+1}(\Up)}{e_{t}(\Up)}\right\}_{t\in [\Lambda]_0}$ is a decreasing and convex sequence. 			
	\end{lem}
	\begin{proof}
	    The proof is relegated to Appendix~\ref{app:achiev_convex}.  
	\end{proof}
The achievability of  \eqref{eq:memory_sharing_delay} implies that, for integer $t$, the straight line between points $(t,T(t,\Up))$ and $(t+1,T(t+1,\Up))$ is also achievable. 
Moreover, we know from~\eqref{eq:Delay3} that  $\left\{T(t,\Up)\right\}_{t\in [\Lambda]_0}=\left\{\frac{e_{t+1}(\Up)}{e_{t}(\Up)}\right\}_{t\in [\Lambda]_0}$, and thus Lemma~\ref{lemma_decreasing_ct_t} implies that $\left\{T(t,\Up)\right\}_{t\in [\Lambda]_0}$ is a convex sequence.  Since the lower convex envelope of a convex sequence is a piece-wise function composed of the segments connecting two successive elements of the sequence $\left\{T(t,\Up)\right\}_{t\in [\Lambda]_0}$ (i.e.,~\eqref{eq:memory_sharing_delay}), Lemma~\ref{lem:achiev} is proven.

%%%%%%%%%%%%%%%%%%%%%%%%%%%%%%%%%%%%%%%%%%%%%
\subsection{Proof of equation~\eqref{eq:gamma_lambda}}\label{app:proof_eq_gamma}
%%%%%%%%%%%%%%%%%%%%%%%%%%%%%%%%%%%%%%%%%%%%%
Let us now prove~\eqref{eq:gamma_lambda}, i.e., that
 $\gamm{\lambda}=\frac{L_\lambda\  e_{t-1}(\Up\setminus{\{L_\lambda\}})}{e_t(\Up)}$ for any $\lambda\in[\Lambda]$. 
To evaluate $\gamm{\lambda}$, we first note that all subfiles are equally-sized and that the placement scheme is \emph{symmetric} with respect to the library files, i.e., the caching strategy does not depend on the file index $n\in[N]$. 
This suggests that  $\gamm{\lambda}$ can be evaluated as the ratio between the number of subfiles (of any file $W^{(n)}$) stored in cache $\lambda$ and the total number of subfiles $S=e_t(\Up)$ into which $W^{(n)}$ is split. The fact that a subfile $W^{(n)}_{\tau,m}$ is placed in $\mathcal{Z}_\lambda$ if and only if $\lambda \in \tau $, together with the fact that $|A_\tau|=\prod_{j=1}^{t}L_{\tau(j)}$, automatically yield the numerator of \eqref{eq:gamma_lambda}.
 \qed

%%%%%%%%%%%%%%%%%%%%%%%%%%%%%%%%%%%%%%%%%%%%%%%%%%%%%%%%
\subsection{Proof of Lemma~\ref{lemma_decreasing_ct_t}}\label{app:achiev_convex}
%%%%%%%%%%%%%%%%%%%%%%%%%%%%%%%%%%%%%%%%%%%%%%%%%%%%%%%%

In the following, we prove Lemma~\ref{lemma_decreasing_ct_t}, which states that the sequence $\{c_t^{(t)}\}_{t\in \{0,\dotsc,\Lambda\}}$ is a decreasing and convex sequence. 
Since the $t$-th elementary symmetric polynomial in $\Up$ is given by $e_t(\Up) \triangleq \sum_{q\in C_{t+1}^{[\Lambda]}}^{} \prod_{j=1}^{t+1} L_{q(j)}$, and $ c_t^{(t)} \triangleq \frac{\sum_{q\in C_{t+1}^{[\Lambda]}}^{} \prod_{j=1}^{t+1} L_{q(j)}}{\ \sum_{q\in C_{t}^{[\Lambda]}}^{} \prod_{j=1}^{t} L_{q(j)}}$, we can write that
    \begin{align}
        c_t^{(t)} &= {e_{1}(\Up)}		&&\text{if $ t  = 0$} \label{eq:ct_0}\\
        c_t^{(t)} &= \frac{e_{t+1}(\Up)}{e_{t}(\Up)}		&&\text{if $1\leq t \leq \Lambda-1$}\\
        \hspace{20ex}~c_t^{(t)} &= 0   &&\text{if $t = \Lambda$.}\hspace{30ex}~\label{eq:ct_Lambda}
    \end{align}
The proof of Lemma~\ref{lemma_decreasing_ct_t} builds on the relation of the coefficients~$c_t^{(t)}$ with the {elementary symmetric polynomials} and the following lemma. 

\begin{lem}\label{convexity_logconcavity}
    Let $\{a_k\}$, $k\in[n]$, be a strictly log-concave sequence satisfying that $a_k>0$ for any $k<n$. Then, the sequence
    $\{b_k \triangleq \frac{a_{k+1}}{a_k}\}$, $k\in[n-1]$, is a decreasing (strictly) convex sequence. 
\end{lem}
\begin{proof}
    Since $\{a_k \mid k\in[n]\}$ is a strictly log-concave sequence, it holds that $a_{k}^2 > a_{k+1}a_{k-1}$. Moreover, since $a_k>0$ for any $k<n$ by definition of  the sequence $\{a_k\}$, we have that 
        \begin{align}\label{eq:logconcave_ab}
            a_{k}^2 > a_{k+1}a_{k-1}
                \quad\iff\quad \frac{a_{k}}{a_{k-1}}>\frac{a_{k+1}}{a_{k}}.
        \end{align}					
    The right-hand side of~\eqref{eq:logconcave_ab} is equivalent to $b_{k-1} > b_k$, which proves that $\{b_j\}$ is a decreasing sequence. Next, we prove that $\{b_j\}$ is also a convex sequence. 
    
    A discrete sequence $\{b_j\}$, $j\in[n-1]$, is convex if and only if
         $2 b_j \leq b_{j-1} + b_{j+1}$ for any $j\in\{2,\dotsc,n-2\}$. 
    Thus, the sequence $\big\{\frac{a_{k+1}}{a_k}\big\}_{k\in[n-1]}$ is convex if and only if 
        \begin{align}\label{eq:convex_a_b_1}
            2 \frac{a_{j+1}}{a_{j}} \leq \frac{a_{j}}{a_{j-1}} + \frac{a_{j+2}}{a_{j+1}},\qquad\forall j\in\{2,\dotsc,n-2\}.
        \end{align}				
    Let us now multiply~\eqref{eq:convex_a_b_1} by the denominators $(a_{j-1}a_{j}a_{j+1})$ to obtain 
        \begin{align}
                2 a_{j-1}a_{j+1}^2 &\leq a_{j}^2a_{j+1} + a_{j-1}a_{j}a_{j+2}\\
                    &< a_{j}^2a_{j+1} + a_{j-1}a_{j+1}^2
        \end{align}				
    which follows from the strict log-concavity of $\{a_k\}$ (i.e., $a_{j}a_{j+2}< a_{j+1}^2$). 
    Re-ordering terms, we obtain 
        \begin{align}
             a_{j-1}a_{j+1}  < a_{j}^2, 						
        \end{align}			
    which is always true because $\{a_k\}$ is strictly log-concave. 
    Consequently, the sequence $\{b_j\}$ is a strictly convex sequence. 
\end{proof}

Continuing with the proof of Lemma~\ref{lemma_decreasing_ct_t}, we note that, given Lemma~\ref{convexity_logconcavity}, we only need to show that the sequence $\{1,\{e_{t}\}_{t\in[\Lambda]},0\}$ is strictly log-concave: If it is strictly log-concave, and upon defining $\{a_k\}=\{1,\{e_{t}\}_{t\in[\Lambda]},0\}$ (such that $b_k \triangleq \frac{a_{k+1}}{a_k} = c^{(t)}_t$),  applying  Lemma~\ref{convexity_logconcavity} yields that  $c^{(t)}_t$ is a strictly convex sequence, which will conclude the proof of Lemma~\ref{lemma_decreasing_ct_t}. 

In order to prove that the sequence $\{1,\{e_{t}\}_{t\in[\Lambda]},0\}$ is strictly log-concave, let us first introduce  the \emph{elementary symmetric means} $E_k$, which are defined as
    \begin{align} \label{eq:def_average_elemt}
        E_k(\Up)  \triangleq \frac{e_k(\Up)}{\binom{n}{k}},
    \end{align}	
where $\binom{n}{k}$ is the number of addends in $e_k(\Up)$ for a set of $|\Up| = n$ elements (in our case, $n=\Lambda$). 
Hereinafter, we omit the dependence of $e_k$ and $E_k$ on the occupancy vector $\Up$ because $\Up$ is the same set for any $e_k$ and $E_k$ considered in the following. 

These elementary symmetric means have a property, which was was proved by~Newton~\cite{Newton1761}, that says that, for any $n$-tuple of non-negative numbers, it holds that the sequence $\{E_k\}_{k \in [n]}$ is a log-concave sequence, and thus
\begin{align} \label{eq:ineq_average_elemt}
    E_k^2 \geq  E_{k-1}E_{k+1},\quad k\in\{2,\dotsc,n-1\},
\end{align}			
where the inequality is strict unless all the elements of the  $n$-tuple coincide. 

In order to prove the log-concavity of our sequence $\{1,\{e_{t}\}_{t\in[\Lambda]},0\}$, we first obtain from~\eqref{eq:def_average_elemt}-\eqref{eq:ineq_average_elemt} that 
    \begin{align}
        E_t^2 \geq  E_{t-1}E_{t+1} 
            \quad\iff\quad  \frac{e_t^2}{\binom{n}{t}^2} \geq \frac{e_{t-1}e_{t+1}}{ \binom{n}{t-1}\binom{n}{t+1}},
    \end{align}					
for any $1<t< n$. 
Then, we can write that 
    \begin{align}
     e_t^2 &\geq \frac{\binom{n}{t}^2}{\binom{n}{t-1}\binom{n}{t+1}}e_{t-1}e_{t+1}\\
                &> e_{t-1}e_{t+1}, \label{eq:logconcave_elemt}
    \end{align}			
which implies that $\{e_{t}\}_{1\leq t \leq \Lambda}$ is strictly log-concave, and where the last step follows from the strict log-concavity of the binomial coefficient~\cite{Stanley2006}. It remains to prove that $e_1^2>1\cdot e_{2} = e_{2}$ and that $e_\Lambda^2> e_{\Lambda-1}\cdot 0 = 0$. The fact that $e_\Lambda^2>0$ is always true because $e_\Lambda=\prod_{j=1}^\Lambda L_j > 0$ because $L_j\geq 1$. On the other hand, to show that $e_1^2> e_{2}$, let us recall the Maclaurin inequalities~\cite{Lin1994}, which state that
    \begin{align}
        E_1 \geq \sqrt{E_2} \geq \sqrt[\leftroot{-3}\uproot{3}3]{E_3} \geq \cdots \geq \sqrt[\leftroot{-3}\uproot{3}n]{E_n},
    \end{align}
with equality holding only if all the $L_j$ for any $j\in [n]$ coincide. 
Let us focus on the first inequality. Since $n=\Lambda$ and $L_j\geq 1$ for any $j$, it follows that
    \begin{align}
        E_1 \geq \sqrt{E_2} 
            &\quad\iff\quad  \frac{e^2_1}{\binom{\Lambda}{1}^2} \geq \frac{e_{2}}{\binom{\Lambda}{2}} %\\
            &\quad\iff\quad  \frac{e_1^2}{\Lambda^2} \geq \frac{e_{2}}{\frac{\Lambda(\Lambda-1)}{2}} %\\
            &\quad\iff\quad  {e_1^2} \geq \frac{2\Lambda}{\Lambda-1} e_{2}. 
    \end{align}			
Since ${e_1^2} \geq \frac{2\Lambda}{\Lambda-1} e_{2} > e_{2}$, we obtain that  the sequence $\{1,\{e_{t}\}_{1\leq t \leq \Lambda},0\}$ is strictly log-concave. Thus, applying  Lemma~\ref{convexity_logconcavity} yields that  $c^{(t)}_t$ is a strictly convex sequence, which concludes the proof of Lemma~\ref{lemma_decreasing_ct_t}.  \qed

%%%%%%%%%%%%%%%%%%%%%%%%%%%%%%%%%%%%%%%%%%%%%%%%%%%%%%%%
\section{Proof of Lemma \ref{lem:long_coefficient}}\label{subsec:proof_of_lemma_long_coeff}
%%%%%%%%%%%%%%%%%%%%%%%%%%%%%%%%%%%%%%%%%%%%%%%%%%%%%%%%

    We derive in the following the value of $c_{\tau,n}^{(p)}$ in the expression 
    $T_{lb,p}(\Z,\Up)
		        = \sum_{n=1}^N\sum_{\tau\in 2^{[\Lambda]}}c_{\tau,n}^{(p)}\frac{|W^{(n)}_\tau|}{N}$ (cf.~\eqref{eq:TLBcmin}), 
	where $T_{lb,p}(\Z,\Up)$ has been defined in~\eqref{eq:lb_long}.
	
	From~\eqref{eq:lb_long} and~\eqref{eq:weights} we can write $T_{lb,p}(\Z,\Up)$ as
        \begin{align}\label{eq:appendixBA_first_equ}
            T_{lb,p}(\Z,\Up)=\frac{1}{\underbrace{|\Dwc|\sum_{\sigma\in\CS_\Lambda}\prod\limits_{j=1}^{p}L_{\sigma(\Lambda-p+j)}}_{\triangleq\CD_p}}\underbrace{\sum_{\Bd\in \Dwc}\sum_{\sigma\in\CS_\Lambda}\CQ^{(p)}_\sigma T_{lb,\sigma}(\Z,\Bd,\BL)}_{\triangleq\CN_p}
        \end{align}
    where~$\CQ^{(p)}_\sigma \triangleq \prod\limits_{j=1}^{p}L_{\sigma(\Lambda-p+j)}$ and
        \begin{equation}\label{eq:def:T_lb_sigma}
          T_{lb,\sigma}(\Z,\Bd,\BL) \triangleq \sum_{\lambda=1}^{\Lambda}\sum_{\ell=1}^{L_{\sigma(\lambda)}}\sum_{\tau_{\lambda}\subseteq [\Lambda]\setminus \{\sigma(1),\dots,\sigma(\lambda)\}}\!\!\!\!|W^{(\Bd_{\sigma(\lambda)}(\ell))}_{\tau_{\lambda}}|. 
        \end{equation}
    First, we rewrite $\CD_p$ (defined in~\eqref{eq:appendixBA_first_equ}) in a more suitable form as
    \begin{align}
        \CD_p&=P(N,K)\sum_{\sigma\in \mathcal{S}_\Lambda}\prod_{j=1}^p L_{\sigma(\Lambda-p+j)}\nonumber\\
        &=P(N,K) \sum_{v\in \CS_{\Lambda-p}} \sum_{q\in \CS_p}\prod_{j=1}^p L_{q(j)}\nonumber\\
        &=P(N,K)(\Lambda-p)!p! \sum_{q\in \Cb{[\Lambda]}{p} }\prod_{j=1}^p L_{q(j)},
    \end{align}
    which follows from basic mathematical manipulations.

    For any $n\in[N]$ and any $\tau\subseteq [\Lambda]$, our goal is now to evaluate the coefficient that multiplies  each $|W^{(n)}_\tau|$ in $\CN_{p}$ from~\eqref{eq:appendixBA_first_equ}, where we denote this coefficient as~$g_{n,\tau}^{(p)}$. 
    We first state the following useful  fact.
    \begin{fact}
    For any $n\in[N]$ and any $\tau\subseteq [\Lambda]$, if $|W^{(n)}_\tau|$ appears in $T_{lb,\sigma}(\Z,\Bd,\Up)$ for some $\Bd\in\Dwc$ and some $\sigma\in S_\Lambda$, then it only appears once for all $\lambda\in[\Lambda],\ \ell\in[L_\lambda]$. 
    \end{fact}
    Next, we need to split the proof in two cases. First, we consider the case when $|\tau|\geq p$, and afterwards we focus on the case $|\tau|< p$.
    In the following, we make use of the notation $\Ld_{q}\triangleq \prod_{j=1}^{|q|} L_{q(j)}$ for any $q\subseteq [\Lambda]$. 
    
    %%%%%%%%%%%%%%%%%%%%%%%%%%%%%%%%%%%%%%%%%%%%%%%
    \subsection{The $|\tau|\geq p$ case}
    %%%%%%%%%%%%%%%%%%%%%%%%%%%%%%%%%%%%%%%%%%%%%%%

    Let us focus on a demand vector $\Bd'$ such that subfile $|W^{(n)}_{\tau}|$ is requested by a certain user $k'\in \CU_{\lambda'}$ associated to cache $\lambda'$. This simply means that $n=d_{k'}$. Afterward, we will consider all possible $\Bd'$. 
    
    %%%%%%%%%%%%%%%%%%%%%%%%%%%%%%%
    \subsubsection{Focusing on a given demand vector $\Bd'$}
    %%%%%%%%%%%%%%%%%%%%%%%%%%%%%%%
    For a specific demand vector $\Bd'$, our objective is now to evaluate the coefficient $g_{n,\tau}^{(p)}$ of $|W^{(d_{k'})}_\tau|$ in
    \begin{equation}\label{eq:minicoeffd'}
        T_{lb}(\Z,\Bd',\BL)\triangleq\sum_{\sigma\in\CS_\Lambda}\CQ^{(p)}_\sigma T_{lb,\sigma}(\Z,\Bd',\BL).
    \end{equation}
    In this respect, we need to identify those permutations $\sigma\in S_\Lambda$ for which  $|W^{(d_{k'})}_\tau|$ appears in $T_{lb,\sigma}(\Z,\Bd',\Up)$.
    From the expression of $T_{lb,\sigma}(\Z,\Bd',\Up)$ in~\eqref{eq:def:T_lb_sigma}, the following proposition holds.
    \begin{prop}\label{prop:validperm}
        The permutations $\sigma$ for which $|W^{(d_{k'})}_\tau|$ appears in $T_{lb,\sigma}(\Z,\Bd',\Up)$ are such that $\lambda'$ appears in the permutation $\sigma$ before any element of the set $\tau$, where $\lambda'$ is the cache to which user $k'$ is associated $(k'\in \CU_{\lambda'})$. 
        We will refer to such permutations as \emph{valid permutations}. 
    \end{prop}
    
    \emph{Example}: Consider $W^{(3)}_{1,2}$, where $d_k=3$ for $k\in\CU_3$ and $\tau=\{1,2\}$. This subfile will not appear in $T_{lb,\sigma}(\Z,\Bd',\Up)$ for permutations %$\sigma$
    $(1,2,3),(2,1,3),(1,3,2)$ and $(2,3,1)$, while it will appear for permutations $(3,1,2)$ and $(3,2,1)$.\qed 
    
    It stems from~\eqref{eq:minicoeffd'} and Proposition~\ref{prop:validperm} that a valid permutation $\sigma$ contributes with a weight~$\CQ_\sigma^{(p)}$  to~$g_{n,\tau}^{(p)}$.  

    With Proposition \ref{prop:validperm} at hand, we notice that, if $\lambda'$ appears in any one of the last $p$ positions of a permutation $\sigma$,  there will certainly be an element of $\tau$ (recall that $|\tau|\geq p$) which will precede $\lambda'$, and thus such permutation $\sigma$ can not be a \emph{valid permutation}. 
    Then, if $|\tau|\geq p$,  any valid permutation does not have $\lambda'$ in the last $p$ positions. In other words, the set of valid permutations is composed of the permutations whose last $p$~positions are given by a set $q$ belonging to $\Cb{[\Lambda]\setminus\{\lambda'\}}{p}$. 
    
    Next, we derive the number of valid permutations for each $q\in\Cb{[\Lambda]\setminus\{\lambda'\}}{p}$. 
    
    \begin{prop}\label{prop:ways_first_pos}
        Let $q\in\Cb{[\Lambda]\setminus\{\lambda'\}}{p}$ be a fixed ordered $p$-tuple. 
        Then, the number of \emph{valid permutations} (as defined in Proposition~\ref{prop:validperm}) whose last  $p$~positions match $q$ is
        \begin{equation}\label{propo4_fact}
            \frac{{\Lambda - p \choose |\tau\setminus q|+1}}{{\Lambda - p - 1 \choose |\tau\setminus q|}}(\Lambda-p-1)!.
        \end{equation}
    \end{prop}
    
    \begin{proof}
        As seen before, Proposition~\ref{prop:validperm} implies that $\lambda'$ cannot be the last $p$ positions of the permutation~$\sigma$. Let us then consider the case when $\lambda'$ is in position $r$ for some $r\in[\Lambda-p]$. 
        Let us also denote the set of elements in the last $p$ positions of $\sigma$ by $q$.
        In this case, we have that in the first $r-1$ positions we cannot place any of the elements in $q$, we cannot place $\lambda'$, and we cannot place any element of $\tau$ that is not in $q$.
        It then follows that, for any $q$, there are
        $$P(\Lambda-p-1-|\tau\setminus q|)(\Lambda-p-r)!$$
        ways in which we can fill the first $\Lambda-p$ positions of $\sigma$ with $\lambda'$ in position $r$. Considering all possible $r$ values, we have 
        \begin{equation}\label{eq:app_ways_xx}
            \sum\limits_{r=1}^{\Lambda-p}P(\Lambda-p-1-|\tau\setminus q|,r-1)(\Lambda-p-r)!
        \end{equation}
        different forms in which we can fill the first $\Lambda-p$ positions of $\sigma$ with $\lambda'$ appearing in the first $\Lambda-p$ positions.  We can manipulate~\eqref{eq:app_ways_xx} to obtain that
         \begin{align}
            \sum\limits_{r=1}^{\Lambda-p}P(\Lambda-p-1-|\tau\setminus q|,r-1)(\Lambda-p-r)!%\nonumber\\
            =&\sum\limits_{r=1}^{\Lambda-p}\frac{(\Lambda-p-1-|\tau\setminus q|)!(\Lambda-p-r)!|\tau\setminus q|!(\Lambda-p-1)!}{(\Lambda-p-|\tau\setminus q|-r)!|\tau\setminus q|!(\Lambda-p-1)!}\nonumber\\
            =&\sum\limits_{r=1}^{\Lambda-p}\frac{{\Lambda - p - r \choose |\tau\setminus q|}}{{\Lambda - p - 1 \choose |\tau\setminus q|}}(\Lambda-p-1)!\nonumber\\
            =&\frac{{\Lambda - p \choose |\tau\setminus q|+1}}{{\Lambda - p - 1 \choose |\tau\setminus q|}}(\Lambda-p-1)!,
        \end{align}
        which concludes the proof of Proposition~\ref{prop:ways_first_pos}.
    \end{proof}
    
    For each such set $q\in\Cb{[\Lambda]\setminus\{\lambda'\}}{p}$, there are $p!$ possible orderings of its elements. 
    Consequently, after recalling that each permutation $\sigma$ has associated a weight~$\CQ^{(p)}_\sigma \triangleq \prod_{j=1}^{p}L_{\sigma(\Lambda-p+j)}$, we can conclude that each $q\in\Cb{[\Lambda]\setminus\{\lambda'\}}{p}$ has a weight in~$ T_{lb}(\Z,\Bd',\BL)$ of
    \begin{equation}\label{eq:ways_last_p_pos}
         p!\sum_{q\in \Cb{[\Lambda]\setminus\{\lambda'\}}{p}}\Ld_q.
    \end{equation}
    Combining equation~\eqref{eq:ways_last_p_pos} and Proposition~\ref{prop:ways_first_pos}, the total weight of subfile~$W^{(d_{k'})}_\tau$ for demand $\Bd'$ is
    \begin{equation}\label{eq:valid_perms}
        p!\sum_{q\in \Cb{[\Lambda]\setminus\{\lambda'\}}{p}  }\Ld_q\frac{{\Lambda - p \choose |\tau\setminus q|+1}}{{\Lambda - p - 1 \choose |\tau\setminus q|}}(\Lambda-p-1)!.
    \end{equation}
    We now proceed to evaluate the total number of demands for which subfile $W^{(d_{k'})}_\tau$ is requested.

    %%%%%%%%%%%%%%%%%%%%%%%%%%%%%%%
    \subsubsection{Joining all possible $\Bd'$}
    %%%%%%%%%%%%%%%%%%%%%%%%%%%%%%%
    
    It is easy to see that the total number of demands with $d_{k'}=n$ is $P(N-1,K-1)$. If user $k'$ is associated to any of the caches in set $\tau$ (i.e., $\lambda'\in\tau$), then subfile $W^{(d_k')}_\tau$ will not be requested, since it is already stored in cache $\lambda'$. Thus, $W^{(n)}_\tau$ will be requested to the server only if $\lambda'\in[\Lambda]\setminus\{\tau\}$.
    Considering all possible demand vectors, it follows that the total number of times that subfile $W^{(n)}_{\tau}$ appears in the demand vector is
    \begin{equation}\label{eq:valid_demands}
        P(N-1,K-1)\sum_{\lambda\in[\Lambda]\setminus\{\tau\}}L_{\lambda}.
    \end{equation}
    From equations \eqref{eq:valid_perms} and \eqref{eq:valid_demands}, we have that the coefficient~$g_{n,\tau}^{(p)}$ corresponding to~$|W^{(n)}_\tau|$ in~$\CN_p$ can be written as
    \begin{equation}
        g_{n,\tau}^{(p)}=P(N-1,K-1)\cdot\sum_{\lambda\in[\Lambda]\setminus\{\tau\}}L_{\lambda}\cdot p!\cdot\sum_{q\in \Cb{[\Lambda]\setminus\{\lambda\}}{p} }\left(\Ld_q\frac{{\Lambda - p \choose |\tau\setminus q|+1}}{{\Lambda - p - 1 \choose |\tau\setminus q|}}(\Lambda-p-1)!\right).
    \end{equation}
    Finally, we obtain the coefficient of any $|W^{(n)}_\tau|$ with $|\tau|\geq p$ in $T_{lb,p}(\Z,\Up)$, i.e.,  $\frac{g_{n,\tau}^{(p)}}{\CD_p}$ (cf.~\eqref{eq:appendixBA_first_equ}), which can be rewritten as
        \begin{align}\label{eq:coeff_tau_greater}
            \frac{g_{n,\tau}^{(p)}}{\CD_p}
            &=\frac{P(N-1,K-1)\cdot\sum_{\lambda\in[\Lambda]\setminus\{\tau\}}L_{\lambda}\cdot p!\cdot\sum_{q\in \Cb{[\Lambda]\setminus\{\lambda\}}{p} }\left(\Ld_q\times\frac{{\Lambda - p \choose |\tau\setminus q|+1}}{{\Lambda - p - 1 \choose |\tau\setminus q|}}(\Lambda-p-1)!\right)}{P(N,K)(\Lambda-p)!p! \sum_{\ell\in \Cb{[\Lambda]}{p} }\Ld_\ell}\nonumber\\
            &\overset{(a)}{=}\frac{1}{N\cdot \sum_{\ell\in \Cb{[\Lambda]}{p}}\Ld_\ell}\sum_{\lambda\in[\Lambda]\setminus\{\tau\}} L_{\lambda}\sum_{q\in \Cb{[\Lambda]\setminus\{\lambda\}}{p} }\Ld_q\cdot \frac{1}{|\tau\setminus q|+1}\nonumber\\
            &\overset{(b)}{=}\frac{\sum_{q\in \Cb{[\Lambda]}{p+1}}\Ld_q\cdot \frac{|q\setminus \tau|}{|\tau\setminus  q|+1}}{N\cdot \sum_{\ell\in \Cb{[\Lambda]}{p}}\Ld_\ell}\nonumber\\
            &\overset{(c)}{=}\frac{\sum_{q\in \Cb{[\Lambda]}{p+1}}\Ld_q\cdot \frac{p+1-|q\cap\tau|}{|\tau|+1-|q\cap \tau|}}{N\cdot \sum_{\ell\in \Cb{[\Lambda]}{p}}\Ld_\ell},%\label{eq:coef_for_simple_case}
        \end{align}
    where $(a)$ follows from basic mathematical manipulations, $(b)$ follows from Property~\ref{prop:prop3on_elementary} and the fact that for any $\lambda\in[\Lambda]\setminus\{\tau\}$ and $q\in\Cb{[\Lambda]\setminus\{\lambda\}}{p}$ we have $|\tau\setminus q| = |\tau\setminus \{q \cup \{\lambda\}\}|$, and $(c)$ follows from the fact that $|q\setminus \tau| = |q| - |q\cap \tau|$ as well as from  $|\tau\setminus q|=|\tau|-|q\cap\tau|$. Defining $c^{(p)}_{n,\tau}\triangleq N\frac{g_{n,\tau}^{(p)}}{\CD_p}$ proves the lemma for the case $\tau\geq p$.

    %%%%%%%%%%%%%%%%%%%%%%%%%%%%%%%%%%%%%%%%%%%%%%%%%%%%%
    \subsection{The $|\tau|<p$ case}
    %%%%%%%%%%%%%%%%%%%%%%%%%%%%%%%%%%%%%%%%%%%%%%%%%%%%%
        We recall that our objective is to obtain the coefficient $c_{\tau}^{(p)}$ that allows us to write $T_{lb,p}(\Z,\Up)$ as 
        $T_{lb,p}(\Z,\Up)  = \sum_{n=1}^N\sum_{\tau\in 2^{[\Lambda]}}c_{\tau}^{(p)}\frac{|W^{(n)}_\tau|}{N}$, where $T_{lb,p}(\Z,\Up)$ has been defined in~\eqref{eq:lb_long}. 
        As for the $|\tau|\geq p$ case, let us focus on a demand vector $\Bd'$ such that subfile $|W^{(n)}_{\tau}|$ is requested by a certain user $k'\in \CU_{\lambda}$ associated to cache $\lambda$.

\begin{figure}[t]\centering%
			\begin{tikzpicture}
				\draw(1,0)--(10,0);
				\foreach \x in {1,1.5,2,2.5,...,10}
						\draw(\x,2.5pt)--(\x,-2.5pt) node[below] {};
				\foreach \x/\xtext in {1/$1$,2/$3$,6/$\Lambda-p$,8/$r$,10/$\Lambda$}
						\draw(\x,5pt)--(\x,-5pt) node[below] {\xtext};
				\node[draw=white] at (6.5,.4) {\footnotesize $r=1$};
				\node[draw=white] at (10,.4) {\footnotesize $r=p$};
				\draw[decorate, decoration={brace, mirror}, yshift=2ex]  (10,0.35) -- node[above=0.4ex] {Last $p$ positions}  (6.5,0.35);
			\end{tikzpicture}
			\caption{Illustration of the meaning of index $r$.}%
			\label{fig:help_R}%
		\end{figure}
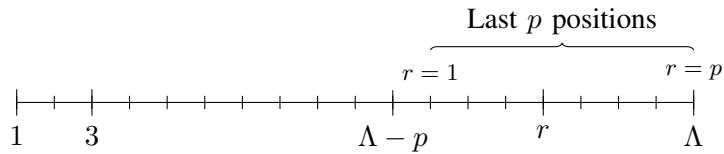

        The number of permutations for which that $\lambda$ is not in the last $p$~positions of the permutation is the same as for the case $|\tau|=j \geq p$, and it is given by~\eqref{eq:coeff_tau_greater}. Let us now denote value in~\eqref{eq:coeff_tau_greater} as $\hat{c}^{(p)}_{\tau}$. %= \frac{g_{n,\tau}^{(p)}}{\CD_p}   
		However, now,  $\lambda$ can also be found in any position up to the $\Lambda-j$ position of the vector describing the permutation, because  $j < p$. In other words, $\lambda$ can appear in some of the last $p$~positions of the permutation $\sigma$. Then, the coefficient $c_{\tau}^{(p)}$ can be written as
			\begin{align}
						c^{(p)}_{\tau}\triangleq \hat{c}^{(p)}_{\tau} + 	\breve{c}^{(p)}_{\tau}
			\end{align}
		where $\breve{c}^{(p)}_{\tau}$ accounts for those permutations in which $\lambda$ appears in one of the last $p$ positions, which are not considered in~$\hat{c}^{(p)}_{\tau}$.  
		In order to obtain $\breve{c}^{(p)}_{\tau}$, let us first fix the position of $\lambda$, and let $r$ denote in which of the last $p$ positions is $\lambda$ located. 
		Hence, $r=1$ implies that $\lambda$ is in the $\Lambda - p+1$ position of the permutation $\sigma$, whereas $r=p$ implies that $\lambda$ is in the last position (see Fig.~\ref{fig:help_R} for a visual explanation).

		If $\lambda$ can be found in any of the last $p$ positions of the permutation~$\sigma$, Proposition~\ref{prop:validperm} implies that all the values in $\tau$ must also be 
		in those last $p$ positions, and in particular in the positions $\{\Lambda-p+r+1,\dots,\Lambda\}$. 
		Now, the remaining $p-j-1$ positions can be filled with the indices of the other $\Lambda - j - 1$ caches. 
		Let us consider a particular set $q$ of indices, $|q|=p$, filling the last $p$ positions. 
		We can write such a set as
			\begin{align}
						q \triangleq \{\lambda\cup\tau\cup s\},
			\end{align}
		where $\lambda\cap\tau=\varnothing$, and $s\in C_{p-j-1}^{[\Lambda]\backslash\{\lambda\cup\tau\}}$. Note that $|[\Lambda]\backslash\{\lambda\cup\tau\}| = \Lambda - j - 1$. 
		Then, the numerator of the coefficient $\breve{c}^{(p)}_{\tau}$ is given by 
			\begin{align}\label{eq:coef_new_B}
				\breve{c}^{(p),num}_{\tau} = 
						\underbrace{P(N-1,K-1)\!\!\sum_{\lambda\in[\Lambda]\backslash\{\tau\}}\!\!\!\! L_\lambda}_{\substack{\text{Times that $W_\tau^n$ is requested, cf.~\eqref{eq:valid_demands}}}}\!
						\underbrace{\sum_{s\in C_{p-j-1}^{[\Lambda]\backslash\{\lambda\cup\tau\}}}}_{\substack{\text{All possible}\\ \text{ combinations}\\ \text{ of cache indices}\\ \text{in last $p$ positions}\\ \text{apart from $\lambda,\tau$}}}
						\!\!\!\!\overbrace{L_\lambda \Ld_\tau \Ld_s}^{\substack{\text{$=\Ld_q$}}}
						&\underbrace{\sum_{r=1}^{p-j}}_{\substack{\text{Possible}\\ \text{positions}\\ \text{of $\lambda$}}}
						\underbrace{P(p-r,j)}_{\substack{\text{Ways of placing }\\ \text{$\tau$'s elements in last}\\ \text{$p-r$ positions}}}
						\underbrace{(p-j-1)!}_{\substack{\text{For every $\tau,\lambda$,}\\ \text{filling the other }\\ \text{$p-j-1$ positions}\\\text{(not occupied by $\tau,\lambda$)}}}
						\underbrace{(\Lambda-p)!}_{\substack{\text{Filling the}\\ \text{first $\Lambda-p$}\\ \text{positions}}}.
			\end{align}
		Adding the denominator (as before), we can simplify as
			\begin{align}
				\breve{c}^{(p)}_{\tau} &= \frac{\breve{c}^{(p),num}_{\tau}}{P(N,K)p!(\Lambda-p)!\sum_{\ell\in C_{p}^{[\Lambda]}}^{} \Ld_{\ell}} \\
						&\overset{(a)}{=} \frac{
								\sum_{\lambda\in[\Lambda]\backslash\{\tau\}} L_\lambda
								\sum_{s\in C_{p-j-1}^{[\Lambda]\backslash\{\lambda\cup\tau\}}}
								L_\lambda \Ld_\tau \Ld_s
								\sum_{r=1}^{p-j}
								P(p-r,j)(p-j-1)!
							}{
								N p!\sum_{\ell\in C_{p}^{[\Lambda]}}^{} \Ld_{\ell}
							}\\
						&\overset{(b)}{=}  \frac{
								\sum_{\lambda\in[\Lambda]\backslash\{\tau\}} L_\lambda
								\sum_{s\in C_{p-j-1}^{[\Lambda]\backslash\{\lambda\cup\tau\}}}
								L_\lambda \Ld_\tau \Ld_s
								\frac{p!}{j+1}
							}{
								N p!\sum_{\ell\in C_{p}^{[\Lambda]}}^{} \Ld_{\ell}
							}, 
			\end{align}
		where in  $(a)$ we have applied that $\frac{P(N-1,K-1)}{P(N,K)}=\frac{1}{N}$ and canceled out $(\Lambda-p)!$, whereas in $(b)$ we have considered that
			\begin{align}
				\sum_{r=1}^{p-j}
						P(p-r,j)(p-j-1)! 
					& = (p-j-1)!\sum_{r=1}^{p-j} \frac{(p-r)!}{(p-r-j)!}	\\					
					& = \frac{p!}{j+1}.
			\end{align}		
		Then, by recalling that $j=|\tau|$, it follows that
			\begin{align}
				\breve{c}^{(p)}_{\tau}
						&= \frac{1}{|\tau|+1}
							\frac{
								\sum_{\lambda\in[\Lambda]\backslash\{\tau\}} L_\lambda
								\sum_{s\in C_{p-j-1}^{[\Lambda]\backslash\{\lambda\cup\tau\}}}
								L_\lambda \Ld_\tau \Ld_s
							}{
								N \sum_{\ell\in C_{p}^{[\Lambda]}}^{} \Ld_{\ell}
							} \\
						& = 
							\frac{%
								1
							}{%
								N\sum_{\ell\in \Cb{[\Lambda]}{p}}^{} \Ld_{\ell} %
							}\cdot
							\frac{1}{|\tau|+1}
								\Ld_{\tau}
								\sum_{s\in C_{p-j}^{[\Lambda]\backslash\{\tau\}}}
								\Big(\Ld_s \cdot \sum_{i=1}^{p-j}L_{s(i)}\Big).\label{eq:last_coef_end}
			\end{align}			
		Thus, from~\eqref{eq:coeff_tau_greater} and~\eqref{eq:last_coef_end}, the total coefficient $c^{(p)}_{\tau} \triangleq \hat{c}^{(p)}_{\tau} + 	\breve{c}^{(p)}_{\tau}$ is then given by
			\begin{align}
				c^{(p)}_{\tau} & =  
							\frac{%
								1
							}{%
								N\sum_{\ell\in \Cb{[\Lambda]}{p}}^{} \Ld_{\ell} %
							}
						\bigg(
							\sum_{q\in \Cb{[\Lambda]}{p+1}}^{} \Ld_{q} \frac{p+1-|q\cap \tau|}{|\tau|+1-|q\cap \tau|}%
							+
							\frac{1}{|\tau|+1}
								\Ld_{\tau}
								\sum_{s\in C_{p-j}^{[\Lambda]\backslash\{\tau\}}}
								\Big(
								\Ld_s \cdot \sum_{i=1}^{p-j}L_{s(i)}\Big)
						\bigg),\label{eq:coef_final_end}
			\end{align}
		which concludes the proof of Lemma~\ref{lem:long_coefficient}.  
		Note that, for the case where $j\geq p$, it holds that $\breve{c}^{(p)}_{\tau}=0$, and hence in this case it holds that $c^{(p)}_{\tau} = \hat{c}^{(p)}_{\tau}$, which allows us to consider~\eqref{eq:coef_final_end} for any value of $|\tau|$.     \qed

%%%%%%%%%%%%%%%%%%%%%%%%%%%%%%%%%%%%%%%%%%%%%%%%%%%%%%%%
\section{Proof of Lemma~\ref{lem:optimal_tau}}\label{app:optimal_tau}
%%%%%%%%%%%%%%%%%%%%%%%%%%%%%%%%%%%%%%%%%%%%%%%%%%%%%%%%

    In this appendix, we obtain the set $\tau$ of cardinality $|\tau| = j$ that minimizes $\Tilde{c}^{(p)}_\tau$ for each $j\in[\Lambda]_0$. 
    For the sake of readability, let us  recover the notation $\Ld_{q}\triangleq \prod_{j=1}^{|q|} L_{q(j)}$, for any set $q$. Let us start by recalling that $\Tilde{c}^{(p)}_\tau$ is defined as 
    \begin{align}\label{eq:c_up}
        \Tilde{c}^{(p)}_\tau\dfn 
            \frac{%
    		    \sum_{q\in \Cb{[\Lambda]}{p+1}}^{}\Ld_{q} \frac{p+1-|q\cap \tau|}{|\tau|+1-|q\cap \tau|}
        	}{%
        		\sum_{\ell\in \Cb{[\Lambda]}{p}}\Ld_{\ell}. %
        	}.      
    \end{align}
	For the case of $j=0$, the only possible set $\tau$ is the empty set, whereas for the case $j=p$ we can see from~\eqref{eq:c_up} that all $\Tilde{c}^{(p)}_\tau$ for any $\tau$ with $|\tau| = p$ have the same value, and thus any $\tau$ for which $|\tau| = p$ is a solution of the optimization problem. We select $\tau^\star_p = \{1,\ 2,\ \dots,\  p\}$ without loss of generality. In the following, we focus on the cases where $j\in\{[\Lambda]\setminus p\}$.
    
    We start by presenting a key lemma where, instead of considering the set $\tau$ that optimizes $\argmin_{\substack{\tau\subseteq [\Lambda],\,|\tau|=j}} \Tilde{c}^{(p)}_\tau$, we consider the problem of finding the \emph{element} $\nu$ that minimizes $\Tilde{c}^{(p)}_\tau$ for a given subset $\phi$ of cardinality $|\phi|=j-1$, such that $\tau \triangleq \{\nu\cup\phi\}$.  
    \begin{lem}\label{lem:optimal_tau_subset}
		Let us consider a fixed set $\phi\subseteq [\Lambda]$ of cardinality $|\phi|=j-1$.  
		Then, for any  $p\in[\Lambda]_0$ and $j\in[\Lambda]$, it holds that
		\begin{align}
			\nu^\star(\phi) \triangleq \argmin_{\substack{\nu\in \{[\Lambda]\setminus\phi\} }} \Tilde{c}^{(p)}_{\{\nu\cup\phi\}} = \argmin_{\substack{\nu\in \{[\Lambda]\setminus\phi\} }}\sgn(p-j)L_{\nu} 
		\end{align}
    \end{lem}
    \begin{proof}
        The proof is relegated to Appendix~\ref{app:fromopt_tau2opt_nu}.
    \end{proof}
    In other words, the cache $\nu^\star$ that minimizes $\Tilde{c}^{(p)}_{\{\nu\cup\phi\}}$ for a given  $\phi$ is: $(i)$ the cache not belonging to $\phi$ with the smallest number of users, when $j< p$; $(ii)$  the cache not belonging to $\phi$ with the biggest number of users, when $j> p$. 
    
    Next, we prove Lemma~\ref{lem:optimal_tau} directly from  Lemma~\ref{lem:optimal_tau_subset} and the assumption that the caches are sorted such that $L_1\geq L_2\geq\dots\geq L_\Lambda$. For this, we split the proof in the cases $j<p$ and $j>p$.  
    
	%%%%%%%%%%%%%%%%%%%%%%%%%%%%%%%%%%%%%%%%%%%%%%%%%%%%%
	\subsubsection{Case $j<p$}
	%%%%%%%%%%%%%%%%%%%%%%%%%%%%%%%%%%%%%%%%%%%%%%%%%%%%%
        From Lemma~\ref{lem:optimal_tau_subset}, we have that 
        \begin{align}\label{eq:case_jmp_first}
            \nu^\star(\phi) = \argmin_{\substack{\nu\in \{[\Lambda]\setminus\phi\} }} \Tilde{c}^{(p)}_{\{\nu\cup\phi\}} = \argmin_{\substack{\nu\in \{[\Lambda]\setminus\phi\} }}L_{\nu} =  \max_{\nu\in\{[\Lambda]\setminus\phi\}} \nu
        \end{align}
        where the last step follows from the ordering  $L_1\geq L_2\geq\dots\geq L_\Lambda$. It remains to prove that~\eqref{eq:case_jmp_first} implies that 
        	$\argmin_{\substack{\tau\subseteq [\Lambda],\, |\tau|=j}} \Tilde{c}_{\tau}^{(p)} 
				=  \{\Lambda -j + 1, \ \Lambda -j + 2,\ \dots,\  \Lambda\}$, which will prove Lemma~\ref{lem:optimal_tau} for $j<p$.
				
		Note that, for any set $\tau$ not including $\Lambda$, i.e., for any $\tau\in C_{j}^{[\Lambda-1]}$, it follows from Lemma~\ref{lem:optimal_tau_subset} that, for any subset of $\tau$ of cardinality $j-1$, which we denote by $\phi \in C_{j-1}^{\tau}$, we have that
		$\Tilde{c}_{\{\Lambda\cup\phi\}}^{(p)} \leq \Tilde{c}_{\tau}^{(p)}$. 
		
		Similarly,  for any set $\tau$ including~$\Lambda$ but not including $\Lambda-1$, i.e., for any $\tau\in \{\Lambda\cup C_{j-1}^{[\Lambda-2]}\}$, it follows from Lemma~\ref{lem:optimal_tau_subset} that, for any subset of $\tau$ of cardinality $j-1$ including $\Lambda$, which we denote by $\phi\in \{\Lambda\cup C_{j-2}^{\tau\setminus\Lambda}\}$, we have that
		$\Tilde{c}_{\{\Lambda-1\cup\phi\}}^{(p)} \leq \Tilde{c}_{\tau}^{(p)}$. We can proceed in the same manner for any possible set $\tau$, taking into account the sets that do not include cache $\lambda$ but include all caches in $\{\lambda +1, \dotsc, \Lambda\}$, which leads to the fact that~\eqref{eq:case_jmp_first} implies that  
		\begin{align}
			\argmin_{\substack{\tau\subseteq [\Lambda] \\ |\tau|=j}} \Tilde{c}_{\tau}^{(p)} =  \{\Lambda -j + 1, \ \Lambda -j + 2,\ \dots,\  \Lambda\}.
		\end{align}
	%%%%%%%%%%%%%%%%%%%%%%%%%%%%%%%%%%%%%%%%%%%%%%%%%%%%%
	\subsubsection{Case $p<j$}
	%%%%%%%%%%%%%%%%%%%%%%%%%%%%%%%%%%%%%%%%%%%%%%%%%%%%%
	    For this case, the only difference is that now $\sgn(p-j) = -1$. This implies that~\eqref{eq:case_jmp_first} becomes 
	   \begin{align}\label{eq:case_jmp_second}
            \nu^\star(\phi) = \argmin_{\substack{\nu\in \{[\Lambda]\setminus\phi\} }} \Tilde{c}^{(p)}_{\{\nu\cup\phi\}} = \argmax_{\substack{\nu\in \{[\Lambda]\setminus\phi\} }}L_{\nu} =  \min_{\nu\in\{[\Lambda]\setminus\phi\}} \nu
        \end{align}
        Hence, we can follow the same steps as for the case $j<p$ but taking into account the sets that do not include cache $\lambda$ but include all caches in $[\lambda - 1]$, which leads to the fact
		\begin{align}
			\argmin_{\substack{\tau\subseteq [\Lambda] \\ |\tau|=j}} \Tilde{c}_{\tau}^{(p)} =  \{1, \ 2,\ \dots,\  j\} 		
		\end{align}			    
		which concludes the proof for $p<j$ and thus the proof of Lemma~\ref{lem:optimal_tau}. \qed
		
	%%%%%%%%%%%%%%%%%%%%%%%%%%%%%%%%%%%%%%%%%%%%%%%%%%%%%
	\subsection{Proof of Lemma~\ref{lem:optimal_tau_subset}} \label{app:fromopt_tau2opt_nu}
	%%%%%%%%%%%%%%%%%%%%%%%%%%%%%%%%%%%%%%%%%%%%%%%%%%%%%
	Let us start by noting that, since $\frac{p+1-|q\cap \tau|}{|\tau|+1-|q\cap \tau|} =  1 + \frac{p-|\tau|}{|\tau|+1-|q\cap \tau|}$, we can re-write~\eqref{eq:c_up} as follows: 
		\begin{align}
			\Tilde{c}^{(p)}_\tau	
				& = \frac{\sum_{q\in \Cb{[\Lambda]}{p+1}}^{}\Ld_{q}}{\sum_{\ell\in \Cb{[\Lambda]}{p}}^{} \Ld_{\ell}} + 
					\frac{p-|\tau|}{\sum_{\ell\in \Cb{[\Lambda]}{p}}^{} \Ld_{\ell}}
					\Big(\underbrace{\sum_{q\in \Cb{[\Lambda]}{p+1}}^{}\Ld_{q}\frac{1}{|\tau|+1-|q\cap \tau|}}_{a^{(p)}_{\tau}}\Big).\label{eq:def_atpproflem2}
		\end{align}
	From this expression, we can see that, for any $\tau$ such that $|\tau| = j$, the only term in~\eqref{eq:def_atpproflem2} that differs with respect to any other $\tau'$ of the same cardinality is the sum denoted by $a^{(p)}_{\tau}$. From this fact, and by taking into account that the sign of $(p-|\tau|)$ is different whether $p>|\tau|$ or not, it follows that 
		\begin{align}
			\argmin_{\substack{\tau\subseteq [\Lambda],\,|\tau|=j}} \Tilde{c}^{(p)}_\tau  
				& = \argmin_{\substack{\tau\subseteq [\Lambda],\,|\tau|=j}} \ \sgn(p-j)a^{(p)}_{\tau}. \label{eq:argmn_gen22lem}
		\end{align}
	In order to prove Lemma~\ref{lem:optimal_tau_subset}, we need to prove that
    	\begin{align}\label{eq:proof_element_optimal0}
			\nu^\star(\phi) \triangleq \argmin_{\substack{\nu\in \{[\Lambda]\setminus\phi\} }} \Tilde{c}^{(p)}_{\{\nu\cup\phi\}} = \argmin_{\substack{\nu\in \{[\Lambda]\setminus\phi\} }}\sgn(p-j)L_{\nu}
		\end{align}
    From~\eqref{eq:def_atpproflem2}--\eqref{eq:argmn_gen22lem}, it follows that
 		\begin{align}
			\nu^\star(\phi) \triangleq \argmin_{\substack{\nu\in \{[\Lambda]\setminus\phi\} }} \Tilde{c}^{(p)}_{\{\nu\cup\phi\}}   
				& = \argmin_{\substack{\nu\in \{[\Lambda]\setminus\phi\} }} \ \sgn(p-j)a^{(p)}_{\{\nu\cup\phi\}}. \label{eq:proof_element_optimal1}
		\end{align}   
	The following lemma is key to obtain~\eqref{eq:proof_element_optimal0} from~\eqref{eq:proof_element_optimal1}. 
	\begin{lem}\label{lem:a_hat}
		Let us consider a given set $\phi\subseteq [\Lambda]$ of cardinality $|\phi|=j-1$.  
		Then, for any  $p\in[\Lambda]_0$ and $j\in[\Lambda]$, it holds that
		\begin{align}
			\nu^\star(\phi) \triangleq \argmin_{\substack{\nu\in \{[\Lambda]\setminus\phi\} }}\  
			\sgn(p-j)a^{(p)}_{\{\nu\cup\phi\}}
			& = \argmin_{\substack{\nu\in \{[\Lambda]\setminus\phi\} }} \ \sgn(p-j)b^{(p)}_{\{\nu\cup\phi\}}
			\label{eq:lemma_a_b_1}
		\end{align}		
		where, for any $\tau \triangleq \{\nu\cup\phi\} \in \Cb{[\Lambda]}{j}$, $b^{(p)}_{\tau}$ is defined as
		\begin{align}
			b^{(p)}_{\tau}
				& \triangleq \sum_{\ell=0}^{\min(p+1,j)}\frac{-1}{(j-\ell)(j-\ell+ 	1)}\sum_{\substack{\mu\subseteq \{\tau\backslash\nu\}\\|\mu| = \ell}}^{}\Ld_{\mu}
				\sum_{\substack{q\in C_{p+1-\ell}^{[\Lambda]\backslash\{\tau\}}}}^{}\Ld_{q}.\label{eq:breaking_problem8a0}
		\end{align}				
	\end{lem}
	\begin{proof}
		The proof is relegated to Appendix~\ref{se:proof_lemma_a_b}. 
	\end{proof}
	Next, we demonstrate that~\eqref{eq:lemma_a_b_1} in Lemma~\ref{lem:a_hat} is equivalent to~\eqref{eq:proof_element_optimal0}. For that, we split the proof for the cases where $p<j$ and $p>j$. 
		
		%%%%%%%%%%%%%%%%%%%%%%%%%%%%%%%%%%%%%%%%%%%%%%%%%%%%%
		\subsubsection{Case $p>j$}
		%%%%%%%%%%%%%%%%%%%%%%%%%%%%%%%%%%%%%%%%%%%%%%%%%%%%%
			Note that the sum $\sum_{\substack{\mu\subseteq \{\tau\backslash\nu\},\;|\mu| = \ell}}^{} \Ld_{\mu}$ in~\eqref{eq:breaking_problem8a0} depends only on the $j-1$ elements on $\tau$ that are assumed to be fixed in this step (i.e., on $\phi$). 
			Furthermore, since it holds that $\sgn(p-j) = 1$, and all the terms in~\eqref{eq:breaking_problem8a0} are negative, we need to maximize $\sum_{\substack{q\in C_{p+1-\ell}^{[\Lambda]\backslash\{\tau\}}}}^{}\Ld_{q}$ in order to minimize~\eqref{eq:breaking_problem8a0}. 
			Since the optimization variable ($\nu$) is the term that we \emph{remove} from the sum (recall that $q\subset [\Lambda]\backslash\{\tau\}$), this maximization is achieved by selecting the cache in $\big\{[\Lambda]\backslash\phi\big\}$ with the smallest~$L$, i.e., it holds that 	
				\begin{align}
						\nu^\star(\phi) 
						    = \argmin_{\substack{\nu\in \{[\Lambda]\backslash\phi\} }}
								b^{(p)}_{\{\nu\cup\phi\}} 
							= \argmin_{\substack{\nu\in \{[\Lambda]\backslash\phi\} }}
								L_{\nu}. 
								\label{eq:breaking_problemtt2b}
				\end{align}			
						
		%%%%%%%%%%%%%%%%%%%%%%%%%%%%%%%%%%%%%%%%%%%%%%%%%%%%%
		\subsubsection{Case $p<j$}
		%%%%%%%%%%%%%%%%%%%%%%%%%%%%%%%%%%%%%%%%%%%%%%%%%%%%%
			Unlike in the previous case, we now have  that $\sgn(p-j) = -1$. With the change of sign, all the addends in~\eqref{eq:breaking_problem8a0} are positive, and thus we now want to maximize~\eqref{eq:breaking_problem8a0}. 
			To do so, we seek to minimize $\sum_{\substack{q\in C_{p+1-\ell}^{[\Lambda]\backslash\{\tau\}}}}^{}\Ld_{q}$. Since the optimization variable ($\nu$) is the term that we \emph{remove} from the sum, this minimization is achieved by selecting the cache in $\big\{[\Lambda]\backslash\phi\big\}$ with the biggest~$L$. 			
			Thus,  applying the same reasoning as for the case $p>j$, we obtain that
				\begin{align}
					\nu^\star(\phi) 
						= \argmax_{\substack{\nu\in \{[\Lambda]\backslash\phi\} }}
								b^{(p)}_{\{\nu\cup\phi\}}
						= \argmax_{\substack{\nu\in \{[\Lambda]\backslash\phi\} }}
								L_{\nu}.
								\label{eq:breaking_problemtt2b2}
				\end{align}				
		    
		Finally, from~\eqref{eq:breaking_problemtt2b} and~\eqref{eq:breaking_problemtt2b2} we obtain~\eqref{eq:proof_element_optimal0}, which  concludes the proof of Lemma~\ref{lem:optimal_tau_subset}. \qed

	%%%%%%%%%%%%%%%%%%%%%%%%%%%%%%%%%%%%%%%%%%%%%%%%%%%%%
	\subsection{Proof of Lemma~\ref{lem:a_hat}}\label{se:proof_lemma_a_b}
	%%%%%%%%%%%%%%%%%%%%%%%%%%%%%%%%%%%%%%%%%%%%%%%%%%%%%		

		We  obtain Lemma~\ref{lem:a_hat} by re-writing the terms inside  $a^{(p)}_{\tau}$ such that some of the terms do not impact the optimization problem, and hence we can remove them.  
				
		%%%%%%%%%%%%%%%%%%%%%%%%%%%%%%%%%%%%%%%%%%%%%%%%%%%%%
		\subsubsection{Obtaining a new expression for $a^{(p)}_{\tau} $}
		%%%%%%%%%%%%%%%%%%%%%%%%%%%%%%%%%%%%%%%%%%%%%%%%%%%%%		
	
		 It follows that, for every $\tau\in[\Lambda]$, the term $a^{(p)}_{\tau}$ that we have defined in~\eqref{eq:def_atpproflem2} can be written as
			\begin{align}
					a^{(p)}_{\tau} 
						&	= \sum_{q\in C_{p+1}^{[\Lambda]}}^{}\Ld_{q}\frac{1}{j+1-|q\cap \tau|} %\\	&	
						= \sum_{m=0}^{\min(j,p+1)}\frac{1}{j+1-m}\sum_{\substack{q\in C_{p+1}^{[\Lambda]}\\ |q\cap \tau| =m}}^{}\Ld_{q}, \label{eq:proof_j_1}
			\end{align}
		where	the $\min(j,p+1)$ term in the summation comes from the fact that $|q\cap \tau| \leq \min(|q|,|\tau|)$. 
		For a given $\tau$, let $\eta$ be a subset of $\tau$ of cardinality $m$, such that $\eta\subseteq \tau$, $|\eta|=m$. 
		The last sum in~\eqref{eq:proof_j_1} can be expanded as
			\begin{align}\label{eq:proof_j_1b}
				\sum_{\substack{q\in C_{p+1}^{[\Lambda]}\\ |q\cap \tau| =m}}^{}\Ld_{q} 
					= \sum_{\substack{\eta\subseteq \tau \\ |\eta|=m}}^{}\sum_{\substack{q\in C_{p+1}^{[\Lambda]}\\ q\cap \tau =\eta}}^{}\Ld_{q}.
			\end{align}		
		Let us consider a particular $\eta\subseteq \tau$, $|\eta|=m$. Then, 
				\begin{align}
					\sum_{\substack{q\in C_{p+1}^{[\Lambda]}\\ q\cap \tau =\eta}}^{}\Ld_{q} 
						= \sum_{\substack{q\in C_{p+1}^{{[\Lambda]}\backslash\{\tau\backslash \eta\}}\\ \eta\subseteq q}}^{}\Ld_{q} 
						= \Ld_{\eta}\sum_{\substack{q\in C_{p+1-m}^{{[\Lambda]}\backslash\{\tau\}}}}^{}\Ld_{q}. \label{eq:proof_j_2}
			\end{align}	
		Let us recall that the term $\sum_{q\in C_{p+1}^{[\Lambda]}}^{}\Ld_{q}$ represents the $(p+1)$-th elementary symmetric polynomial for the set $\Up\dfn\{L_\lambda\}_{\lambda=1}^{\Lambda}$, and that the  elementary symmetric polynomials satisfy Property~\ref{prop:recursive_e_t}. 
		Hence, for any set of positive integers $\Omega$ and any integer $i\notin\Omega$, we can rewrite~\eqref{eq:eq_prop1} in Property~\ref{prop:recursive_e_t} using the above notation as 
			\begin{align}
			    L_i\sum_{q\in C_{p}^{\Omega}}^{}\Ld_{q} = \sum_{q\in C_{p+1}^{\{\Omega\cup i\}}}^{}\Ld_{q} - \sum_{q\in C_{p+1}^\Omega}^{}\Ld_{q}. \label{eq:key_prop}
			\end{align}
		This is equivalent to saying that the sum over all the terms $\Ld_{q}$ (with $|q|=p+1$) that include $L_i$ is equal to the sum over all the terms $\Ld_{q}$ (with $|q|=p+1$) minus the sum over all the terms $\Ld_{q}$ ($|q|=p+1$) that do not include $L_i$. This intuitive relation will prove important for the derivation. 
		
		Note that $\Ld_{\eta} = \prod_{\substack{i\in \eta}} L_i$. Then, we can successively apply~\eqref{eq:key_prop} to~\eqref{eq:proof_j_2}  for all the $L_i$, $i\in \eta$, $|\eta|=m$, such that~\eqref{eq:proof_j_2} is expanded in $2^m$ sums as  
			\begin{align}
					\sum_{\substack{q\in C_{p+1}^{[\Lambda]}\\ q\cap \tau =\eta}}^{}\Ld_{q}
						& = \sum_{\kappa \subseteq \eta }^{} (-1)^{|\kappa|}\sum_{\substack{q\in C_{p+1}^{{[\Lambda]}\backslash\{\tau\backslash \{\eta\backslash s\}\}}}}^{}\Ld_{q} %\\ &
						 = \sum_{k=0}^m (-1)^{k} 
						\sum_{\substack{\kappa \subseteq \eta\\ |\kappa| = k}}^{} 
						\sum_{\substack{q\in C_{p+1}^{{[\Lambda]}\backslash\{\tau\backslash \{\eta\backslash \kappa\}\}}}}^{}\Ld_{q}. 
					\label{eq:proof_j_3}
			\end{align}		
		Thus, from~\eqref{eq:proof_j_1},~\eqref{eq:proof_j_1b}, and~\eqref{eq:proof_j_3}, we obtain that
			\begin{align}
					a^{(p)}_{\tau} 
						&	= \sum_{m=0}^{\min(j,p+1)}\frac{1}{j+1-m}
								\sum_{\substack{\eta\subseteq \tau \\ |\eta|=m}}^{}
								\sum_{k=0}^m (-1)^{k} 
						\sum_{\substack{\kappa \subseteq \eta\\ |\kappa| = k}}^{} 
						\sum_{\substack{q\in C_{p+1}^{{[\Lambda]}\backslash\{\tau\backslash \{\eta\backslash \kappa\}\}}}}^{}\Ld_{q}. \label{eq:completely_expanded}
			\end{align}
		Since $\{\tau\backslash \{\eta\backslash \kappa\}\}$ can be the same set for different $\eta,\kappa$, let us count how many times  the term $\sum_{q\in C_{p+1}^{{[\Lambda]}\backslash\{w\}}}^{}\Ld_{q}$ appears in~\eqref{eq:completely_expanded} for a certain $w\subseteq [\tau]$, $|w| = i$. 			
		Let us fix $m$ (i.e., the cardinality of $|\eta|$) and $j = |\tau|$.  
		It follows that $|w| = |\{\tau\backslash \{\eta\backslash \kappa\}\} | = j -(m-k)$. 
		We have that $w = \tau\backslash \{\eta\backslash \kappa\} = \{\tau\backslash \eta\} \cup \kappa$. This implies that $\kappa\subseteq w$. 
		Furthermore, for any $\kappa\subseteq w$, there exists a distinct and unique $\eta$ such that $w = \{\tau\backslash \eta\} \cup \kappa$. Since there are $\binom{|w|}{|\kappa|} =\binom{i}{k}=\binom{i}{i+m-j}$ possible $\kappa\subseteq w$ of cardinality $k$,  each of the terms $\sum_{q\in C_{p+1}^{{[\Lambda]}\backslash\{w\}}}^{}\Ld_{q}$ appears in~\eqref{eq:completely_expanded} exactly $\binom{i}{i+m-j}$ times for a particular $m$, $i$ and $j$.

		Let~$\Theta_{\backslash i}$  denote the sum over all $q\in C_{p+1}^{{[\Lambda]}\backslash\{w\}}$ for any $w$ of cardinality $|w|=i$. This~$\Theta_{\backslash i}$ takes the form 
			\begin{align}
					\Theta_{\backslash i} \triangleq
						\sum_{\substack{w \subseteq \tau \\ |w| = i}}^{} 
						\sum_{\substack{q\in C_{p+1}^{{[\Lambda]}\backslash\{w\}}}}^{}\Ld_{q}. 
					\label{eq:proof_j_3b}
			\end{align}	
		From~\eqref{eq:proof_j_3b} and the fact that the term $\sum_{q\in C_{p+1}^{{[\Lambda]}\backslash\{w\}}}^{}\Ld_{q}$ appears in~\eqref{eq:completely_expanded} exactly $\binom{i}{i+m-j}$ times for a particular $m$, $i$ and $j$, and after applying $k = i +m-j$,  it follows that   
			\begin{align}
					  \sum_{k=0}^m (-1)^{k} 
							\sum_{\substack{\eta\subseteq \tau \\ |\eta|=m}}^{}
							\sum_{\substack{\kappa \subseteq \eta\\ |\kappa| = k}}^{} 
							\sum_{\substack{q\in C_{p+1}^{{[\Lambda]}\backslash\{\tau\backslash \{\eta\backslash \kappa\}\}}}}^{}\Ld_{q} 
						= \sum_{i=j-m}^{j} (-1)^{i+m-j} 
							\binom{i}{i+m-j}\Theta_{\backslash i}, 
							\label{eq:completely_expanded1cd}
			\end{align}		
		where we have substituted $k = i +m-j$. 
  
        Then, we can apply~\eqref{eq:completely_expanded1cd} into~\eqref{eq:completely_expanded} to obtain that
			\begin{align}
					a^{(p)}_{\tau} 
						&	= \sum_{m=0}^{\min(j,p+1)}
						    \sum_{i=j-m}^{j}\frac{1}{j+1-m} (-1)^{i+m-j} 
								\binom{i}{i+m-j}\Theta_{\backslash i}. \label{eq:completely_expanded2aa}
			\end{align}		
		This expression of $a^{(p)}_{\tau}$ can be further simplified. 
		Before continuing, let us take a look at the term $\Theta_{\backslash i}$. We show in the following that not all the components of $\Theta_{\backslash i}$ will impact the optimization of  $a^{(p)}_{\tau}$. 
		
		%%%%%%%%%%%%%%%%%%%%%%%%%%%%%%%%%%%%%%%%%%%%%%%%%%%%%
		\subsubsection{Reducing $\Theta_{\backslash i}$ to its meaningful components}
		%%%%%%%%%%%%%%%%%%%%%%%%%%%%%%%%%%%%%%%%%%%%%%%%%%%%%		
			We recall that, as expressed in~\eqref{eq:lemma_a_b_1} in Lemma~\ref{lem:a_hat}, our goal is to consider a single element $\nu$ for a given set $\phi$, such that $\tau\triangleq\{\nu\cup \phi\}$, and obtain
				\begin{align}
					\argmin_{\substack{\nu\in \{[\Lambda]\backslash\phi\}}}
							\sgn(p-j)a^{(p)}_{\tau}. \label{eq:breaking_problem2}
				\end{align}			
			In order to continue from~\eqref{eq:completely_expanded2aa}, let us focus on the term  $\Theta_{\backslash i}\triangleq\sum_{\substack{w \subseteq \tau,\ |w| = i}}^{}\sum_{\substack{q\in C_{p+1}^{{[\Lambda]}\backslash\{w\}}}}^{}\Ld_{q}$, which has been defined in~\eqref{eq:proof_j_3b}. $\Theta_{\backslash i}$ is composed of $\binom{j}{i}$ sums, one for each $w\subseteq\tau : |w|=i$. 
			Interestingly, if  $w\subseteq \tau$ is actually a subset of $\phi$ ($w\subseteq \phi$), the term $\sum_{\substack{q\in C_{p+1}^{{[\Lambda]}\backslash\{w\}}}}^{}\Ld_{q}$ is the same no matter which value in $\{[\Lambda]\backslash \phi\}$ is selected as $\nu$. Thus, such terms are irrelevant for the optimization problem. 
			
			Let us then consider the remaining cases that do impact the optimization problem, and let us denote the sum of the $\binom{j-1}{i-1}$ subsets $w$ in~\eqref{eq:proof_j_3b}  which contain $\nu$  as $\Theta_{\backslash i-1}^{\backslash \nu} $, i.e.,
				\begin{align}
						\Theta_{\backslash i-1}^{\backslash \nu} \triangleq
							\sum_{\substack{\chi \subseteq \phi \\ |\chi| = i-1}}^{} 
							\sum_{\substack{q\in C_{p+1}^{{[\Lambda]}\backslash\{\chi,\nu\}}}}^{}\Ld_{q} 
						\label{eq:proof_j_4}
				\end{align}		
			such that we can define the term that impacts the optimization problem as
				\begin{align}
					b^{(p)}_{\tau}
						 &	\dfn \sum_{m=0}^{\min(j,p+1)}
									\sum_{i=j-m}^{j}\frac{1}{j+1-m} (-1)^{i+m-j} 
									\binom{i}{i+m-j}\Theta_{\backslash i-1}^{\backslash \nu}, 
								\label{eq:breaking_problem2bss}
				\end{align}					
			where $\tau\triangleq \{\phi\cup\nu\}$ and $b^{(p)}_{\tau}$ is obtained by substituting $\Theta_{\backslash i}$ in~\eqref{eq:completely_expanded2aa} %${a}^{p}_{\tau}$ 
			by $\Theta_{\backslash i-1}^{\backslash \nu}$. Then, it follows that
				\begin{align}
						\argmin_{\substack{\nu\in \{[\Lambda]\backslash\phi\} }} \sgn(p-j)a^{(p)}_{\tau}
						= \argmin_{\substack{\nu\in \{[\Lambda]\backslash\phi\} }} \sgn(p-j)b^{(p)}_{\tau}.
								\label{eq:breaking_problem2b}
				\end{align}			
			Next, we simplify $\Theta_{\backslash i-1}^{\backslash \nu}$ to later apply this result into  $b^{(p)}_{\tau}$ and obtain Lemma~\ref{lem:a_hat}.

		%%%%%%%%%%%%%%%%%%%%%%%%%%%%%%%%%%%%%%%%%%%%%%%%%%%%%
		\subsubsection{Simplifying the term $\Theta_{\backslash i-1}^{\backslash \nu}$}
		%%%%%%%%%%%%%%%%%%%%%%%%%%%%%%%%%%%%%%%%%%%%%%%%%%%%%		
		A combination of $p+1$ elements in ${[\Lambda]}\backslash\{\nu,\chi\}$, where $\chi\subseteq \phi$ and $|\chi|=i-1$,   can be expressed as the concatenation of $\ell$ elements of $\{\tau\backslash\{\nu,\chi\}\} = \{\phi\backslash\chi\}$ and $p+1-\ell$ elements of  $\big\{{[\Lambda]}\backslash\{\nu,\chi\}\big\}\backslash\big\{\tau\backslash\{\nu,\chi\}\big\}$ $ = {[\Lambda]}\backslash\{\tau\}$, for any $\ell\in[j-i]_0$. Consequently, it follows that
			\begin{align}
					\sum_{\substack{q\in C_{p+1}^{{[\Lambda]}\backslash\{\nu,\chi\}}}}^{}\Ld_{q} 
						& = \sum_{\ell = 0}^{j-i}
						\sum_{\substack{\mu\subseteq \{\phi\backslash\chi\}\\|\mu| = \ell}}^{}\Ld_{\mu}   
						\sum_{\substack{q\in C_{p+1-\ell}^{{[\Lambda]}\backslash\{\tau\}}}}^{}\Ld_{q}.
					\label{eq:proof_j_5}
			\end{align}					
		Applying~\eqref{eq:proof_j_5} into~\eqref{eq:proof_j_4} yields 
			\begin{align}
					\Theta_{\backslash i-1}^{\backslash \nu}  
						& = 
							\sum_{\substack{\chi \subseteq \phi \\ |\chi| = i-1}}^{} 
							\sum_{\ell = 0}^{j-i}
							\sum_{\substack{\mu\subseteq \{\phi\backslash\chi\}\\|\mu| = \ell}}^{}\Ld_{\mu} \sum_{\substack{q\in C_{p+1-\ell}^{{[\Lambda]}\backslash\{\tau\}}}}^{}\Ld_{q}\\
						& = 
							\sum_{\ell = 0}^{j-i}
							\Big(\underbrace{\sum_{\substack{\chi \subseteq \phi \\ |\chi| = i-1}}^{} 
							\sum_{\substack{\mu\subseteq \{\phi\backslash\chi\}\\
							|\mu| = \ell}}^{}\Ld_{\mu}}_{E_{j-1,\ell}}   \Big)
							\sum_{\substack{q\in C_{p+1-\ell}^{{[\Lambda]}\backslash\{\tau\}}}}^{}\Ld_{q}.
							\label{eq:breaking_problem4}
			\end{align}			
		Next, we want to count how many times the last sum ($\sum_{\substack{q\in C_{p+1-\ell}^{{[\Lambda]}\backslash\{\tau\}}}}^{}\Ld_{q}$) appears in~$\Theta_{\backslash i-1}^{\backslash \nu}$. 
		Consider some given $i$ and $\ell$. 
		In the term $E_{j-1,\ell}$ in~\eqref{eq:breaking_problem4}, a specific $\Ld_\mu$ appears $\binom{j-1-\ell}{i-1}$ times. Then, it holds that 
			\begin{align}
					\Theta_{\backslash i-1}^{\backslash \nu}   
						&= 
						\sum_{\ell = 0}^{j-i}\binom{j-1-\ell}{i-1}\sum_{\substack{\mu\subseteq \phi\\|\mu| = \ell}}^{}\Ld_{\mu}
						\sum_{\substack{q\in C_{p+1-\ell}^{{[\Lambda]}\backslash\{\tau\}}}}^{}\Ld_{q}. 
						\label{eq:breaking_problem5}
			\end{align}
		In the following, we incorporate in~\eqref{eq:breaking_problem2bss} the value of $\Theta_{\backslash i-1}^{\backslash \nu}$  obtained in~\eqref{eq:breaking_problem5}  to derive~\eqref{eq:breaking_problem8a0} and hence~Lemma~\ref{lem:a_hat}.

		%%%%%%%%%%%%%%%%%%%%%%%%%%%%%%%%%%%%%%%%%%%%%%%%%%%%%
		\subsubsection{Obtaining~\eqref{eq:breaking_problem8a0}}
		%%%%%%%%%%%%%%%%%%%%%%%%%%%%%%%%%%%%%%%%%%%%%%%%%%%%%		
		 Let us introduce the notation $\varpi\triangleq\min(j,p+1)$ for ease of readability. Then, we continue from~\eqref{eq:breaking_problem2bss}  as
			\begin{align}
				b^{(p)}_{\tau}
					&\dfn \sum_{m=0}^{\varpi}
					\sum_{i=j-m}^{j}\frac{1}{j+1-m} (-1)^{i+m-j} 
					\binom{i}{i+m-j}\Theta_{\backslash i-1}^{\backslash \nu} \label{aux_eq_solv_appe}\\
					&	\overset{(a)}{=} \sum_{i=j-\varpi}^{j}
					\sum_{m'=j-\varpi}^{i}\frac{1}{m'+1} (-1)^{i-m'} 
					\binom{i}{i-m'}\Theta_{\backslash i-1}^{\backslash \nu} \\
					&	\overset{(b)}{=} \sum_{i=j-\varpi}^{j}
					\frac{(-1)^{i-j+\varpi}}{i+1}\binom{i}{j-\varpi}
					\Theta_{\backslash i-1}^{\backslash \nu},  
					\label{eq:breaking_problem2bsszg2}
			\end{align}	
		where $(a)$ follows from interchanging the summations in~\eqref{aux_eq_solv_appe} and applying the change of variable $m' = j-m$, and where $(b)$ follows from solving the inner summation. 		
		
		Let us now substitute in~\eqref{eq:breaking_problem2bsszg2} the expression of $\Theta_{\backslash i-1}^{\backslash \nu}$ provided in~\eqref{eq:breaking_problem5}, which leads to
			\begin{align}
					b^{(p)}_{\tau}
						& = \sum_{i=j-\varpi}^{j}
							\frac{(-1)^{i-j+\varpi}}{i+1}\binom{i}{j-\varpi}
							\sum_{\ell = 0}^{j-i}\binom{j-1-\ell}{i-1}\sum_{\substack{\mu\subseteq \phi\\|\mu| = \ell}}^{}\Ld_{\mu}
						\sum_{\substack{q\in C_{p+1-\ell}^{{[\Lambda]}\backslash\{\tau\}}}}^{}\Ld_{q}.
			\end{align}
		By interchanging the summations (since $\sum_{i=j-\varpi}^{j}\sum_{\ell = 0}^{j-i} f(i,\ell) = \sum_{\ell = 0}^{\varpi}\sum_{i=j-\varpi}^{j-\ell} f(i,\ell)$), we have that 
			\begin{align}
				b^{(p)}_{\tau}
					&	= \sum_{\ell = 0}^{\varpi}\sum_{i=j-\varpi}^{j-\ell}
					F_{\ell,p,j}
					\sum_{\substack{\mu\subseteq \phi\\|\mu| = \ell}}^{}\Ld_{\mu}
					\sum_{\substack{q\in C_{p+1-\ell}^{{[\Lambda]}\backslash\{\tau\}}}}^{}\Ld_{q}, \label{eq:inner_sum_ast}
			\end{align}
		where $F_{\ell,p,j}\triangleq \frac{(-1)^{i-j+\varpi}}{i+1}\binom{i}{j-\varpi}\binom{j-1-\ell}{i-1}$. Let us consider $\sum_{i=j-\varpi}^{j-\ell}F_{\ell,p,j}$. 
		To simplify the notation, let us define $h\triangleq j - \varpi$ and $g= \varpi-\ell$. Thus, it follows that %6
			\begin{align}
				\sum_{i=j-\varpi}^{j-\ell}F_{\ell,p,j} 
					& = \sum_{i=h}^{h+g}
						\frac{(-1)^{i-h}}{i+1}
						\binom{i}{h}
						\binom{h+g-1}{i-1} \\ 
					& =	\frac{-1}{(h+g+1)(h+g)}  
					  = \frac{-1}{(j-\ell)(j-\ell+1)}.
				\label{eq:inner_sum_asts}
			\end{align}
		Incorporating~\eqref{eq:inner_sum_asts} into~\eqref{eq:inner_sum_ast} yields
			\begin{align}
				b^{(p)}_{\tau}
					&	= \sum_{\ell = 0}^{\varpi}
					\frac{-1}{(j-\ell)(j-\ell+1)}
					\sum_{\substack{\mu\subseteq \phi\\|\mu| = \ell}}^{}\Ld_{\mu}
					\sum_{\substack{q\in C_{p+1-\ell}^{{[\Lambda]}\backslash\{\tau\}}}}^{}\Ld_{q}, \label{eq:inner_sum_assst}
			\end{align}		
		which concludes the proof of~\eqref{eq:breaking_problem8a0}, and consequently it also concludes the proof of Lemma~\ref{lem:a_hat}. \qed

%%%%%%%%%%%%%%%%%%%%%%%%%%%%%%%%%%%%%%%%%%%%%%%%%%%%%%%%
\section{Proof of Proposition~\ref{lem:monotone_c}}\label{app:monotone_c_proof}
%%%%%%%%%%%%%%%%%%%%%%%%%%%%%%%%%%%%%%%%%%%%%%%%%%%%%%%%

    In the following, we prove that the sequence $\big\{\Tilde{c}_{\tau^\star_j}^{(p)}\big\}$ is a decreasing sequence in $j\in [\Lambda]_0 = \{0\cup [\Lambda]\}$, where we recall that $\Tilde{c}_{\tau}^{(p)}$ is given by 
        $$\Tilde{c}^{(p)}_{\tau}\triangleq 
            \frac{%
					\sum_{q\in \Cb{[\Lambda]}{p+1}}^{} \frac{p+1-|q\cap {\tau}|}{|\tau|+1-|q\cap {\tau}|}\prod_{j=1}^{p+1} L_{q(j)}
			}{%
				\sum_{\ell\in \Cb{[\Lambda]}{p}} \prod_{j=1}^{p} L_{\ell(j)} %
			},$$
    and where $\tau^\star_j \triangleq \argmin_{{\tau\subset [\Lambda]_0,\,|\tau| = j}} \Tilde{c}^{(p)}_{\tau}$. 
	Since the denominator of $\Tilde{c}^{(p)}_{\tau}$ is the same for any $\tau$, we focus on the numerator. First, let us denote the numerator of $\Tilde{c}^{(p)}_{\tau}$ by~$A({p,\tau})$, such that 
		\begin{align}\label{eq:def_num_c}
			A({p,\tau}) \triangleq \sum_{q\in \Cb{[\Lambda]}{p+1}}^{}  \frac{p+1-|q\cap {\tau}|}{|\tau|+1-|q\cap {\tau}|}\prod_{j=1}^{p+1} L_{q(j)}. 
		\end{align}	
    Hence, we need to prove that for any $0\leq j\leq\Lambda-1$ it holds that 
		\begin{align}
			A({p,\;\tau_j^\star}) > A({p,\;\tau_{j+1}^\star}).
		\end{align}

%%%%%%%%%%%%%%%%%%%%%%%%%%%%%%%%%%%%%%%%%%%%%%%%%%%%%%%%%%%%%%%%%%%%%%%%%%%
	Let us consider an arbitrary $j$, $0\leq j \leq \Lambda -1$. 
	We select a set~$\tau'$ with cardinality $j+1$ that includes $\tau_j^\star$, and we write $\tau'$ as $\tau' = \{\tau_j^\star \cup r\}$, where $r\in\{[\Lambda]\setminus \tau_j^\star\}$. 
	Then, it follows from~\eqref{eq:def_num_c} that		
		\begin{align}\label{eq:num_c_tau_prime}
			A({p,\tau'}) &= \sum_{q\in C_{p+1}^{[\Lambda]}}^{}  \frac{p+1-|q\cap \{\tau_j^\star \cup r\}|}{(j+1)+1-|q\cap \{\tau_j^\star \cup r\}|} \prod_{j=1}^{p+1} L_{q(j)}. 
		\end{align}			
	Note that $\prod_{j=1}^{p+1} L_{q(j)}$ is independent of~$\tau^\prime, \tau_j^\star$.
	Furthermore, it holds that 
		\begin{align}\label{eq:ineq_mplus}
			\frac{p+1-|q\cap \{\tau_j^\star \cup r\}|}{(j+1)+1-|q\cap \{\tau_j^\star \cup r\}|} 
				< \frac{p+1-|q\cap\tau_j^\star|}{j+1-|q\cap \tau_j^\star|} 
		\end{align}				
	for any $p\in\Lambda$, $r\in\{[\Lambda]\setminus \tau_j^\star\}$,  $q\in C_{p+1}^{[\Lambda]}$. 
	Merging~\eqref{eq:num_c_tau_prime} and~\eqref{eq:ineq_mplus} yields
		\begin{align}
			A({p,\tau'}) 
			& <  \sum_{q\in \Cb{[\Lambda]}{p+1}}^{} \frac{p+1-|q\cap\tau_j^\star|}{|\tau_j^\star|+1-|q\cap \tau_j^\star|} \prod_{j=1}^{p+1} L_{q(j)} %\\ &
			= A({p,\tau_j^\star}). 
		\end{align}				
	By definition, $A({p,\tau'}) \geq A({p,\tau_{j+1}^\star})$ for any $\tau'$ such that $|\tau'|=j+1$. Thus,
		\begin{align}
			A({p,\tau_{j+1}^\star}) \leq A({p,\tau'}) <  A({p,\tau_j^\star}), 
		\end{align}					
	which concludes the proof of Proposition~\ref{lem:monotone_c}. \qed

% %%%%%%%%%%%%%%%%%%%%%%%%%%%%%%%%%%%%%%%%%%%%%%%%%%%%%%%%
% \section{~~}
% %%%%%%%%%%%%%%%%%%%%%%%%%%%%%%%%%%%%%%%%%%%%%%%%%%%%%%%%
% \input{appendixA}

\ifCLASSOPTIONcaptionsoff
  \newpage
\fi

\bibliographystyle{IEEEtran}
\bibliography{IEEEabrv,my_bibliography2021}

\end{document}